\documentclass[%
 aip,
 amsmath,amssymb,
reprint
]{revtex4-1}

\usepackage{graphicx}
\usepackage{dcolumn}
\usepackage{bm}

\usepackage[utf8]{inputenc}
\usepackage[T1]{fontenc}
\usepackage{mathptmx}
\usepackage{etoolbox}

\usepackage{amsmath} 
\usepackage{amsfonts}  
\usepackage{mathrsfs} 
\usepackage{xifthen} 
\usepackage[]{xcolor}
\usepackage{hyperref}
\usepackage{microtype}
\usepackage{mathrsfs}
\usepackage{scalerel}

\DeclareMathAlphabet{\Mymathbb}{U}{bbold}{m}{n}
\DeclareMathAlphabet{\mathpzc}{OT1}{pzc}{m}{it}
\DeclareMathOperator*{\dprime}{\prime \prime}

\newtheorem{problem}{Problem}

\newcommand{\T}{\mbox{\tiny T}}

\newcommand{\timeSymbol}{t}
\newcommand{\Time}{\timeSymbol} 
\newcommand{\TimeF}{T} 

\newcommand{\FirstSymbol}{\mathcal{G}}
\newcommand{\First}[1][]{
  \ifthenelse{\equal{#1}{}}
  {\FirstSymbol}
  {\FirstSymbol_{\scriptscriptstyle{#1}}}
}

\newcommand{\ConstGeom}[1][]{
   \ifthenelse{\equal{#1}{}}       
   {K_{\First}\,}
   {K_{\First,#1}\,}
}

\newcommand{\FirstForm}[1][]{
  \ifthenelse{\equal{#1}{}}       
  {\operatorname{I_{\point}}}             
  {\operatorname{I_{#1}}}             
}
\newcommand{\GradSymbol}{\operatorname{\mathbf{\nabla}}}
\newcommand{\Grad}{\GradSymbol}
\newcommand{\Div}{\Grad\cdot}
\newcommand{\Lap}{\Delta}
\newcommand{\GradSurf}{\GradSymbol_{\SurfDomain}}

\newcommand{\DivSurf}{\GradSurf\cdot}

\newcommand{\DivP}{\operatorname{div}_{\!\ProjMat}\,}
\newcommand{\LapSurf}{\Lap_{\SurfDomain}}

\newcommand{\GradSurfConv}[1]{\GradSymbol_{#1}}

\newcommand{\nablaP}{\nabla_{\boldsymbol{P}}}
\newcommand{\divP}{\operatorname{div}_{\boldsymbol{P}}}
\newcommand{\divS}{\operatorname{div}_{\cal{S}}}

\newcommand{\Der}[1][]
{
  \ifthenelse{\equal{#1}{}}
  {\partial}
  {\partial_{\scriptscriptstyle{#1}}}
}
\newcommand{\DerT}{\Der_{\Time}\,}

\newcommand{\DerTot}[2][t]
{
  \ifthenelse{\equal{#2}{}}
  {\frac{d #2}{dt}}
  {\frac{d #2}{d #1}}
}
\newcommand{\Derd}[2][]{
  \ifthenelse{\equal{#1}{}}
  {\Diffsymbol{#2}}
  {\Diffsymbol_{{#1}}{#2}}
}
\newcommand{\Diffsymbol}{\operatorname{d}\!}
\newcommand{\Diff}[2][]{
  \ifthenelse{\equal{#1}{}}
  {\Diffsymbol{#2}}
  {\Diffsymbol{#2}_{{#1}}}
}
\newcommand{\DirDerSymb}{D}
\newcommand{\DirDer}[2][]
{
  \ifthenelse{\equal{#1}{}}
  {\DirDerSymb^{#2}}
  {\DirDerSymb^{#2}_{#1}}
}

\newcommand{\REALsymbol}{\mathbb{R}}
\newcommand{\REAL}[1][]{
  \ifthenelse{\equal{#1}{}}
  {\REALsymbol}
  {{\REALsymbol}^{#1}}
}
\newcommand{\EUCLsymbol}{\mathbb E}
\newcommand{\EUCL}[1][]{
  \ifthenelse{\equal{#1}{}}
  {\EUCLsymbol}
  {{\EUCLsymbol}^{#1}}
}
\newcommand{\NATURALsymbol}{\mathbb N}
\newcommand{\NATURAL}[1][]{
  \ifthenelse{\equal{#1}{}}
  {\NATURALsymbol}
  {{\NATURALsymbol}^{#1}}
}

\newcommand{\ABS}[2][]
{
  \ifthenelse{\equal{#1}{}}
  {\left| #2 \right|}
  {\left| #2 \right|_{#1}}
}

\newcommand{\NORM}[2][]
{
  \ifthenelse{\equal{#1}{}}
  {\left\| #2 \right\|}
  {\left\| #2 \right\|_{#1}}
}
\newcommand{\vertiii}[1]{{\left\vert\kern-0.25ex\left\vert\kern-0.25ex\left\vert #1 \right\vert\kern-0.25ex\right\vert\kern-0.25ex\right\vert}}
\newcommand{\BrNORM}[2][]
{
  \ifthenelse{\equal{#1}{}}
  {\vertiii{#2}}
  {\vertiii{#2}_{#1}}
}

\newcommand{\SCAL}[3][]
{
  \ifthenelse{\equal{#1}{}}
  {\left\langle{#2},{#3}\right\rangle}
  {\left\langle{#2},{#3}\right\rangle_{#1}}
}
\newcommand{\SCALF}[3][]
{
  \ifthenelse{\equal{#1}{}}
  {\left({#2},{#3}\right)}
  {\left({#2},{#3}\right)_{#1}}
}

\newcommand{\scalprodSurf}[3][]
{
  \ifthenelse{\equal{#1}{}}
  {\left\langle {#2},{#3} \right\rangle_{\scriptscriptstyle\SurfDomain}}
  {\left\langle {#2},{#3} \right\rangle_{{#1}}}
}

\newcommand{\InnerApprox}[2]{\left({#1}\,,\,{#2}\right)_{\SurfDomain[\meshparam]}}

\newcommand{\point}[1][]
{
  \ifthenelse{\equal{#1}{}}
  {\mathbf{p}}
  {\mathbf{p}_{#1}}
}

\newcommand{\midPoint}{\mathbf{m}}

\newcommand{\From}{:}

\newcommand{\RegionSymb}{R}
\newcommand{\Region}[1][]
{
  \ifthenelse{\equal{#1}{}}
  {\RegionSymb}
  {\RegionSymb_{#1}}
}

\newcommand{\IT}{[0,\TimeF]} 
\newcommand{\Interval}[1][]
{
  \ifthenelse{\equal{#1}{}}
  {I}
  {I_{#1}}
}

\newcommand{\SurfDomainsymb}{\mathcal{S}}
\newcommand{\SurfDomain}[1][]{
  \ifthenelse{\equal{#1}{}}
  {\SurfDomainsymb}
  {{\SurfDomainsymb_{#1}}}
}
\newcommand{\tSurfDomain}[1][]{
  \ifthenelse{\equal{#1}{}}
  {\tilde{\SurfDomainsymb}}
  {\tilde{\SurfDomainsymb}_{#1}}
}
\newcommand{\SurfDomainBndsymb}{\partial\Gamma}
\newcommand{\SurfDomainBnd}[1][]{
  \ifthenelse{\equal{#1}{}}
  {\SurfDomainBndsymb}
  {\SurfDomainBndsymb_{#1}}
}
\newcommand{\ClosedSurfDomain}[1][]{
  \ifthenelse{\equal{#1}{}}
  {\Closedsymb{\SurfDomain}}
  {\Closedsymb{\SurfDomain[#1]}}
}

\newcommand{\heightsymb}{\mathcal{H}}
\newcommand{\height}[1][]{
  \ifthenelse{\equal{#1}{}}
  {\heightsymb}
  {\heightsymb_{#1}}
}

\newcommand{\BSMsymbol}{\mathcal{B}}

\newcommand{\BSM}[1][]
{
  \ifthenelse{\equal{#1}{}}
  {\BSMsymbol}
  {\BSMsymbol_{#1}}
}

\newcommand{\Surf}{\mathcal{S}}

\newcommand{\SurfBndsymb}{\partial\Surf}
\newcommand{\SurfBnd}[1][]{
  \ifthenelse{\equal{#1}{}}
  {\SurfBndsymb}
  {\SurfBndsymb_{#1}}
}
\newcommand{\Closedsymb}[1]{\bar{#1}}
\newcommand{\ClosedSurf}[1][]{
  \ifthenelse{\equal{#1}{}}
  {\Closedsymb{\Surf}}
  {\Closedsymb{\Surf[#1]}}
}

\newcommand{\Vector}[1]{\mathbf{#1}}

\newcommand{\press}{p}
\newcommand{\tpress}{\tilde{\press}}

\newcommand{\tension}{\sigma}
\newcommand{\ttension}{\tilde{\tension}}
\newcommand{\densitySymb}{\rho}
\newcommand{\density}[1][]{
  \ifthenelse{\equal{#1}{}}
  {\densitySymb}
  {\densitySymb_{\scriptscriptstyle{#1}}}
}

\newcommand{\concSymbol}{\phi}
\newcommand{\CHsol}{\concSymbol}
\newcommand{\CHsolApprox}{\CHsol_{\meshparam}}
\newcommand{\pressApprox}{\press_{\meshparam}}

\newcommand{\dWellSymbol}{W}
\newcommand{\dWell}{\dWellSymbol}
\newcommand{\mobilitySymbol}{m}
\newcommand{\mobility}{\mobilitySymbol}

\newcommand{\chempot}{\mu}
\newcommand{\chempotApprox}{\mu_{\meshparam}}

\newcommand{\bendingSymb}{\kappa}
\newcommand{\bendStiff}[1][]{
  \ifthenelse{\equal{#1}{}}
  {\bendingSymb}
  {\bendingSymb_{#1}}
}
\newcommand{\bendStiffGauss}[1][]{
  \ifthenelse{\equal{#1}{}}
  {\overline{\bendingSymb}}
  {\overline{\bendingSymb}_{#1}}
}

\newcommand{\Reynolds}{\operatorname{Re}}

\newcommand{\AspRatio}{\epsilon}

\newcommand{\interfaceparam}{\AspRatio}

\newcommand{\param}{\boldsymbol{X}}

\newcommand{\MapUsymb}{\phi}
\newcommand{\MapU}[1][]
{
  \ifthenelse{\equal{#1}{}}
    {\MapUsymb}
    {\MapUsymb_{#1}}
}
\newcommand{\MapLinsymb}{F}
\newcommand{\MapLin}[1][]
{
  \ifthenelse{\equal{#1}{}}
  {\ensuremath{\MapLinsymb}}
  {\ensuremath{\MapLinsymb_{#1}}}
}

\newcommand{\MapVsymb}{\psi}
\newcommand{\MapV}[1][]
{
  \ifthenelse{\equal{#1}{}}
    {\MapVsymb}
    {\MapVsymb_{#1}}
}
\newcommand{\Transsymb}{\Phi}
\newcommand{\Trans}[1][]
{
  \ifthenelse{\equal{#1}{}}
    {\Transsymb} 
    {\Transsymb_{\scriptscriptstyle{#1}}}
}
\newcommand{\InvMapsymb}{\Psi}
\newcommand{\InvMap}[1][]
{
  \ifthenelse{\equal{#1}{}}
    {\InvMapsymb}
    {\InvMapsymb_{\scriptscriptstyle{#1}}}
}

\newcommand{\fsymb}{f}
\newcommand{\scalFun}[1][]
{
  \ifthenelse{\equal{#1}{}}
  {\fsymb}
  {\fsymb_{#1}}
}
\newcommand{\tscalFun}[1][]
{
  \ifthenelse{\equal{#1}{}}
  {\tilde{\fsymb}}
  {\tilde{\fsymb}_{#1}}
}
\newcommand{\bscalFun}[1][]
{
  \ifthenelse{\equal{#1}{}}
  {\bar{\fsymb}}
  {\bar{\fsymb}_{#1}}
}
\newcommand{\gsymb}{g}
\newcommand{\scalFung}[1][]
{
  \ifthenelse{\equal{#1}{}}
  {\gsymb}
  {\gsymb_{#1}}
}
\newcommand{\Fsymb}{F}
\newcommand{\FvecFun}[1][]
{
  \ifthenelse{\equal{#1}{}}
  {\Fsymb}
  {\Fsymb_{#1}}
}
\newcommand{\tFvecFun}[1][]
{
  \ifthenelse{\equal{#1}{}}
  {\tilde{\Fsymb}}
  {\tilde{\Fsymb}_{#1}}
}
\newcommand{\Ffunc}[2][]
{
  \ifthenelse{\equal{#1}{}}
  {\ensuremath{\Fsymb_{#2}}}
  {\ensuremath{\Fsymb^{#1}_{#2}}}
}
\newcommand{\Gsymb}{g}
\newcommand{\Gfun}[1][]
{
  \ifthenelse{\equal{#1}{}}
  {\ensuremath{\Gsymb}}
  {\ensuremath{\Gsymb_{{#1}}}}
}
\newcommand{\bGfun}[1][]
{
  \ifthenelse{\equal{#1}{}}
  {\ensuremath{\bar{\Gsymb}}}
  {\ensuremath{\bar{\Gsymb}_{{#1}}}}
}

\newcommand{\PrincipalK}[1][]
{
  \ifthenelse{\equal{#1}{}}
  {k}
  {k_{#1}}
}

\newcommand{\vecsymb}{u}
\newcommand{\vecFun}[1][]{
  \ifthenelse{\equal{#1}{}}
  {\Vector{\vecsymb}}
  {\vecsymb^{#1}}
}
\newcommand{\tvecFun}[1][]{
  \ifthenelse{\equal{#1}{}}
  {\tilde{\vecsymb}}
  {\tilde{\vecsymb}_{#1}}
}

\newcommand{\vvsymb}{v}
\newcommand{\vv}[1][]{
  \ifthenelse{\equal{#1}{}}
  {\mathbf{\vvsymb}}
  {\vvsymb^{#1}}
}
\newcommand{\wwsymb}{w}
\newcommand{\ww}[1][]
{
  \ifthenelse{\equal{#1}{}}
  {\mathbf{\wwsymb}}
  {\wwsymb^{#1}}
}
\newcommand{\uusymb}{u}
\newcommand{\uu}[1][]
{
  \ifthenelse{\equal{#1}{}}
  {\mathbf{\uusymb}}
  {\uusymb^{#1}}
}

\newcommand{\VecFieldSymbol}{X}
\newcommand{\VecField}[1][]
{
  \ifthenelse{\equal{#1}{}}
  {\VecFieldSymbol}
  {\VecFieldSymbol^{#1}}
}
\newcommand{\VecFieldYSymbol}{Y}
\newcommand{\VecFieldY}[1][]
{
  \ifthenelse{\equal{#1}{}}
  {\VecFieldYSymbol}
  {\VecFieldYSymbol^{#1}}
}

\newcommand{\xvsymb}{x}
\newcommand{\xv}[1][]
{
  \ifthenelse{\equal{#1}{}}
  {\mathbf{\xvsymb}}
  {\mathbf{\xvsymb}_{\scriptscriptstyle{#1}}}
}
\newcommand{\xvcomp}[1][]{
  \ifthenelse{\equal{#1}{}}
  {\xvsymb}
  {\xvsymb^{\scriptscriptstyle{#1}}}
}
\newcommand{\xcg}[1][]{
  \ifthenelse{\equal{#1}{}}
  {\xvcomp[1]}
  {\xvcomp[1]_{\scriptscriptstyle{#1}}}
}
\newcommand{\ycg}[1][]{
  \ifthenelse{\equal{#1}{}}
  {\xvcomp[2]}
  {\xvcomp[2]_{\scriptscriptstyle{#1}}}
}
\newcommand{\zcg}[1][]{
  \ifthenelse{\equal{#1}{}}
  {\xvcomp[3]}
  {\xvcomp[3]_{\scriptscriptstyle{#1}}}
}

\newcommand{\svsymb}{s}
\newcommand{\sv}[1][]{
  \ifthenelse{\equal{#1}{}}
  {\mathbf{\svsymb}}
  {\mathbf{\svsymb}_{\scriptscriptstyle{#1}}}
}
\newcommand{\svcomp}[1][]
{
  \ifthenelse{\equal{#1}{}}
  {\svsymb}
  {\svsymb^{\scriptscriptstyle{#1}}}
}
\newcommand{\xcl}[1][]{
  \ifthenelse{\equal{#1}{}}
   {\svcomp[1]}
   {\svcomp[1]_{\scriptscriptstyle{#1}}}
}
\newcommand{\ycl}[1][]{
  \ifthenelse{\equal{#1}{}}
   {\svcomp[2]}
   {\svcomp[2]_{\scriptscriptstyle{#1}}}
}
\newcommand{\zcl}[1][]{
  \ifthenelse{\equal{#1}{}}
   {\svcomp[3]}
   {\svcomp[3]_{\scriptscriptstyle{#1}}}
}

\newcommand{\ProjSymb}{\operatorname{\pi}}
\newcommand{\ProjFun}[2][]{
  \ifthenelse{\equal{#1}{}}
  {\ProjSymb\left(#2\right)}
  {\ProjSymb_{\scriptscriptstyle{#1}}\left(#2\right)}
}
\newcommand{\Prm}[1][]
{
  \ifthenelse{\equal{#1}{}}
  {\operatorname{pr}}
  {\operatorname{pr}_{\scriptscriptstyle{#1}}}
}
\newcommand{\TanPlane}[2][]
{
  \ifthenelse{\equal{#1}{}}
  {T_{\scriptscriptstyle{\point}}#2}
  {T_{\scriptscriptstyle{#1}}#2}
}

\newcommand{\SubsetSymbol}{\mathcal{U}}

\newcommand{\SubsetU}[1][]
{
  \ifthenelse{\equal{#1}{}}
  {{U}}
  {{U}_{#1}}
}
\newcommand{\SubsetV}[1][]
{
  \ifthenelse{\equal{#1}{}}
  {{V}}
  {{V}_{#1}}
}
\newcommand{\SubsetW}[1][]
{
  \ifthenelse{\equal{#1}{}}
  {{W}}
  {{W}_{#1}}
}
\newcommand{\NeighSymbol}{\mathcal{N}}
\newcommand{\Neigh}[1][]
{
  \ifthenelse{\equal{#1}{}}
  {\NeighSymbol_{\point}}
  {\NeighSymbol_{#1}}
}
\newcommand{\NeighSurf}[1][]
{
  \ifthenelse{\equal{#1}{}}
  {\SubsetSymbol_{\point}}
  {\SubsetSymbol_{#1}}
}
\newcommand{\NormSymb}{\boldsymbol{\nu}}
\newcommand{\normalvec}{\NormSymb}
\newcommand{\normalvecApprox}{\NormSymb_{\meshparam}}
\newcommand{\normalSurf}[1][]
{
  \ifthenelse{\equal{#1}{}}
  {\NormSymb}
  {\NormSymb(#1)}
}
\newcommand{\normalInterp}[1][]
{
  \ifthenelse{\equal{#1}{}}
   {\tilde{\NormSymb}}
   {\tilde{\NormSymb}_{\scriptscriptstyle{#1}}}
}

\newcommand{\normalEdge}{\mathbf{\nu}} 
\newcommand{\basisCC}{t}
\newcommand{\basisGC}{e}
\newcommand{\vecBaseGC}[1][]
{
  \ifthenelse{\equal{#1}{}}
  {\mathbf{\basisGC}}
  {\mathbf{\basisGC}_{#1}}
}
\newcommand{\vecBasePhys}[1][]
{
  \ifthenelse{\equal{#1}{}}
  {\mathbf{\basisGC}}
  {\mathbf{\basisGC}_{#1}}
}

\newcommand{\vecBaseCCcv}[1][]
{
  \ifthenelse{\equal{#1}{}}
  {\mathbf{\basisCC}}
  {\mathbf{\basisCC}_{#1}}
}

\newcommand{\tvecBaseCCcv}[1][]
{
  \ifthenelse{\equal{#1}{}}
  {\tilde{\mathbf{\basisCC}}}
  {\tilde{\mathbf{\basisCC}}_{#1}}
}
\newcommand{\hvecBaseCCcv}[1][]
{
  \ifthenelse{\equal{#1}{}}
  {\hat{\mathbf{\basisCC}}}
  {\hat{\mathbf{\basisCC}}_{#1}}
}
\newcommand{\vecBaseCCctrv}[1][]
{
  \ifthenelse{\equal{#1}{}}
  {\mathbf{\basisCC}}
  {\mathbf{\basisCC}^{#1}}
}

\newcommand{\first}[1]{
  \IfEqCase{#1}{
    {1}{\operatorname{E}}
    {2}{\operatorname{F}}
    {3}{\operatorname{G}}
  }
  [\PackageError{first}{Undefined option to first: #1}{}]%
}
\newcommand{\SecondFormSymbol}{\ensuremath{\operatorname{II}}}
\newcommand{\SecondForm}[1][]
{
  \ifthenelse{\equal{#1}{}}
  {\SecondFormSymbol_{\point}}
  {\SecondFormSymbol_{#1}}
}
\newcommand{\second}[1]{
  \IfEqCase{#1}{
    {1}{\operatorname{e}}
    {2}{\operatorname{f}}
    {3}{\operatorname{g}}
  }
  [\PackageError{first}{Undefined option to first: #1}{}]
}
\newcommand{\WeigSymbol}{\mathcal{W}}
\newcommand{\Weig}[1][]
{
  \ifthenelse{\equal{#1}{}}
  {\WeigSymbol}
  {\WeigSymbol_{#1}}
}

\newcommand{\velSymbol}{\boldsymbol{u}}
\newcommand{\vectvel}[1][]
{
   \ifthenelse{\equal{#1}{}}
   {\mathbf{\velSymbol}}
   {\mathbf{\velSymbol}(#1)}
}
\newcommand{\vectvelApprox}[1][]
{
   \ifthenelse{\equal{#1}{}}
   {\mathbf{\velSymbol}_{\meshparam}}
   {\mathbf{\velSymbol}_{\meshparam}(#1)}
}

\newcommand{\velcompContr}[2][i]
{
   \ifthenelse{\equal{#2}{}}
   {\velSymbol^{#1}}
   {\velSymbol^{#1}(#2)}}
\newcommand{\velcompPhys}[2][i]
{
   \ifthenelse{\equal{#2}{}}
   {\velSymbol_{(#1)}}
   {\velSymbol_{(#1)}(#2)}}

\newcommand{\velSymbolRP}{v}
\newcommand{\velcompContrRP}[2][i]
{
   \ifthenelse{\equal{#2}{}}
   {\velSymbolRP^{#1}}
   {\velSymbolRP^{#1}(#2)}}
\newcommand{\velRP}[1][]
{
  \ifthenelse{\equal{#1}{}}
  {\velSymbolRP}
  {\velSymbolRP_{#1}}
}

\newcommand{\velcompApprox}[2][i]
{
   \ifthenelse{\equal{#2}{}}
   {\velSymbol^{#1)}}
   {\velSymbol^{#1}_{(#2)}}
}

\newcommand{\coord}{\mathbf{x}}
\newcommand{\coordC}{x}

\newcommand{\VelSymbol}{U}
\newcommand{\vectVel}[1][]
{
   \ifthenelse{\equal{#1}{}}
   {\vec{\VelSymbol}}
   {\vec{\VelSymbol}(#1)}
}
\newcommand{\Velcomp}[2][i]
{
   \ifthenelse{\equal{#2}{}}
   {\VelSymbol^{#1}}
   {\VelSymbol^{#1}(#2)}
}
\newcommand{\VprimoSymbol}{\tilde{u}}
\newcommand{\Vprimo}[1][]
{
   \ifthenelse{\equal{#1}{}}
   {\VprimoSymbol}
   {\VprimoSymbol(#1)}
}
\newcommand{\VprimoComp}[2][i]
{
   \ifthenelse{\equal{#2}{}}
   {\VprimoSymbol^{#1}}
   {\VprimoSymbol^{#1}(#2)}
}
\newcommand{\ttvelSymbol}{\tilde{\Mymathbb{u}}}
\newcommand{\ttvel}[1][]
{
   \ifthenelse{\equal{#1}{}}
   {\mathbf{\ttvelSymbol}}
   {\mathbf{\ttvelSymbol}(#1)}
}
\newcommand{\ttvelComp}[2][i]
{
   \ifthenelse{\equal{#2}{}}
   {\ttvelSymbol^{#1}}
   {\ttvelSymbol^{#1}(#2)}
}

\newcommand{\MatAlphaSymbol}{\mathbb{A}}
\newcommand{\MatAlpha}[1][]{%
  \ifthenelse{\equal{#1}{}}
  {\MatAlphaSymbol}
  {\MatAlphaSymbol_{#1}}
}

\newcommand{\QSymbol}{q}
\newcommand{\Qdisch}[1][]
{
   \ifthenelse{\equal{#1}{}}
   {\mathbf{\QSymbol}}
   {\mathbf{\QSymbol}(#1)}
}
\newcommand{\Qcomp}[2][i]
{
   \ifthenelse{\equal{#2}{}}
   {\QSymbol^{#1}}
   {\QSymbol^{#1}(#2)}
}
\newcommand{\Qvect}[1][]
{
   \ifthenelse{\equal{#1}{}}
   {\mathbf{\QSymbol}}
   {\mathbf{\QSymbol}(#1)}
}
\newcommand{\FricSymbol}{f}
\newcommand{\vectFric}[1][]
{
   \ifthenelse{\equal{#1}{}}
   {\mathbf{\FricSymbol}}
   {\mathbf{\FricSymbol}_{\scriptscriptstyle{#1}}}
}
\newcommand{\Friccomp}[2][i]
{
   \ifthenelse{\equal{#2}{}}
   {\FricSymbol_{#1}}
   {\FricSymbol_{#1}(#2)}}
\newcommand{\BFsymbol}{\tau}
\newcommand{\BottomFriction}[1][]{
  \ifthenelse{\equal{#1}{}}
  {\BFsymbol_{b}}
  {\BFsymbol_{b}^{#1}}
}

\newcommand{\ProjMatSymb}{\boldsymbol{P}}
\newcommand{\ProjMat}[1][]{
  \ifthenelse{\equal{#1}{}}
  {\ProjMatSymb}
  {\ProjMatSymb_{#1}}
}
\newcommand{\IDSymbol}{\boldsymbol{I}}
\newcommand{\IDtens}[1][]{
  \ifthenelse{\equal{#1}{}}
  {\IDSymbol}
  {\IDSymbol(#1)}
}
\newcommand{\IDMat}{\IDSymbol}

\newcommand{\tensSymbol}{\boldsymbol{T}}
\newcommand{\tenscompSymbol}{\tau}
\newcommand{\tens}[1][]{
  \ifthenelse{\equal{#1}{}}
  {\tensSymbol}
  {\tensSymbol(#1)}
}
\newcommand{\tenscomp}[2][ij]
{
  \ifthenelse{\equal{#2}{}}
  {\tenscompSymbol^{#1}}
  {\tenscompSymbol^{#1}(#2)}
}
\newcommand{\tensrow}[2][i]
{
  \ifthenelse{\equal{#2}{}}
  {\tensSymbol^{(#1)}}
  {\tensSymbol^{(#1)}(#2)}
}
\newcommand{\TensSymbol}{\mathbf{T}}
\newcommand{\Tens}[1][]{
  \ifthenelse{\equal{#1}{}}
  {\TensSymbol}
  {\TensSymbol_{#1}}
}

\newcommand{\TensCompSymbol}{\TensSymbol}
\newcommand{\TensComp}[2][ij]
{
  \ifthenelse{\equal{#2}{}}
  {\TensCompSymbol^{#1}}
  {\TensCompSymbol^{#1}(#2)}
}

\newcommand{\tensPrimoSymbol}{\tilde{\mathbf{\tau}}}
\newcommand{\tensPrimo}[1][]{
  \ifthenelse{\equal{#1}{}}
  {\tensPrimoSymbol}
  {\tensPrimoSymbol(#1)}
}
\newcommand{\tensPrimoCompSymbol}{\tensPrimoSymbol}
\newcommand{\tensPrimoComp}[2][ij]
{
  \ifthenelse{\equal{#2}{}}
  {\tensPrimoCompSymbol^{#1}}
  {\tensPrimoCompSymbol^{#1}(#2)}
}
\newcommand{\MCxl}[1][]{\ifthenelse{\equal{#1}{}}{h_{(1)}}{h_{(1),#1}}} 
\newcommand{\MCyl}[1][]{\ifthenelse{\equal{#1}{}}{h_{(2)}}{h_{(2),#1}}} 
\newcommand{\MCzl}[1][]{\ifthenelse{\equal{#1}{}}{h_{(3)}}{h_{(3),#1}}}

\newcommand{\MPsymb}{h}

\newcommand{\smallestH}[1][]
{
  \ifthenelse{\equal{#1}{}}
    {{l}}
    {{l}_{#1}}
}
\newcommand{\meshparam}[1][]
{
  \ifthenelse{\equal{#1}{}}
    {\MPsymb}
    {\MPsymb_{\scriptscriptstyle{#1}}}
}
\newcommand{\InradiusSymbol}{r}
\newcommand{\Inradius}[1][]
{
  \ifthenelse{\equal{#1}{}}
  {\InradiusSymbol}
  {\InradiusSymbol_{\scriptscriptstyle{#1}}}
}
\newcommand{\Tsymb}{\mathcal{T}}
\newcommand{\Triang}[1][]
{
  \ifthenelse{\equal{#1}{}}
    {\Tsymb}
    {\Tsymb_{#1}}
}
\newcommand{\TriangH}[1][]
{
  \ifthenelse{\equal{#1}{}}
    {\Tsymb_{\meshparam}}
    {\Tsymb_{#1}}
}
\renewcommand{\Triang}{\TriangH}
\newcommand{\Edgesymb}{\sigma}
\newcommand{\Edge}[1][]{
  \ifthenelse{\equal{#1}{}}
    {\Edgesymb}
    {\Edgesymb_{#1}}
}
\newcommand{\EdgeH}[1][]{
  \ifthenelse{\equal{#1}{}}
    {\Edgesymb_{\meshparam}}
    {\Edgesymb_{\meshparam,#1}}
}
\newcommand{\NEdge}[1][]{
  \ifthenelse{\equal{#1}{}}
    {N_{\Edgesymb}}
    {N_{\Edgesymb({#1})}}
}
\newcommand{\Cellsymb}{T}
\newcommand{\Cell}[1][]{
  \ifthenelse{\equal{#1}{}}
    {\Cellsymb}
    {\Cellsymb_{#1}}
}
\newcommand{\tCell}[1][]{
  \ifthenelse{\equal{#1}{}}
    {\tilde{\Cellsymb}}
    {\tilde{\Cellsymb}_{#1}}
}
\newcommand{\CellH}[1][]{
  \ifthenelse{\equal{#1}{}}
    {\Cellsymb_{\meshparam}}
    {\Cellsymb_{\meshparam,#1}}
}
\newcommand{\areaSymb}{\mathcal{A}}
\newcommand{\CellArea}[1][]
{
  \ifthenelse{\equal{#1}{}}
    {\areaSymb_{\Cell}}
    {\areaSymb_{#1}}
}
\newcommand{\CellHArea}[1][]
{
  \ifthenelse{\equal{#1}{}}
    {\areaSymb_{\CellH}}
    {\areaSymb_{\meshparam,#1}}
}

\newcommand{\NCell}[1][]{
  \ifthenelse{\equal{#1}{}}
    {N_{\Cellsymb}}
    {N_{\Cellsymb({#1})}}
}

\newcommand{\lengthSymb}{l}
\newcommand{\interfaceLength}{\lengthSymb}
\newcommand{\edgeLength}[1][]
{
  \ifthenelse{\equal{#1}{}}
  {\lengthSymb_{\Edge}}
  {\lengthSymb_{#1}}
}
 
\newcommand{\edgeHLength}[1][]
{
  \ifthenelse{\equal{#1}{}}
  {\lengthSymb_{\EdgeH}}
  {\lengthSymb_{\meshparam,#1}}
}

\newcommand{\Sourcesymb}{\mathbf{S}}
\newcommand{\Source}[1][]
{
  \ifthenelse{\equal{#1}{}}
    {\Sourcesymb}
    {\Sourcesymb_{#1}}
}

\newcommand{\SourceEdge}[1][]{
  \ifthenelse{\equal{#1}{}}
    {\Sourcesymb_{ij}}
    {\Sourcesymb_{#1}}
}

\newcommand{\FluxEdgesymb}{\mathbf{F}}
\newcommand{\FluxEdge}[1][]{
  \ifthenelse{\equal{#1}{}}
    {\FluxEdgesymb_{ij}}
    {\FluxEdgesymb_{#1}}
}

\newcommand{\numFlux}[1][]
{
  \ifthenelse{\equal{#1}{}}
  {\tilde{\fsymb}}
  {\tilde{\fsymb}_{#1}}
}

\newcommand{\FluxFuncNormSymbol}{\mathbf{F}}
\newcommand{\FluxFuncNorm}[1][]{
  \ifthenelse{\equal{#1}{}}
    {\FluxFuncNormSymbol^{\normalEdge}}
    {\FluxFuncNormSymbol^{\normalEdge}_{#1}}    
}

\newcommand{\JacobianSymbol}{\mathbf{A}}
\newcommand{\Jacobian}[1][]{
  \ifthenelse{\equal{#1}{}}
    {\JacobianSymbol}
    {\JacobianSymbol_{#1}}    
}
\newcommand{\EValSymbol}{\lambda}
\newcommand{\EVal}[1][]{
  \ifthenelse{\equal{#1}{}}
  {\EValSymbol}
  {\EValSymbol_{#1}}
}
\newcommand{\EVecSymbol}{\mathbf{r}}
\newcommand{\EVec}[2][]{
  \ifthenelse{\equal{#1}{}}
  {\EVecSymbol^{(#2)}}
  {\EVecSymbol^{(#2)}_{#1}}
}

\newcommand{\midPointEdge}[1][]
{
  \ifthenelse{\equal{#1}{}}
  {\midPoint_{\scriptscriptstyle\Edge}}
  {\midPoint_{\scriptscriptstyle\Edge[#1]}}
}
\newcommand{\gpPointEdgeDG}[1][]
{
  \ifthenelse{\equal{#1}{}}
  {\point_{\scriptscriptstyle\Edge}}
  {\point_{\scriptscriptstyle\Edge,#1}}
}
\newcommand{\midPointCell}[1][]
{
  \ifthenelse{\equal{#1}{}}
  {\midPoint_{\scriptscriptstyle\Cell}}
  {\midPoint_{\scriptscriptstyle\Cell[#1]}}
}

\newcommand{\energySymbol}{\mathcal{F}}
\newcommand{\energy}[1][]
{
  \ifthenelse{\equal{#1}{}}
  {\energySymbol}
  {\energySymbol_{#1}}
}
\newcommand{\energyPF}{\energy[PF]}

\newcommand{\speedRS}[2][]
{
  \ifthenelse{\equal{#2}{}}
  {S_{#2}}
  {S_{#2}^{#1}}
}
\newcommand{\Interpolant}{I_{\meshparam}}

\newcommand{\ContSymbol}{C}
\newcommand{\Cont}[1][]{
  \ifthenelse{\equal{#1}{}}
  {\ContSymbol^{0}}
  {\ContSymbol^{#1}}
}
\newcommand{\Cinf}[1][]{
  \ifthenelse{\equal{#1}{}}
  {\ContSymbol^{\infty}}
  {\ContSymbol^{\infty}(#1)}
}
\newcommand{\SobSymbol}{W}
\newcommand{\HilbSymbol}{H}
\newcommand{\Hilb}[1][]{
  \ifthenelse{\equal{#1}{}}
  {\HilbSymbol^{1}}
  {\SobSymbol^{1}_{#1}}  
}
\newcommand{\Sob}[2][]{
  \ifthenelse{\equal{#1}{}}
  {\HilbSymbol^{#2}}
  {\SobSymbol^{#2,#1}}  
}

\newcommand{\LspaceSymb}{L}
\newcommand{\Lspace}[1][]{
  \ifthenelse{\equal{#1}{}}
  {\LspaceSymb^{2}}
  {\LspaceSymb^{#1}}  
}

\newcommand{\TestSpSymbol}{{V}}
\newcommand{\TestSpace}[1][]{
  \ifthenelse{\equal{#1}{}}
  {\TestSpSymbol({\SurfDomain})}
  {\TestSpSymbol_{#1}({\SurfDomain})}
}

\newcommand{\TestSpaceEmbedded}[1][]{
  \ifthenelse{\equal{#1}{}}
  {\TestSpSymbol(\TriangH(\SurfDomain))}
  {\TestSpSymbol_{#1}(\TriangH(\SurfDomain))}
}
\newcommand{\TestSpaceIntrinsic}[1][]{
  \ifthenelse{\equal{#1}{}}
  {\TestSpSymbol(\Triang(\SurfDomain))}
  {\TestSpSymbol_{#1}(\Triang(\SurfDomain))}
}
\newcommand{\TestSpaceChart}[1][]{
  \ifthenelse{\equal{#1}{}}
  {\TestSpSymbol(\Triang(\SubsetU))}
  {\TestSpSymbol_{#1}(\Triang(\SubsetU))}
}
\newcommand{\TestSpaceCell}[1][]{
  \ifthenelse{\equal{#1}{}}
  {\TestSpSymbol_{\meshparam}(\Cell)}
  {\TestSpSymbol_{\meshparam}(\Cell_{#1})}
}

\newcommand{\TestSpaceApprox}[1][]{
  \ifthenelse{\equal{#1}{}}
  {\TestSpSymbol_{\meshparam}}
  {\TestSpSymbol_{\meshparam}(#1)}
}
\newcommand{\TestSpaceVec}[1][]{
  \ifthenelse{\equal{#1}{}}
  {\mathbf{\TestSpSymbol}_{\meshparam}}
  {\mathbf{\TestSpSymbol}_{\meshparam}(#1)}
}

\newcommand{\TestSpGammaSymbol}{{P}}
\newcommand{\TestSpaceGamma}[1][]{
  \ifthenelse{\equal{#1}{}}
  {\TestSpGammaSymbol_{\meshparam}}
  {\TestSpGammaSymbol_{\meshparam}(#1)}
}

\newcommand{\TestSpCHSymbol}{{W}}
\newcommand{\TestSpaceCH}[1][]{
  \ifthenelse{\equal{#1}{}}
  {\TestSpCHSymbol_{\meshparam}}
  {\TestSpCHSymbol_{\meshparam}(#1)}
}

\newcommand{\TestSpPressSymbol}{Q}
\newcommand{\TestSpacePress}[1][]{
  \ifthenelse{\equal{#1}{}}
  {\TestSpPressSymbol_{\meshparam}}
  {\TestSpPressSymbol_{\meshparam}(#1)}
}

\newcommand{\PolySymb}{\mathcal{P}}
\newcommand{\PC}[1]{\PolySymb{_{#1}}}

\newcommand{\GaussCurv}{\mathcal{K}}
\newcommand{\calH}{\mathcal{H}}
\newcommand{\meanCurv}[1][]
{
  \ifthenelse{\equal{#1}{}}
  {\calH}
  {\calH_{#1}}
}
\newcommand{\meanCurvApprox}{\meanCurv[\meshparam]}
\newcommand{\shapeOp}{\mathcal{B}}
\newcommand{\AngleSymbol}{\theta}
\newcommand{\DevAngle}[1][]
{
  \ifthenelse{\equal{#1}{}}
  {\AngleSymbol}
  {\AngleSymbol_{\scriptscriptstyle{#1}}}
}
\newcommand{\relheightSymb}{\pi}
\newcommand{\relheight}[1][]
{
  \ifthenelse{\equal{#1}{}}
  {\relheightSymb_{\scriptscriptstyle{\SurfDomain}}}
  {\relheightSymb_{\scriptscriptstyle{#1}}}
}

\newcommand{\StressTens}{\boldsymbol{\sigma}}

\newcommand{\DiffEig}[1][]
{ \ifthenelse{\equal{#1}{}}
  {d}
  {d_{#1}}
}

\newcommand{\BilinearStiffSymbol}{a}
\newcommand{\BilinearStiff}[3][]
{
  \ifthenelse{\equal{#1}{}}
  {\BilinearStiffSymbol(#2,#3)}    
  {\BilinearStiffSymbol_{#1}(#2,#3)}    
}
\newcommand{\BilinearAdvSymbol}{b}
\newcommand{\BilinearAdv}[3][]
{
  \ifthenelse{\equal{#1}{}}
  {\BilinearAdvSymbol(#2,#3)}    
  {\BilinearAdvSymbol_{#1}(#2,#3)}    
}
\newcommand{\BilinearMassSymbol}{m}
\newcommand{\BilinearMass}[3][]
{
  \ifthenelse{\equal{#1}{}}
  {\BilinearMassSymbol(#2,#3)}    
  {\BilinearMassSymbol_{#1}(#2,#3)}    
}
\newcommand{\BilinearReactSymbol}{c}
\newcommand{\BilinearReact}[3][]
{
  \ifthenelse{\equal{#1}{}}
  {\BilinearReactSymbol(#2,#3)}    
  {\BilinearReactSymbol_{#1}(#2,#3)}    
}

\newcommand{\TestSymbol}{v}
\newcommand{\Test}[1][]
{
  \ifthenelse{\equal{#1}{}}
  {\TestSymbol}  
  {\TestSymbol_{\scriptscriptstyle{#1}}}
}

\newcommand{\nNodes}[1][]
{
  \ifthenelse{\equal{#1}{}}
  {N^{\scriptscriptstyle{dof}}}
  {N^{\scriptscriptstyle{dof}}_{\scriptscriptstyle{#1}}}
}

\newcommand{\TestApprox}[1][]
{
  \ifthenelse{\equal{#1}{}}
  {\TestSymbol_{\scriptscriptstyle{\meshparam}}}  
  {\TestSymbol_{\scriptscriptstyle{\meshparam,#1}}}
}

\newcommand{\bTestApprox}[1][]
{
  \ifthenelse{\equal{#1}{}}
  {\bar{\TestSymbol}_{\scriptscriptstyle{\meshparam}}}  
  {\bar{\TestSymbol}_{\scriptscriptstyle{\meshparam,#1}}}
}

\newcommand{\TestWSymbol}{w}
\newcommand{\TestW}[1][]
{
  \ifthenelse{\equal{#1}{}}
  {\TestWSymbol}  
  {\TestWSymbol_{\scriptscriptstyle{#1}}}
}

\newcommand{\TestWApprox}[1][]
{
  \ifthenelse{\equal{#1}{}}
  {\TestWSymbol_{\scriptscriptstyle{\meshparam}}}  
  {\TestWSymbol_{\scriptscriptstyle{\meshparam,#1}}}
}
\newcommand{\bTestWApprox}[1][]
{
  \ifthenelse{\equal{#1}{}}
  {\bar{\TestWSymbol}_{\scriptscriptstyle{\meshparam}}}  
  {\bar{\TestWSymbol}_{\scriptscriptstyle{\meshparam,#1}}}
}

\newcommand{\TestCHSymbol}{\psi}
\newcommand{\TestCHApprox}[1][]
{
  \ifthenelse{\equal{#1}{}}
  {\TestCHSymbol_{\scriptscriptstyle{\meshparam}}}  
  {\TestCHSymbol_{\scriptscriptstyle{\meshparam,#1}}}
}
\newcommand{\TestCH}[1][]
{
  \ifthenelse{\equal{#1}{}}
  {\TestCHSymbol}  
  {\TestCHSymbol_{\scriptscriptstyle{#1}}}
}
\newcommand{\TestCHmuSymbol}{\xi}
\newcommand{\TestCHmuApprox}[1][]
{
  \ifthenelse{\equal{#1}{}}
  {\TestCHmuSymbol_{\scriptscriptstyle{\meshparam}}}  
  {\TestCHmuSymbol_{\scriptscriptstyle{\meshparam,#1}}}
}

\newcommand{\TestVecSymbol}{\boldsymbol{v}}
\newcommand{\TestVecApprox}[1][]
{
  \ifthenelse{\equal{#1}{}}
  {\TestVecSymbol_{\scriptscriptstyle{\meshparam}}}  
  {\TestVecSymbol_{\scriptscriptstyle{\meshparam,#1}}}
}
\newcommand{\TestPressSymbol}{q}
\newcommand{\TestPress}[1][]
{
  \ifthenelse{\equal{#1}{}}
  {\TestPressSymbol}  
  {\TestPressSymbol_{\scriptscriptstyle{#1}}}
}
\newcommand{\TestPressApprox}[1][]
{
  \ifthenelse{\equal{#1}{}}
  {\TestPressSymbol_{\scriptscriptstyle{\meshparam}}}  
  {\TestPressSymbol_{\scriptscriptstyle{\meshparam,#1}}}
}

\newcommand{\viscositySymb}{\nu}
\newcommand{\viscosity}[1][]
{
  \ifthenelse{\equal{#1}{}}
  {\viscositySymb}
  {\viscositySymb_{\scriptscriptstyle{#1}}}
}

\newcommand{\ResidualSymbol}{R}
\newcommand{\Residual}[1][]
{
  \ifthenelse{\equal{#1}{}}
  {\ResidualSymbol}
  {\ResidualSymbol_{#1}}
}

\newcommand{\curvature}[1][]
{
  \ifthenelse{\equal{#1}{}}
  {\kappa}
  {\kappa_{#1}}
}

\newcommand{\QuadRule}[2][]
{
  \ifthenelse{\equal{#1}{}}
  {Q(#2)}
  {Q_{#1}(#2)}
}

\makeatletter
\def\@email#1#2{%
 \endgroup
 \patchcmd{\titleblock@produce}
  {\frontmatter@RRAPformat}
  {\frontmatter@RRAPformat{\produce@RRAP{*#1\href{mailto:#2}{#2}}}\frontmatter@RRAPformat}
  {}{}
}%
\makeatother

\begin{document}
\microtypesetup{activate=true}
\title{The interplay of geometry and coarsening in multicomponent lipid vesicles under the influence of hydrodynamics}

\author{Elena Bachini}
\affiliation{Institute of Scientific Computing, TU Dresden, Germany}
\author{Veit Krause}
\affiliation{Institute of Scientific Computing, TU Dresden, Germany}
\author{Axel Voigt}
\affiliation{Institute of Scientific Computing, TU Dresden, Germany}
\affiliation{Center for Systems Biology Dresden, Germany}  
\affiliation{Cluster of Excellence, Physics of Life, TU Dresden, Germany} 


\date{\today}

\begin{abstract}
We consider the impact of surface hydrodynamics on the interplay
between curvature and composition in coarsening processes on model
systems for biomembranes. This includes scaling laws and equilibrium
configurations, which are investigated by computational studies of a
surface two-phase flow problem with additional phase-depending bending
terms. These additional terms geometrically favor specific
configurations. We find that as in 2D the effect of hydrodynamics
strongly depends on the composition. In situations where the composition
allows a realization of a geometrically favored configuration, the
hydrodynamics enhances the evolution into this configuration. We
restrict our model and numerics to stationary surfaces and validate
the numerical approach with various benchmark problems and convergence
studies.
\end{abstract}

\maketitle

\section{Introduction}
\label{sec:intro}

The interplay between curvature and composition is a ubiquitous
structural feature for biomembranes and it plays a key role in
biological functions and synthetic membrane-based applications
(Ref.~\onlinecite{MG_N_2005}). Various simplified membrane models have been
developed to mimic the heterogeneous organization of biomembranes in
order to understand this complex interplay. Among them, there are
multicomponent giant unilamellar vesicles (GUV), which feature liquid-ordered and disordered phases. The complexity of these synthetic
systems is strongly reduced and they are ideally suited to study phase
separation and coarsening of the lipid phases.

A comprehensive overview of such results is given in
Ref.~\onlinecite{Stanichetal_BPJ_2013}. In combination with
theoretical models and simulations
(Refs.~\onlinecite{FHH_JCP_2010,CB_JCP_2011}) the different, sometimes
contradicting, scaling results for the coarsening process could be
clarified. Depending on the Saffmann-Delbr\"uck number
(Ref.~\onlinecite{SD_PNAS_1975}), which is a hydrodynamic length
relating the viscosity of the membrane and the surrounding bulk fluid
$l_H = \eta_m / \eta_b$, the hydrodynamics is effectively 2D (3D) on
spatial scales smaller than (greater than) $l_H$. Here, we consider the
situation of $l_H \to \infty$ that allows to neglect the influence of
the bulk fluid.  Besides the influence of hydrodynamics, which is
mainly studied on flat membranes
(Refs.~\onlinecite{FHH_JCP_2010,CB_JCP_2011}), composition plays an
important role. Experimental and simulation studies have addressed the
interplay between curvature and composition, see
Refs.~\onlinecite{BHW_N_2003,BDWJ_BPJ_2005,WD_JMB_2008,LRV_PRE_2009,ES_SIAMJAM_2010,EFDG_ARBP_2010,GKRR_MMMAS_2016,GS_CF_2018,Zimmermannetal_CMAME_2019}
for various approaches. However, a detailed investigation of the role
of membrane curvature on spatial arrangements of lipid phases has only
been considered recently using multicomponent scaffolded lipid
vesicles (SLV)
(Refs.~\onlinecite{FRKG_PRE_2018,FRKG_PRE_2019,RFGK_NC_2020}). In
these systems, non-spherical shapes are stabilized and the effect of
spatially varying curvature on phase separation and coarsening can be
considered. The results demonstrate the influence of curvature on the
spatial arrangement of the lipid phases and the phase
diagram. However, the focus of the modeling is on the equilibrium
state and the influence of hydrodynamics is not considered in these
approaches. Due to recent numerical developments in surface
hydrodynamics
(Refs.~\onlinecite{Arroyoetal_PRE_2009,Nitschkeetal_JFM_2012,Reutheretal_PF_2018,Fries_IJNMF_2018,ABWPK_PRE_2019,Olshanskiietal_VJM_2022}),
here we combine these different investigations and study the influence
of membrane curvature on phase separation and coarsening under the
influence of hydrodynamics.

We consider a mesoscale modeling approach building on
Ref.~\onlinecite{JL_PRL_1993}, which provides an extension of the classical
Canham/Helfrich model (Refs.~\onlinecite{C_JTB_1970,H_ZN_1973}) to two lipid phases,
e.g., liquid-ordered and disordered phases. The corresponding energy
reads:
\begin{equation}
\energy[JL] = \sum_{i=1}^2 \int_{\SurfDomain[i]} \bendStiff[i]
(\meanCurv - \meanCurv[0,i])^2 + \bendStiffGauss[i]\, \GaussCurv \;
\Diff\SurfDomain + \tension \int_\Gamma \, \Diff s \,,
\label{eq:energy_JL}
\end{equation}
with bending rigidity $\bendStiff[i]$ and Gaussian bending rigidity
$\bendStiffGauss[i]$, spontaneous curvatures $\meanCurv[0,i]$, mean and
Gaussian curvatures $\meanCurv$ and $\GaussCurv$, and interfacial line
tension $\tension$. Here, $\SurfDomain[i]$, $i=1,2$, denotes the surface area of the
$i-$th lipid phase, for which $\SurfDomain[1] \cup \SurfDomain[2] = \SurfDomain$
and $\SurfDomain[1] \cap \SurfDomain[2] = \Gamma$. We consider $\SurfDomain
\subset \REAL^3$ a closed regular surface without boundary. The model
can be extended by considering Lagrange multipliers to ensure area conservation or
local inextensibility of the two phases. As long as
$\bendStiffGauss[1] = \bendStiffGauss[2]$ and topological changes in
a possible evolution of $\SurfDomain$ are prevented, the Gaussian curvature
term in Eq.~\eqref{eq:energy_JL} only contributes a constant to the
energy and will therefore be neglected in the
following. Eq.~\eqref{eq:energy_JL} (with $\bendStiffGauss[i] = 0$)
can be cast in a phase-field approximation, obtaining:
\begin{align}
\energyPF = &\int_{\SurfDomain} \bendStiff(\CHsol) (\meanCurv -
\meanCurv[0](\CHsol))^2 \; \Diff\SurfDomain \nonumber \\
&+ \ttension \int_{\SurfDomain} \frac{\interfaceparam}{2} \NORM{\GradSurf\CHsol}^2 + \frac{1}{\interfaceparam} \dWell(\CHsol) \, \Diff\SurfDomain \,,
\label{eq:energy_PF}
\end{align}
with phase-field function $\CHsol \in [-1,1]$, double-well potential
$\dWell(\CHsol) = \frac{1}{4}(\CHsol^2 - 1)^2$, interface thickness
$\interfaceparam$, rescaled interfacial line tension $\ttension =
\tension \frac{3}{2\sqrt{2}}$, and surface gradient
$\GradSurf$. The bending rigidity $\bendStiff(\CHsol)$ and
spontaneous curvature $\meanCurv[0](\CHsol)$ are smoothly interpolated
between the values of the two phases, such that:
\begin{equation}\label{eq:bendstiff}
  f(\CHsol)=
  \begin{cases}
    f_1 & \mbox{if } \CHsol = 1\\
\displaystyle    \frac{f_1+f_2}{2}
    + \frac{f_1-f_2}{4}\CHsol(3-\CHsol^2) & \mbox{if } -1< \CHsol< 1\\
    f_2 & \mbox{if } \CHsol = -1
  \end{cases}\,,
\end{equation}
for $f = \bendStiff, \meanCurv[0]$. A connection between
Eqs.~\eqref{eq:energy_PF} and \eqref{eq:energy_JL} can be established
by formal matched asymptotic or $\Gamma$-convergence if $\interfaceparam \to
0$, following
Ref.~\onlinecite{ES_SIAMJAM_2010,GKRR_MMMAS_2016,EHS_IFB_2022}. Possibilities to
also consider the Gaussian curvature term and constraints on area and
local inextensibility in the phase-field context have been discussed
in Refs.~\onlinecite{AEV_JCP_2014,ES_SIAMJAM_2010,HMLLRV_IJBB_2013}.

While the minimization of $\energy[JL]$ and $\energyPF$ with respect
to $\SurfDomain$ and $\Gamma$, or $\SurfDomain$ and $\CHsol$,
respectively, has been discussed in the literature (see,
e.g., Refs.~\onlinecite{LXV_FDMP_2007,LRV_PRE_2009,ES_SIAMJAM_2010,HMLLRV_IJBB_2013,GN_IMAJNA_2021,EHS_IFB_2022}),
here we consider the special case of a stationary surface
$\SurfDomain$. This reduces the problem to the evolution of $\Gamma$
or $\CHsol$ on a given surface and is mathematically related to
so-called geodesic evolution equations and their phase-field
approximations. The conserved evolution of $\CHsol$ is a surface
Cahn-Hilliard model with additional curvature terms:
\begin{align}
\DerT \CHsol = &\,\DivSurf (\mobility \GradSurf
\chempot)\,, \label{eq:sCH1} \\
\chempot = &\,\bendStiff^\prime(\CHsol) (\meanCurv -
\meanCurv[0](\CHsol))^2 - 2 \bendStiff(\CHsol) (\meanCurv -
\meanCurv[0](\CHsol)) \meanCurv[0]^\prime(\CHsol) \nonumber\\
&\,
+ \ttension \left(-
\interfaceparam \LapSurf \CHsol + \frac{1}{\interfaceparam} \dWell^\prime(\CHsol) \right)\,,
\label{eq:sCH2}
\end{align}
with mobility $\mobility$, chemical potential $\chempot$, surface
divergence $\DivSurf$, and Laplace-Beltrami operator $\LapSurf =
\DivSurf\GradSurf$. Eqs.~\eqref{eq:sCH1} and \eqref{eq:sCH2} are
defined on $\SurfDomain$ and need to be supplemented with appropriate
initial conditions for $\CHsol$. For $\interfaceparam \to 0$ one
obtains the corresponding surface Mullins-Sekerka type model
(Ref.~\onlinecite{GKRR_MMMAS_2016}), for which equilibrium solutions have been
analyzed in detail in Refs.~\onlinecite{FRKG_PRE_2018,FRKG_PRE_2019}. The
corresponding experimental setting for the considered case of a
stationary surface are the mentioned scaffolded lipid vesicles (SLV) with two lipid
phases (Ref.~\onlinecite{RFGK_NC_2020}). They provide an ideal test case to study
curvature-induced effects in phase separation and coarsening of lipid
domains and can be used for validation.

With recent developments in the modeling and simulation of surface fluids
(Refs.~\onlinecite{Nitschkeetal_JFM_2012,Reutheretal_MMS_2015,Nitschkeetal_TFI_2017,Reutheretal_MMS_2018,Jankuhnetal_IFB_2018,Fries_IJNMF_2018,Reutheretal_PF_2018,Torres_Sanchezetal_JFM_2019,Reutheretal_JFM_2020,Ledereretal_IJNME_2020}),
it becomes feasible to also consider the surface viscosity of the
liquid phases in such systems and study the evolution toward the
equilibrium solutions. This leads to surface two-phase flow
problems. The corresponding model to the Cahn-Hilliard like
Eqs.~\eqref{eq:sCH1} and \eqref{eq:sCH2} reads:
\begin{eqnarray}
\label{eq:sNSCH1}
\DerT \vectvel + \GradSurfConv{\vectvel}\vectvel &=& 
-\GradSurf \tpress + \frac{2}{\Reynolds} \divS \StressTens(\vectvel) +
\chempot\GradSurf\CHsol\,, \\
\label{eq:sNSCH2}
\DivSurf\vectvel&=&0\,, \\
\label{eq:sNSCH3}
\DerT\CHsol + \GradSurfConv{\vectvel}\CHsol&=&
\DivSurf\left(\mobility\GradSurf\chempot\right)\,, \\
\label{eq:sNSCH4}
\chempot&=&\bendStiff^\prime(\CHsol) (\meanCurv -
\meanCurv[0](\CHsol))^2 \nonumber \\
&&- 2 \bendStiff(\CHsol) (\meanCurv - \meanCurv[0](\CHsol))
\meanCurv[0]^\prime(\CHsol) \nonumber\\
&&+ \ttension \left(- \interfaceparam \LapSurf \CHsol + \frac{1}{\interfaceparam} \dWell^\prime(\CHsol) \right)\,,
\end{eqnarray}
with tangential surface velocity $\vectvel$, rescaled surface pressure
$\tpress$, surface stress tensor
$\StressTens(\vectvel)=\tfrac{1}{2}(\GradSurf\vectvel+\GradSurf\vectvel^{\T})$,
surface divergence for tensor fields $\divS$, and surface Reynolds
number $\Reynolds$. The convective terms are defined by
$[\GradSurfConv{\vectvel}\vectvel]_i = \GradSurf u_i \cdot
\vectvel$, $i = 0,1,2$, and $\GradSurf \CHsol \cdot
\vectvel$. Eqs.~\eqref{eq:sNSCH1}-\eqref{eq:sNSCH4} are defined on
$\SurfDomain$ and supplemented by initial conditions for $\CHsol$ and
$\vectvel$. We consider $\vectvel(\xv,t) = (u_1(\xv,t),
u_2(\xv,t), 0)$ with components corresponding to the local
basis vectors $\mathbf{e}_1(\xv), \mathbf{e}_2(\xv)$ and
surface normal $\normalvec(\xv)$. As we have the relation:
\begin{align*}
2 \divS \StressTens(\vectvel) &= \divS(\GradSurf \vectvel) + \divS
(\GradSurf \vectvel)^T\\
&= - \LapSurf^{dR} \vectvel +
\GaussCurv\vectvel + \GradSurf (\GradSurf \cdot \vectvel) +
\GaussCurv\vectvel \\
&= - \LapSurf^{dR} \vectvel + 2
\GaussCurv\vectvel\,,
\end{align*}
with Laplace-deRham operator
$\LapSurf^{dR}$, the surface Navier-Stokes part corresponds
to previous formulations on stationary surfaces
(Refs.~\onlinecite{Arroyoetal_PRE_2009,Reutheretal_MMS_2015,Yavarietal_JNS_2016,Kobaetal_QAM_2017,Reutheretal_MMS_2018,Miura_QAM_2018,Kobaetal_QAM_2018,Jankuhnetal_IFB_2018,Pruessetal_JEE_2021}).

The model is a surface Navier-Stokes-Cahn-Hilliard like equation, a
generalization of the classical ``Model H''
(Refs.~\onlinecite{HH_RMP_1977,GPV_MMMAS_1996}) to surfaces. For simplicity, we only
consider the case of equal density and equal viscosity for the two
phases. More general approaches emerge by considering the models
compared in Ref.~\onlinecite{AV_IJNMF_2012} and extend them to surfaces. We refer to
Ref.~\onlinecite{Olshanskiietal_VJM_2022} for an extension to surfaces of the
thermodynamically consistent model proposed in Ref.~\onlinecite{AGG_MMMAS_2012}. The combination of the asymptotic analysis in
Ref.~\onlinecite{GKRR_MMMAS_2016} for surface Cahn-Hilliard equations with
matched asymptotic for Navier-Stokes-Cahn-Hilliard equations
(Refs.~\onlinecite{A_ARMA_2009,AGG_MMMAS_2012,MPCMC_JFM_2013}) should allow to
obtain the corresponding sharp interface surface two-phase flow
problems. Previous numerical studies of
Eqs.~\eqref{eq:sNSCH1}-\eqref{eq:sNSCH4}, such as
Refs.~\onlinecite{Nitschkeetal_JFM_2012,ABWPK_PRE_2019,Olshanskiietal_VJM_2022},
are restricted to simply connected surfaces, special geometries or
only consider lower order methods. None of the previous approaches
takes into account the additional curvature terms emerging from the bending energy.

Here we propose a discretization based on surface finite elements
(SFEM). We consider a higher-order surface approximation and solve
Eqs.~\eqref{eq:sNSCH1}-\eqref{eq:sNSCH4} using an operator splitting
approach. The scalar-valued convected surface Cahn-Hilliard equation
is solved by the scalar SFEM (Ref.~\onlinecite{DE_AN_2013}) and for the
surface Navier-Stokes equations the vector-valued SFEM
(Ref.~\onlinecite{NNV_JCP_2019}) is used. Discretization in time considers a
semi-implicit finite difference approach. We validate each part and
demonstrate convergence with potential optimal order.

The overall scheme is used to study curvature-induced effects in phase
separation and coarsening. Considering a geometry with regions of two
distinct mean curvatures and choosing $\kappa_i$ or $\meanCurv[{0,i}]$
correspondingly allows to determine the spatial arrangement of the
lipid phases. Phase-dependent properties in the bending rigidity
$\kappa$ influence the coarsening process: the lipid phase with lower
bending rigidity is guided toward regions of higher mean curvature,
while the lipid phase with higher bending rigidity is guided toward
regions of lower mean curvature. This behavior is known for the
equilibrium configuration and has been demonstrated experimentally
together with a geometric influence of the phase diagram
(Ref.~\onlinecite{RFGK_NC_2020}). The same effect can be achieved by
specifying the spontaneous curvatures $\meanCurv[{0,i}]$ to values
that match the mean curvature of the different regions.  Not only the
phase diagram and equilibrium arrangements are influenced by the
underlying geometry, but also the evolution toward this state changes. As
in 2D and 3D, the coarsening process can be influenced by the flow and
can lead to faster coarsening. However, the theoretical prediction for
viscous hydrodynamic scaling of the interface length $\interfaceLength
\sim \Time^{-1}$ could not be observed. Depending on the composition
and the Reynolds number $\Reynolds$, we obtain $\interfaceLength \sim
\Time^{-\alpha}$ with $\frac{1}{3} \leq \alpha \leq \frac{2}{3}$,
which is in agreement with theoretical and simulation results in 2D
(Refs.~\onlinecite{FHH_JCP_2010,CB_JCP_2011}) and simulations in
Ref.~\onlinecite{ABWPK_PRE_2019}. Moreover, it has been measured in
various experiments on giant unilamellar vesicles (GUV), see
Ref.~\onlinecite{Stanichetal_BPJ_2013} for a review. We also consider
the topology and geometry of the phases. When islands are formed, they
become more circular if flow is considered. This is one of the reasons
for the slowdown of the coarsening process under flow for low
compositions. Using the zeroth Betti number as a topological measure
for connectivity of the phases
(Refs.~\onlinecite{Mickelinetal_PRL_2018,RV_PF_2021}), and the
interfacial shape distribution (ISD) as a geometrical measure of the
isolated patches (Refs.~\onlinecite{MAV_MMT_2003,KTV_PM_2010}), the
scaling behavior can be quantified further. The additional curvature
terms emerging from the bending energy not only have a global effect
on the dynamics but also influence the flow field locally. We
quantify this behavior by establishing a relation between surface
mean curvature and velocity in phase-separated regions.

The paper is structured as follows. A detailed description of the
numerical approach together with convergence studies for each
subproblem is given in Section~\ref{sec:2}, while a systematic
investigation of the results on phase separation and coarsening is
presented in Section~\ref{sec:3}. In Section~\ref{sec:4}, we discuss
the results, draw conclusions, and give an outlook on possible model
extensions.

\section{Numerical approach}
\label{sec:2}

In this section we describe the discrete setting, starting from the
approximation of the surface, the discrete functional spaces and,
finally, we present the discrete version of
Eqs.~\eqref{eq:sNSCH1}-\eqref{eq:sNSCH4}. We discuss
implementational aspects, and  validate the numerical approach by
considering benchmark problems for each subproblem.

\subsection{Surface approximation}
We assume that the smooth surface $\SurfDomain$ is approximated by a
discrete $k$-th order approximation $\SurfDomain[\meshparam]$.  If
$k=0$, $\SurfDomain[\meshparam]^0$ is formed by the union of
non-intersecting flat triangles with vertices on $\SurfDomain$. For
$k\ge 1$, we consider the bijective map $\param\From
\SurfDomain[\meshparam]^0\to \SurfDomain$ and
$\Interpolant^k(\param)=\SurfDomain[\meshparam]^k$.
We can write that
$\TriangH(\SurfDomain)=\cup_{i=1}^{\NCell}\Cell[i]=\SurfDomain[\meshparam]^k$. In
the following, we use $k = 2$.
Surface quantities, when needed, are computed from the approximate
surface $\SurfDomain[\meshparam]$. In particular, this is done for
the approximate value of the surface normal $\normalvecApprox$ and the mean curvature $\meanCurvApprox$.
A more detailed description of the construction of higher-order
surface approximation can be found in Refs.~\onlinecite{Demlow2009,praetorius2020dunecurvedgrid}. 
An adaptive mesh is used to refine elements along the 
interface described implicitly by $\CHsol(\xv,t) = 0$. \\

\subsection{Surface finite element functional spaces}
We consider the surface finite element spaces defined by:
\begin{align*}
\TestSpaceApprox(\SurfDomain[\meshparam]) &= \{\TestApprox\in\Cont[0](\SurfDomain[\meshparam])\cap\Hilb[0] \;|\;\TestApprox{|_{\Cell}}\in\PC{k_{\Test}}(\Cell)\}\,,\\
\TestSpacePress(\SurfDomain[\meshparam]) &= \{\TestPressApprox\in\Lspace_0(\SurfDomain[\meshparam])\;|\;\TestPressApprox{|_{\Cell}}\in\PC{k_{\TestPress}}(\Cell)\}\,,\\
\TestSpaceCH(\SurfDomain[\meshparam]) &= \{\TestCHApprox\in\Cont[0](\SurfDomain[\meshparam]) \;|\;\TestCHApprox{{|_{\Cell}}}\in\PC{k_{\TestCH}}(\Cell)\}\,,
\end{align*}
where $k_{\Test}, k_{\TestPress}, k_{\TestCH}$ denote the polynomial
orders. We extend the space $\TestSpaceApprox$ to
$\TestSpaceVec(\SurfDomain[\meshparam])=(\TestSpaceApprox)^3$, the
space of (three-dimensional) vector functions.  Classical
$\PC{2}-\PC{1}$ Taylor-Hood elements are used for the Navier-Stokes
equations, with polynomial order 2 for the velocity $\vectvel$ and 1
for pressure $\tpress$. $\PC{2}$ is used for discretizing the
phase-field function $\CHsol$ and the chemical potential $\chempot$.

\subsection{Discrete formulation}
We can now write the semi-discrete formulation of
Eqs.~\eqref{eq:sNSCH1}-\eqref{eq:sNSCH4}. The surface Cahn–Hilliard
like and surface Navier–Stokes like equations are solved separately in
an operator splitting approach.
At every time step $\Time^n\in\IT$, we first solve the surface
Cahn-Hilliard like Eqs.~\eqref{eq:sNSCH3} and \eqref{eq:sNSCH4} and
then solve the surface Navier-Stokes like Eqs.~\eqref{eq:sNSCH1} and
\eqref{eq:sNSCH2}.
For the time discretization, we apply an implicit Euler scheme
for both systems.

In the following, we denote by $\InnerApprox{\cdot}{\cdot}$ the
approximation by the quadrature rule  of the $L^2$-inner product on
the approximated surface $\SurfDomain[\meshparam]$. 
Note that, the quadrature order has to be chosen high enough such that test
and trial functions and area elements are well integrated.

\begin{widetext}
\begin{problem}[Semi-discrete surface Cahn-Hilliard like problem]
\label{pb:ch}
Find $(\CHsolApprox,\chempotApprox)\in\TestSpaceCH\times\TestSpaceCH$ such that:
\begin{align*}
\InnerApprox{\DerT\CHsolApprox^n}{\TestCHApprox}
+ \InnerApprox{\GradSurfConv{\vectvel^{n-1}}\CHsolApprox^n}{\TestCHApprox}
=&
- \mobility \InnerApprox{\GradSurf\chempotApprox^{n}}{\GradSurf\TestCHApprox}\,,
\\[0.7 em]
\InnerApprox{\chempotApprox^n}{\TestCHmuApprox}
=& \InnerApprox{
  \bendStiff^\prime(\CHsolApprox^{n-1})\left(\meanCurvApprox-\meanCurv[0](\CHsolApprox^{n-1})\right)^2}{\TestCHmuApprox}
\\ & 
- 2 \InnerApprox{\bendStiff(\CHsolApprox^{n-1})\meanCurv[0]^{\prime}(\CHsolApprox^{n-1})\left(\meanCurvApprox-\meanCurv[0](\CHsolApprox^{n-1})\right)}{\TestCHmuApprox} \\
&
+ \tilde{\sigma} \interfaceparam \InnerApprox{\GradSurf\CHsolApprox^n}{\GradSurf\TestCHmuApprox} 
+ \frac{\tilde{\sigma}}{\interfaceparam} \InnerApprox{\dWell^\prime(\CHsolApprox^n)}{\TestCHmuApprox}\,,
\end{align*}
for all $(\TestCHApprox,\TestCHmuApprox)\in \TestSpaceCH\times\TestSpaceCH$.
\end{problem}
\end{widetext}
This is a classical SFEM approach (Ref.~\onlinecite{DE_AN_2013}) for scalar-valued
surface PDEs. Note that, as standard for Cahn-Hilliard equations, we
linearize the derivative of the double-well potential by a Taylor
expansion of order one, i.e., $\dWell^\prime(\CHsolApprox^n)\approx
\dWell^\prime(\CHsolApprox^{n-1}) +
\dWell^{\dprime}(\CHsolApprox^{n-1})(\CHsolApprox^n-\CHsolApprox^{n-1}) =
-2(\CHsolApprox^{n-1})^3 +
\left(3(\CHsolApprox^{n-1})^2-1\right)\CHsolApprox^n $. Following
Ref.~\onlinecite{MPCMC_JFM_2013}, we consider $\mobility = \gamma
\interfaceparam^2$ with $\gamma > 0$.

For the surface Navier-Stokes like problem, we drop the requirement of
$\vectvel$ being tangential and enforce this condition by a penalty
approach. This allows to consider a global coordinate system, the one
of the embedding space $\REAL^3$, and apply the classical scalar SFEM approach
(Ref.~\onlinecite{DE_AN_2013})  to each component, see
Ref.~\onlinecite{NNV_JCP_2019} for a general discussion and
Refs.~\onlinecite{NNPV_JNS_2018,Jankuhnetal_IFB_2018,Reutheretal_PF_2018,HLL_IMAJNA_2020}
for specific realizations. For this setting, we consider
$\vectvel(\xv,\Time) = (u_1(\xv,\Time), u_2(\xv,\Time),
u_3(\xv,\Time))$ with components corresponding to the global basis
vectors $\mathbf{e}_1, \mathbf{e}_2$ and $\mathbf{e}_3$ of
$\REAL^3$. Under the assumption that the added penalization enforces
$\vectvel \cdot \normalvec = 0$, Eqs.~\eqref{eq:sNSCH1} and
\eqref{eq:sNSCH2} read:
\begin{align}
\label{eq:sNSCH1a}
\DerT \vectvel + \GradSurfConv{\vectvel}\vectvel =& \,
-\GradSurf \tpress + \frac{2}{\Reynolds} \divP \StressTens(\vectvel)
\nonumber \\
& \, +
\chempot\GradSurf\CHsol
+ \beta(\meshparam) (\vectvel \cdot \normalvec
) \normalvec \,,\\
\label{eq:sNSCH2a}
\nablaP \cdot \vectvel=&\,0 \,, 
\end{align}
with $\StressTens = \frac{1}{2} (\nablaP \vectvel + \nablaP
\vectvel^T)$ using the tangential gradient $\nablaP \vectvel =
\ProjMat \nabla \vectvel^e \ProjMat$, where $\vectvel^e$ is an
extension of $\vectvel$ constant in normal direction, $\ProjMat =
\IDMat - \normalvec \otimes \normalvec$ is the surface projection and
$\Grad$ is the gradient of the embedding space $\REAL^3$. The
tangential divergence is defined as $\nablaP \cdot \vectvel =
tr[\ProjMat \nabla \vectvel^e]$ and $\divP$ denotes the corresponding
tangential divergence for tensor fields. The term $\nablaP \vectvel$
is a pure tangential tensor field and relates to the covariant
operator by $\nablaP \vectvel = \GradSurf (\ProjMat \vectvel) -
(\vectvel \cdot \normalvec) \shapeOp$, where $\shapeOp$ is the shape
operator with $B_{ij} = - \GradSurf^i \nu_j$. Similarly, it holds
$\divP \vectvel = \divS (\ProjMat \vectvel) - (\vectvel \cdot
\normalvec) \meanCurv$, see Ref.~\onlinecite{Brandneretal_SIAMJSC_2022}. The
additional term in Eq.~\eqref{eq:sNSCH1a} considers a penalization of
the normal component of $\vectvel$, with $\beta(\meshparam)$ a
parameter depending on the mesh size. Therefore, if $\vectvel$ is
tangential, Eqs.~\eqref{eq:sNSCH1a} and \eqref{eq:sNSCH2a} coincide
with Eqs.~\eqref{eq:sNSCH1} and \eqref{eq:sNSCH2}. If tangentiality is
only approximated, Eqs.~\eqref{eq:sNSCH1a} and \eqref{eq:sNSCH2a}
provide an approximation to Eqs.~\eqref{eq:sNSCH1} and
\eqref{eq:sNSCH2}, which has been considered previously for surface
approximations with $k=0$ and shown to converge experimentally
(Refs.~\onlinecite{Reutheretal_PF_2018,RV_PF_2021,NV_CICP_2022}) with order $1$ for
the $L^2$ error in $\vectvel$. More detailed results only exist for
the surface Stokes problem
(Refs.~\onlinecite{Brandneretal_SIAMJSC_2022,HP_unpublished}).
These studies show
optimal order of convergence of order $3$ for the $L^2$ error in
$\vectvel$ for surface approximations with $k=2$, if the additional
terms $(\vectvel \cdot \normalvec) \shapeOp$ and $(\vectvel \cdot
\normalvec) \meanCurv$ are considered with $\normalvec$ approximated
by order $2$ and $\normalvec$ in the penalization term approximated by
order $3$ and $\beta(\meshparam) = \frac{\beta}{\meshparam^2}$. These results are analytically known
only for the surface vector-Laplace problem
(Refs.~\onlinecite{HLL_IMAJNA_2020,JR_IMAJNA_2021}), but expected also for the
surface Stokes problem (Refs.~\onlinecite{JORZ_JNM_2021,HP_unpublished}).
However,
if only the error in the tangential component of $\vectvel$ is
considered, the additional requirement on the higher-order
approximation in $\normalvec$ can be dropped, see
Ref.~\onlinecite{HP_IMAJNA_2022}. With these results, we consider the following
problem.

\begin{widetext}
\begin{problem}[Semi-discrete surface Navier-Stokes like problem]
\label{pb:ns}
Find $(\vectvelApprox,\pressApprox)\in\TestSpaceVec\times\TestSpacePress$ such that:
\begin{align*}
\InnerApprox{\DerT \vectvelApprox^n}{\TestVecApprox} + \InnerApprox{\GradSurfConv{\vectvel^{n-1}}\vectvelApprox^n}{\TestVecApprox} = &
\InnerApprox{\pressApprox^n}{\DivSurf\TestVecApprox} - \frac{2}{\Reynolds} \InnerApprox{
   (\StressTens(\vectvelApprox^{n})}{\GradSurf\TestVecApprox} \\
 &+ \InnerApprox{\chempotApprox^n\GradSurf\CHsolApprox^n}{\TestVecApprox} + \frac{\beta}{\meshparam^{2}} \InnerApprox{ (\vectvelApprox^{n}\cdot\normalvecApprox)\normalvecApprox}{\TestVecApprox}\,,\\
\InnerApprox{\Div\vectvelApprox^{n}}{\TestPressApprox} = & \;0\,,
\end{align*}
for all $(\TestVecApprox,\TestPressApprox)\in\TestSpaceVec\times\TestSpacePress$.
\end{problem}
\end{widetext}

\subsection{Implementational aspects}
The above-described discretization is implemented within the finite
element toolbox AMDiS (Refs.~\onlinecite{Vey_CVS_2007,Witkowski_ACM_2015}), using the
Dune-CurvedGrid library (Ref.~\onlinecite{praetorius2020dunecurvedgrid}) to handle
the approximation of the surfaces. This environment allows
for a straightforward parallelization by using the PETSc library. We
use a Richardson iteration method with full $LU$ preconditioner and
MUMPS as direct linear solver.

\subsection{Validation of numerical approach}
We first consider Problem~\ref{pb:ch} and Problem~\ref{pb:ns} separately, ignoring both the
coupling terms and the additional curvature
terms. This reduces the problems to the surface Cahn-Hilliard and the
surface Navier-Stokes equations, respectively.

\subsubsection{Surface Cahn-Hilliard equation}
\label{sec:sCH}

Different benchmark problems have been proposed for the surface
Cahn-Hilliard model on a sphere. In Ref.~\onlinecite{R_AML_2016} a rotational
symmetric solution of the surface Mullins-Sekerka problem is
constructed. As this is the sharp interface limit of the surface
Cahn-Hilliard equation (Ref.~\onlinecite{GKRR_MMMAS_2016}), it allows to study
the convergence of solutions of Problem~\ref{pb:ch} in the interface thickness
$\interfaceparam\to 0$.  We only consider this qualitatively. A quantitative
convergence study is beyond the scope of this paper. For theoretical
results in this direction in 2D we refer to Ref.~\onlinecite{FP_NM_2003}.

\begin{figure*}
  \centering
  \includegraphics[width=\textwidth,
  trim={20 0 0 0},clip]{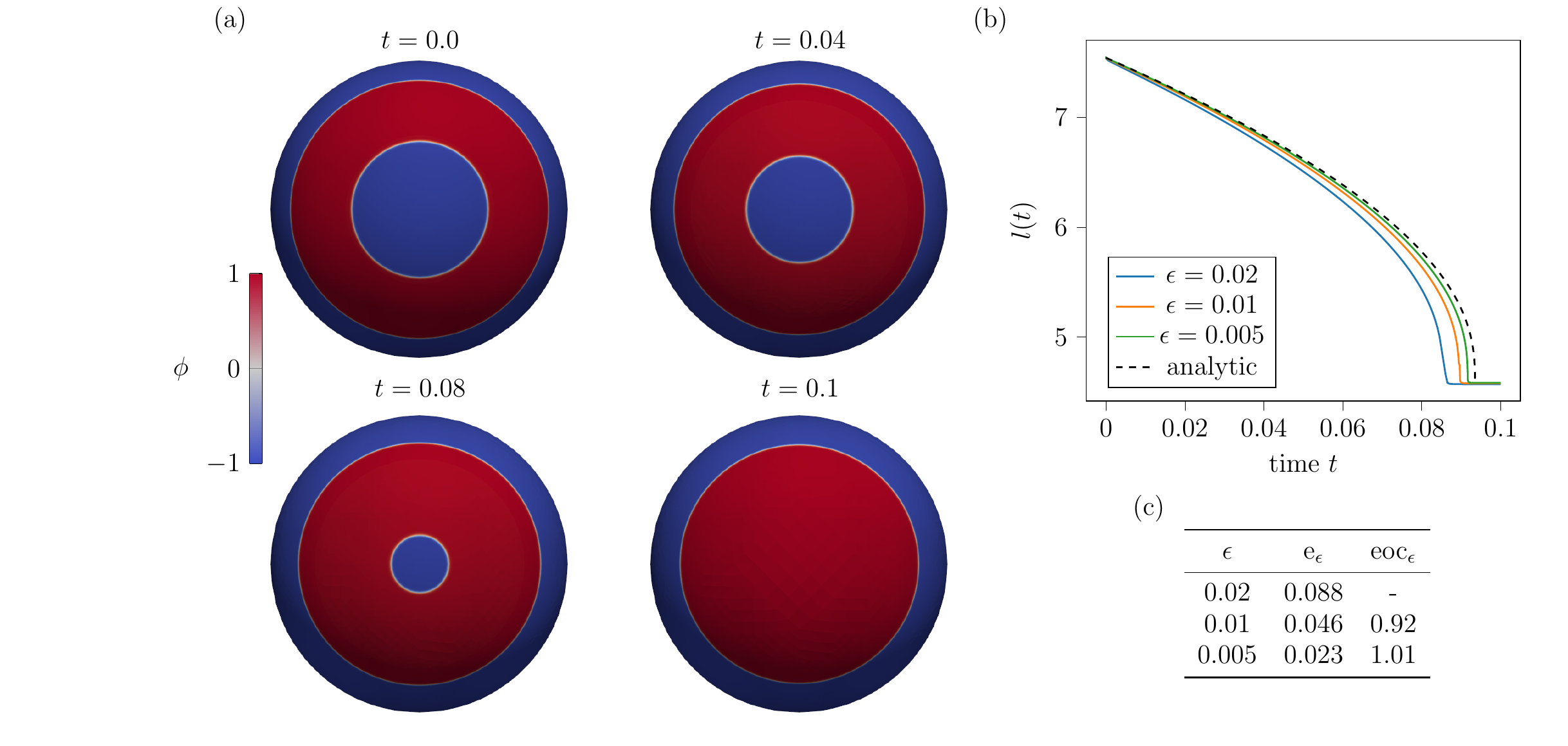}
  \caption{Numerical solution for the surface Cahn-Hilliard model on
    benchmark problem (Ref.~\onlinecite{R_AML_2016}). (a) Numerical solution at
    times $\Time=0.0, 0.04, 0.08$,$0.1$ for $\interfaceparam =
    0.005$. (b) Comparison between computed lengths in the cases of
    different values of $\interfaceparam$ and analytical solution. (c)
    Experimental order of convergence in the parameter
    $\interfaceparam$, in the time interval $[0,0.1]$. The error considers the difference in interface length over time.}
  \label{fig:ch-benchmark1}
\end{figure*}

We consider the unit sphere $\SurfDomain=S^2$ ($R=1$) with an annular
domain $\SurfDomain[1]$ for the phase $\CHsol=1$ defined by two
concentric circles with geodesic radii $R_{\max}=0.8$ and
$R_{\min}=0.4$ as initial condition (see
Fig.~\ref{fig:ch-benchmark1}-(a), top-left panel).  In
Eqs.~\eqref{eq:sCH1} and \eqref{eq:sCH2}, we set the parameters
$\mobility=1$, $\ttension=\frac{3}{2 \sqrt{2}}$ and
$\bendStiff=\meanCurv_0=0$. We consider $\interfaceparam=0.02, 0.01,
0.005$ to study the convergence in $\interfaceparam$. As in our
general setting, $\PC{2}$ elements are used for discretizing the phase-field function $\CHsol$ and the chemical potential $\chempot$. The mesh is
adaptively refined to guarantee a resolution with at least 3 degrees
of freedom around the interface,
i.e., $\CHsol\in[-0.95,0.95]$, and
the time step $\tau$ is chosen as $\tau\sim\meshparam^2$.

Fig.~\ref{fig:ch-benchmark1}-(a) shows the evolution at
$\Time=0.0, 0.04, 0.08$, $0.1$ for $\interfaceparam = 0.005$. Both
circular interfaces shrink until the smaller one collapses and a final
configuration of a single geodesic circle for $\SurfDomain[1]$ is
reached. Fig.~\ref{fig:ch-benchmark1}-(b) shows the lengths of the
interface computed as $\interfaceLength_{\interfaceparam} = \energyPF[\CHsol]$ for different
$\interfaceparam$ and the analytic length $\interfaceLength_*$ of the surface
Mullins-Sekerka problem, see Ref.~\onlinecite{R_AML_2016} for details.
For each value of
$\interfaceparam$, we define the error:
\begin{equation}
  \label{eq:err-ch-benchmark1}
  \mbox{e}_{\interfaceparam}=\NORM[\Lspace(0,T)]{\interfaceLength_{\interfaceparam}-\interfaceLength_*}\,,
\end{equation}
where the integration in time is approximated by a trapezoidal rule
over the time interval [0,T], with $T=0.1$.
The experimental order of convergence is then defined by:
\begin{equation*}
  \mbox{eoc}_{\interfaceparam}=\log\left(\frac{\mbox{e}_{\interfaceparam,i-1}}{\mbox{e}_{\interfaceparam,i}}\right)/\log\left(\frac{\interfaceparam_{i-1}}{\interfaceparam_{i}}\right)\,.
\end{equation*}
The results are reported in Fig.~\ref{fig:ch-benchmark1}-(c) and
indicate first-order convergence in $\interfaceparam$.

\begin{figure*}
  \centering
  \includegraphics[width=\textwidth,
  trim={20 0 0 0},clip]{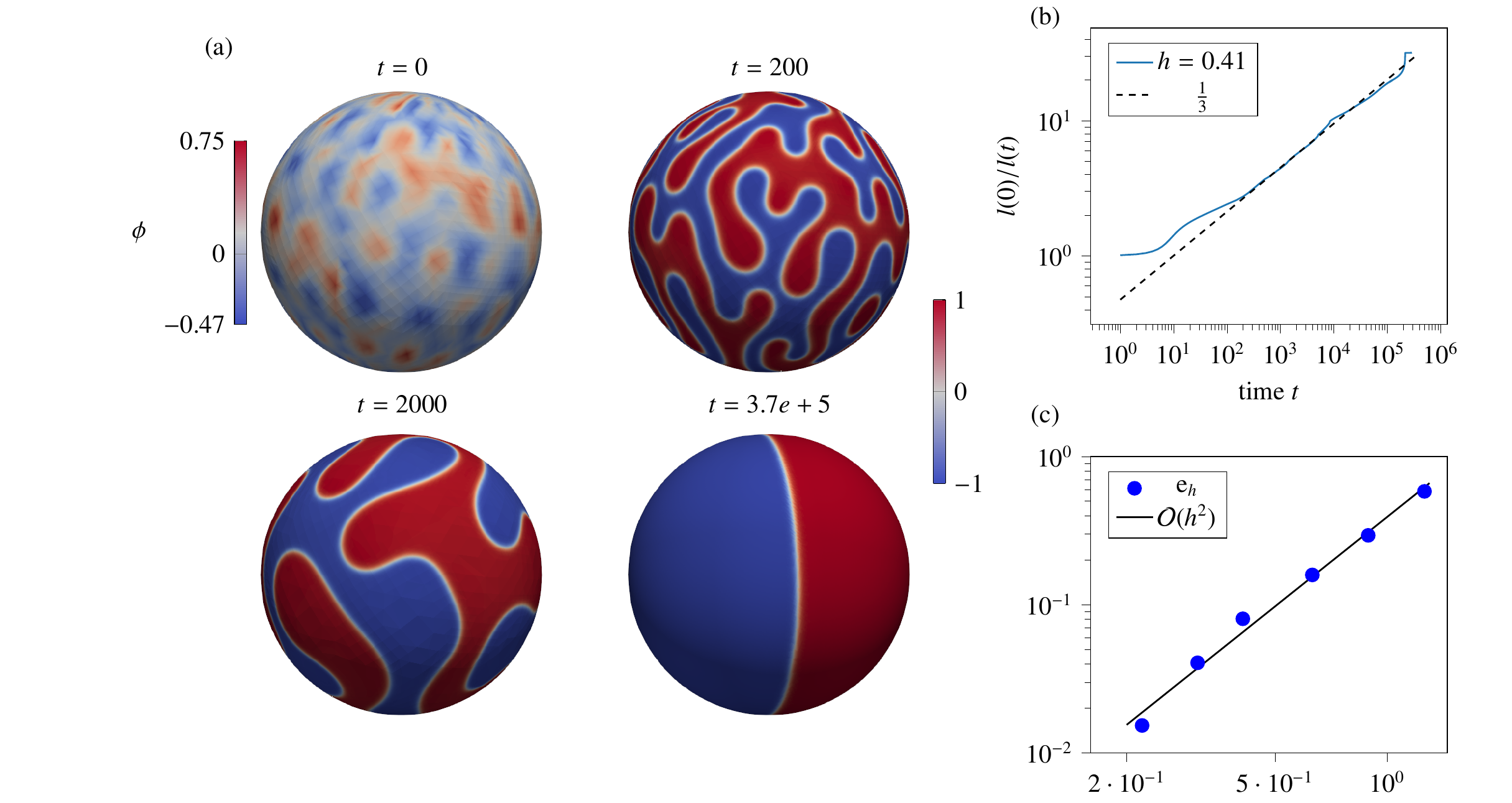}
  \caption{Numerical solution for the surface Cahn-Hilliard model on
    benchmark problem (Ref.~\onlinecite{Jokisaarietal_CMATS_2017}). (a)
    Approximate solution at times $\Time=0, 200, 2000, \Time=3.7e+5$. (b) Inverse of the approximate interface length over
    time (normalized with the initial length), and interpolated line
    of order $1/3$. (c)  Experimental order of convergence over
    refinement in the mesh parameter $\meshparam$ in the time interval $[0,200]$.
  } 
  \label{fig:ch-benchmark2} 
\end{figure*}

More realistic situations of phase separation and coarsening on a
sphere are considered in Ref.~\onlinecite{Jokisaarietal_CMATS_2017}. We consider
the proposed benchmark problem with a sphere of radius $R=100$ and
initial condition defined by:
\begin{align*}
  \CHsol(\theta, \varphi) = c_0 &+
  \varepsilon_{\SurfDomain}\big[\cos(8\theta)\cos(15\varphi)
    + \left(\cos(12\theta)\cos(10\varphi)\right)^2 \\
    & 
    +\cos(2.5\theta-1.5\varphi)\cos(7\theta-2\varphi)\big]\,,
\end{align*}
with $c_0=0.5$, $\varepsilon_{\SurfDomain}=0.05$ and $\theta=\cos^{-1}(z/R)$ and
$\varphi=\tan^{-1}(y/x)$ the polar and azimuth angles in a spherical coordinate
system. Fig.~\ref{fig:ch-benchmark2}-(a), top-left panel, shows the
initial solution.
In Eqs.~\eqref{eq:sCH1} and \eqref{eq:sCH2} we set $\mobility=5$,
$\ttension=\frac{3}{2 \sqrt{2}}$ and
$\bendStiff=\meanCurv[0]=0$. We further consider a modified double-well
potential $\dWell(\CHsol)=\frac{2}{5}(\CHsol^2-1)^2$ and
$\interfaceparam=2$ to obtain the same model and parameter setting as
in Ref.~\onlinecite{Jokisaarietal_CMATS_2017}. Fig.~\ref{fig:ch-benchmark2}-(a)
shows the evolution at $\Time = 0, 200, 2000$, $3.7e+5$, and
Fig.~\ref{fig:ch-benchmark2}-(b) shows the scaling of the interface
length $\interfaceLength$ over time, on a mesh with $\meshparam=0.41$ at the interface. Note that, for this benchmark problem, we
consider a first-order surface approximation and $\PC{1}$ elements for the
phase-field function $\CHsol$ and the chemical potential $\chempot$.
As already demonstrated numerically in Refs.~\onlinecite{RV_CMS_2006,WBV_PCCP_2012}, the
coarsening process is purely driven by line tension and the spherical
geometry has no effect on the scaling behavior. As theoretically
predicted for 2D, we obtain $\interfaceLength \sim \Time^{-1/3}$.

With analogous definitions as in the previous benchmark problem, we
compute the error over different mesh refinements for a fixed
$\interfaceparam=2$, and the corresponding experimental order of
convergence. We start from the same initial condition (see
Fig.~\ref{fig:ch-benchmark2}-(a), top-left panel) and then consider
subsequent refinements of the region $\CHsol\in[-0.95,0.95]$, with mesh
sizes at the interface of $\meshparam= 1.26, 0.89, 0.63, 0.41, 0.31,
0.22$ and $\meshparam_*= 0.11$, which we set as reference level.
We consider a final time $T=200$ and choose a fixed time step $\tau=1$ for all simulations. The numerical
solutions look qualitatively equivalent to the one
presented in
Fig.~\ref{fig:ch-benchmark2}-(a). Fig.~\ref{fig:ch-benchmark2}-(c) shows the
error values and indicate second-order convergence in $\meshparam$.

\subsubsection{Surface Navier-Stokes equation}

\begin{figure*}
  \centering
  \includegraphics[width=\textwidth]{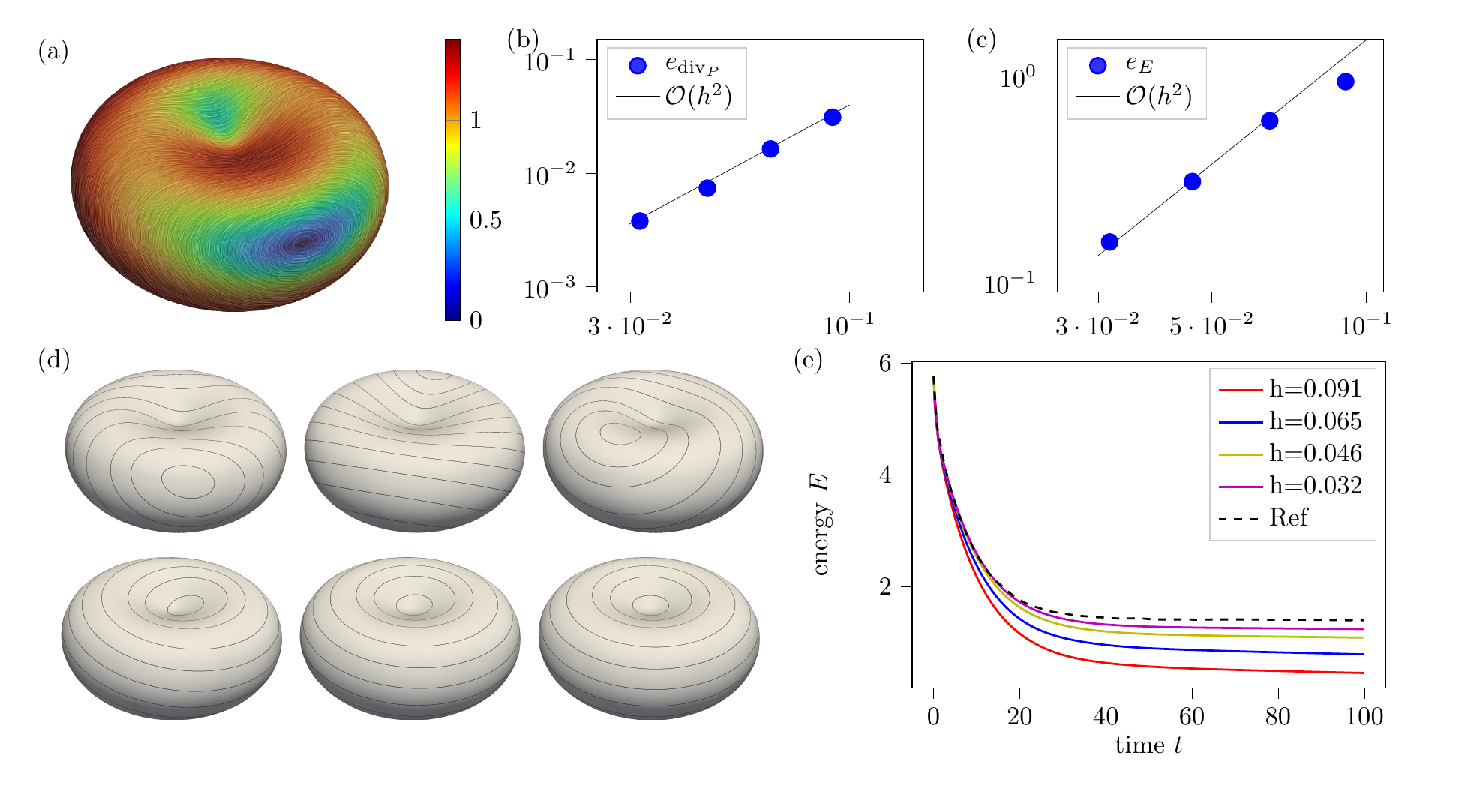}
  \caption{Numerical solution for the surface Navier-Stokes
    equation. (a) Initial value of the velocity highlighted by a
    LIC-filter color code by the magnitude of the velocity. (b)
    Convergence study of the incompressibility error $\mbox{e}_{\divP}$. (c)
    Convergence study of the energy error $\mbox{e}_{E}$, with the
    corresponding DEC solution in Ref.~\onlinecite{Nitschkeetal_TFI_2017} as
    reference. (d) Relaxation of the velocity field shown as snapshots
    of the streamlines for $\Time={0,10,20,50,75,100}$ with the final Killing
    field. (e) Evolution of the kinetic energy $E$ for different refinement levels together with the reference solution.}
  \label{fig:ns-conv}
\end{figure*}

Different numerical methods have been proposed to solve the
incompressible surface Navier-Stokes equations on stationary
surfaces. One of the possibilities is to consider
the vorticity/stream function formulation, where the problem is first
transformed to a system of surface scalar-valued PDEs and then solved
with established approaches such as surface FEM
(Refs.~\onlinecite{Nitschkeetal_JFM_2012,Reutheretal_MMS_2015,Reutheretal_MMS_2018}),
trace FEM (Ref.~\onlinecite{Brandneretal_SIAMJSC_2022}), and spectral
methods (Ref.~\onlinecite{Grossetal_JCP_2018}). Formulations in
$\vectvel, \press$ variables have to deal with tangential vector
fields and have been considered using discrete exterior calculus (DEC)
(Ref.~\onlinecite{Nitschkeetal_TFI_2017}), surface FEM
(Refs.~\onlinecite{Fries_IJNMF_2018,Reutheretal_PF_2018,Reutheretal_JFM_2020,KV_arxiv_2022}),
and trace FEM
(Refs.~\onlinecite{Brandneretal_SIAMJSC_2022,Olshanskiietal_VJM_2022}). Also
Lattice-Boltzmann approaches have been developed to solve the surface
Navier-Stokes equations (Ref.~\onlinecite{ABWPK_PRE_2019}). A precise
benchmark problem has not yet been proposed. Here, we consider a
setting that has been used to compare different numerical methods, see
Ref.~\onlinecite{Nitschkeetal_TFI_2017}, and which takes into account
geometric and topological implications. The surface $\SurfDomain$ is
given by the level-set function
$\varphi(\coord)=(a^2+\coordC_1^2+\coordC_2^2+\coordC_3^2)^2-4a^2(\coordC_1^2+\coordC_2^2)-c^4$,
with $\coord=(\coordC_1,\coordC_2,\coordC_3)\in\REAL^3$ and $a=0.72$
and $c=0.75$ \footnote{We note a typing error in
Ref.~\onlinecite{Nitschkeetal_TFI_2017} which has been confirmed by
the authors.}. The initial velocity is defined by
$\vectvel_0=\normalvec\times(0,1,1)$, and it is shown in
Fig.~\ref{fig:ns-conv}-(a). We consider the relaxation toward a
Killing field and compute the kinetic energy
$E=\frac{1}{2}\int_{\SurfDomain}\NORM{\vectvel}^2 \,
\Diff\SurfDomain$. This energy
 dissipates and is bounded from below by the energy of the resulting
Killing field. For $\Reynolds=10$, the evolution is shown by streamlines in
Fig.~\ref{fig:ns-conv}-(d). The final solution has two vortices
located on the symmetry axis. Besides this qualitative agreement with
Ref.~\onlinecite{Nitschkeetal_TFI_2017}, we analyze the
incompressibility error
$e_{\divP}=\NORM[L^\infty(L^2)]{\DivP\vectvel}$ and the energy error
$e_E=\NORM[L^\infty]{E-E_{ref}}$ with respect to $h$ and $\tau$, where
$\tau\sim \meshparam^2$.  We take the DEC
solution of Ref.~\onlinecite{Nitschkeetal_TFI_2017} as a reference
$E_{ref}$ for
$\Reynolds=10$. Fig.~\ref{fig:ns-conv}-(b) indicates second-order
convergence for the incompressibility error $e_{\divP}$, which
corresponds to the result in Ref.~\onlinecite{Brandneretal_SIAMJSC_2022}
for the surface Stokes model and is assumed to be
optimal. Fig.~\ref{fig:ns-conv}-(c) also indicates second-order
convergence for the energy error $e_E$.  The evolution of the kinetic
energy $E$ is shown in Fig.~\ref{fig:ns-conv}-(e) for different
refinement levels, together with the reference solution taken from
Ref.~\onlinecite{Nitschkeetal_TFI_2017}.

\subsubsection{Coupling}
For the coupling, we rely on established approaches in 2D for
Navier-Stokes-Cahn-Hilliard equations. They are well tested on
benchmark problems (Ref.~\onlinecite{AV_IJNMF_2012}) and can be
one-to-one extended to surfaces. We therefore simply iterate Problem~\ref{pb:ch}
and Problem~\ref{pb:ns} once for each time step. The additional curvature terms
are of lower order and should not influence the numerical properties
of the algorithm.

\section{Results}
\label{sec:3}

\begin{figure}
  \centering
  \includegraphics[width=\columnwidth,
    trim={100 0 110 0},clip
  ]{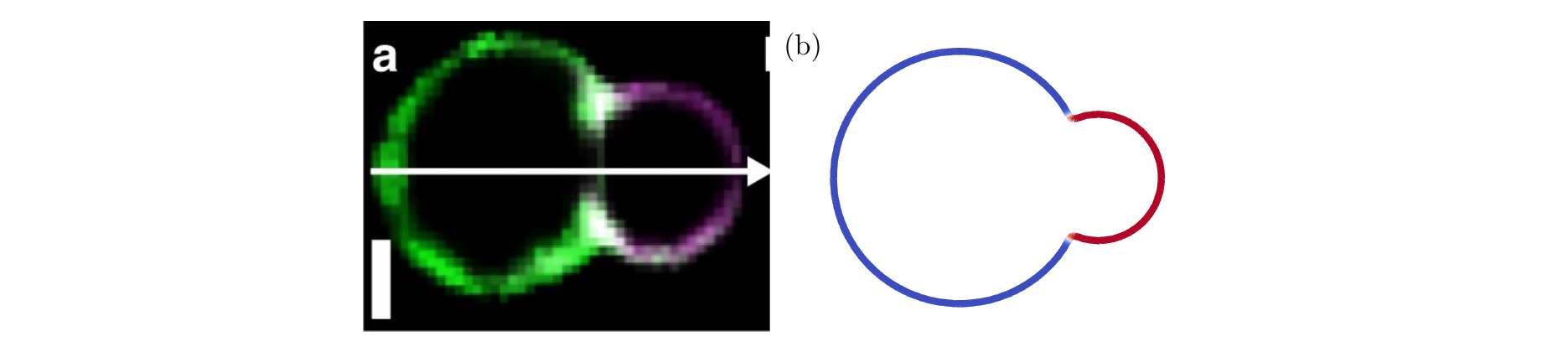}
  \caption{Equilibrium configuration. Images of
    the equatorial plane for the phase-separated state. (a) Fluorescence
    intensity along the symmetry axis of a scaffolded lipid vesicle (SLV). The image is taken with
    permission from Ref.~\onlinecite[Fig. 3]{RFGK_NC_2020}. (b) Our
    simulation result for the final phase separated state on the same
    equatorial plane.}
  \label{fig:exp_sim}
\end{figure}

We consider a set of numerical experiments designed to investigate the
influence of flow and of the additional curvature terms emerging from
the bending contribution on coarsening in
two-phase lipid membranes. We consider a geometry similar to the
asymmetric dumbbell-shape (``snowman-shape'') in
Ref.~\onlinecite{RFGK_NC_2020}. It is formed by the union of two
spheres with different radii: a sphere of radius $R_1 = 1$ intersects a
sphere of radius $R_2 = 0.5$ with a smoothed
intersection. Fig.~\ref{fig:exp_sim} shows a section along the
symmetry axis and the qualitative comparison with an experimental result
on a scaffolded lipid vesicle (SLV) taken from Ref.~\onlinecite{RFGK_NC_2020}..

\begin{figure*}
  \centering
  \includegraphics[width=\textwidth]{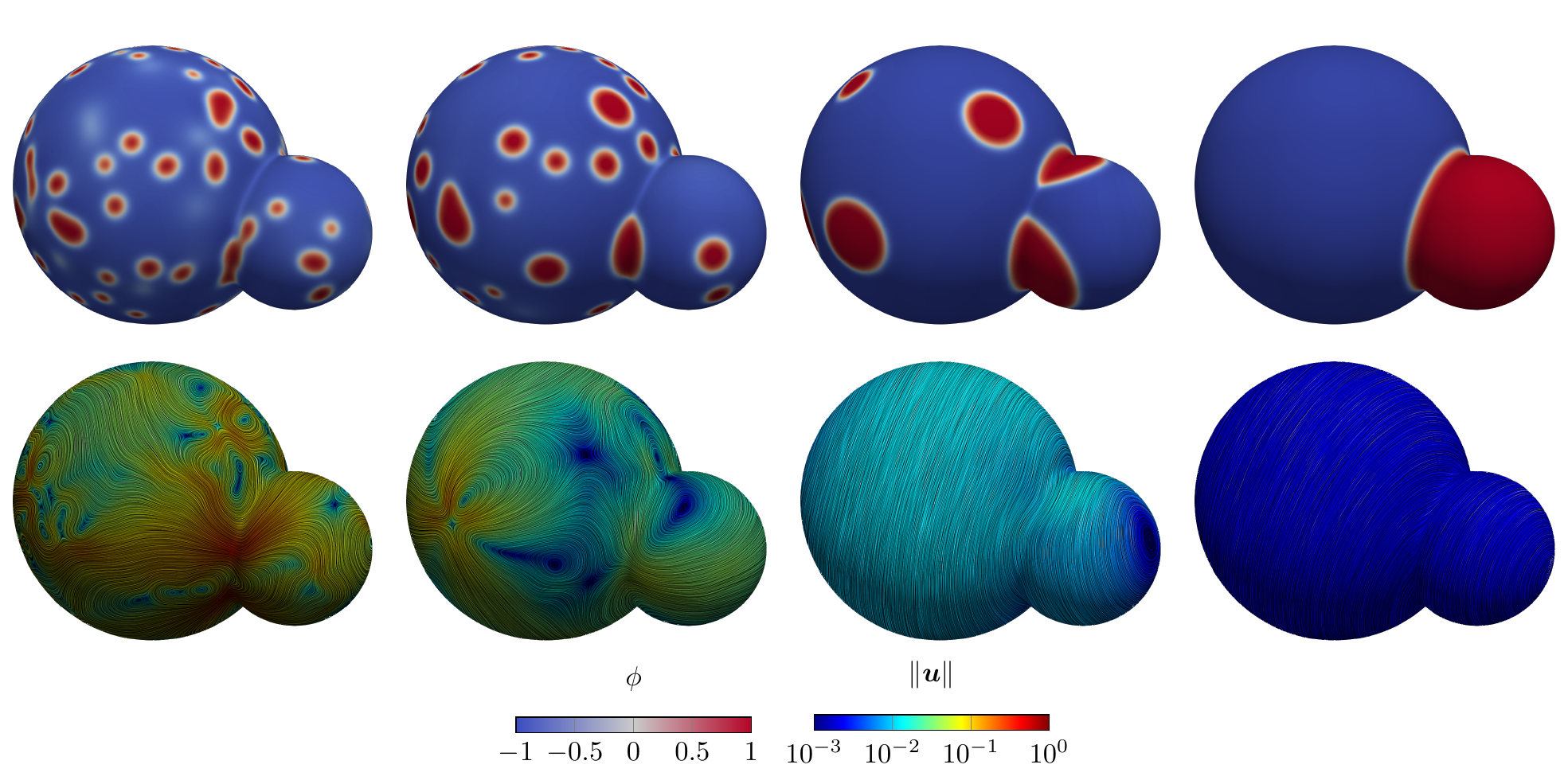}
  \caption{Snapshots of the evolution of the two-phase flow problem at
    time instances $\Time=0.2,1.0,10,1200$ (from left to right). Top
    row: phase-field function $\CHsol$. Bottom row: velocity
    $\vectvel$, visualized by a LIC filter and color-coding according
    to the magnitude of the velocity in logarithmic scale.}
  \label{fig:sim_evo}
\end{figure*}

The considered geometry is ideally suited to explore the influence of
the additional curvature terms. This influence has already been
considered in Ref.~\onlinecite{RFGK_NC_2020} for equilibrium
configurations and effects of the geometry on the phase
diagram. Therefore, here we only focus on the dynamics. We consider
$\bendStiff(\CHsol)$ as defined in Eq.~\eqref{eq:bendstiff}, with
$\bendStiff[1]=0.001$ $\bendStiff[2]=0.1$ and specify
$\meanCurv_0(\CHsol)=\meanCurv_0=0$. We vary the Reynolds number,
$\Reynolds = 3, 1$ and $0.1$, and we consider three sets of initial
configurations. In particular, we set
uniform distributed circles for the phase $\CHsol=1$, and $\CHsol=-1$
elsewhere, such that the mean value is $0.5$, $0.35$ and $0.16$, in
order to represent a composition of $50:50$, $65:35$ and $84:16$, respectively. The
latter corresponds to the case where the phase $\CHsol = -1$ occupies the
large sphere and the phase $\CHsol = 1$ the small sphere, as considered in
the qualitative comparison with the experimental data
(Fig.~\ref{fig:exp_sim}) and in the illustration of the coarsening
process and surface flow (Fig.~\ref{fig:sim_evo}). In addition, we
consider for comparison the evolution without flow, by solving Problem~\ref{pb:ch} with $\vectvel^{n-1} = 0$. The results are marked as
CH. Fig.~\ref{fig:sim_evo} shows snapshots of the evolution of the
phase-field function $\CHsol$ and the velocity $\vectvel$ for $\Reynolds = 1$
and composition $84:16$, as an example. We obtain a classical
coarsening process, with the merging and coarsening of islands. However,
due to the additional curvature terms resulting from the
bending terms and the phase-specific bending rigidity $\bendStiff$, the
process is guided toward a final configuration with the phase $\CHsol =
-1$ occupying the large sphere and the phase $\CHsol = 1$ occupying the
small sphere. The surface flow is influenced by the geometry, but
primarily by changes of the lipid patches with large magnitudes
associated with merging events. At late times, the velocity is almost
zero.

\begin{figure*}
  \centering
  \includegraphics[width=\textwidth]{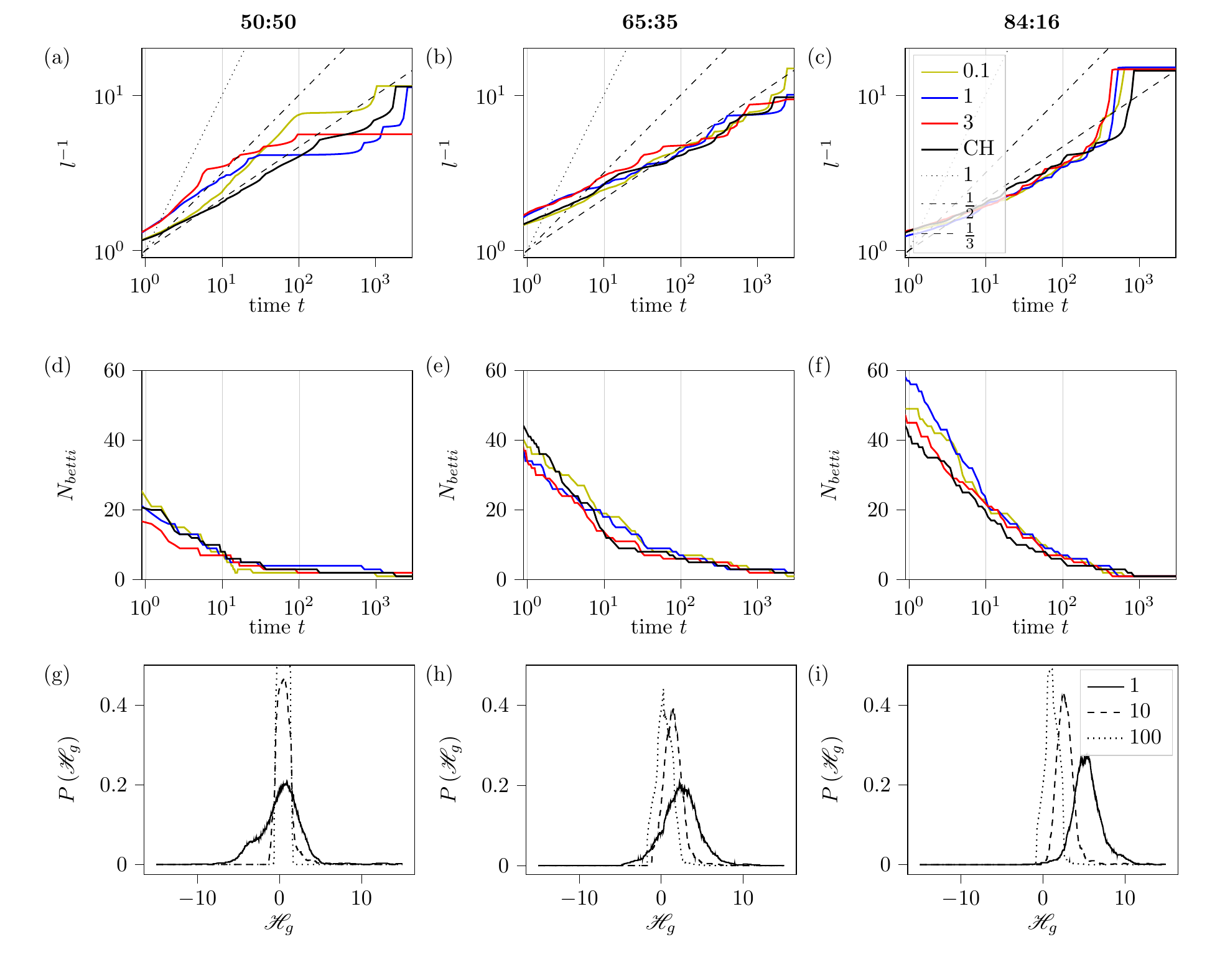}
  \caption{Scaling of interface length, topological properties, and
    geometric features for different compositions and Reynolds numbers
    $\Reynolds$. (a)-(c) Scaling of the interface length together with
    theoretically predicted exponents in 2D. The legend corresponds to
    all panels and indicates the values for $\Reynolds$. (d)-(f) Evolution of
    zeroth Betti number. (g)-(i) Interfacial shape distribution
    (ISD) for the geodesic curvature ${\cal{H}}_g$ at different time
    instances $t=1, 10, 100$ for $\Reynolds=1$. The corresponding time
    instances are highlighted in panels (a)-(f).}
  \label{fig:variationRe}
\end{figure*}

We first address the scaling behavior of the interface length. In
Fig.~\ref{fig:variationRe}, panels (a)-(c) show the results for the
considered compositions and Reynolds numbers $\Reynolds$, together
with the theoretically proposed scaling exponents for 2D: $-1$ for
viscous hydrodynamic scaling, $-1/2$ for inertial hydrodynamic
scaling, and $-1/3$ for diffusive scaling. We never observe fast
coarsening associated with viscous hydrodynamic scaling. Only for a
composition of $50:50$ we observe a dependency of the scaling exponent
on $\Reynolds$, at least within some time period. In
Ref.~\onlinecite{WY_PRL_1998}, authors argue that self-similar
growth is absent under the influence of flow and the scaling exponent
results from an interplay of fast viscous hydrodynamic scaling and
slow diffusive scaling, leading to values between $-2/3$ and $-1/3$
depending on $\Reynolds$. Our results on the dependency of the
exponent on $\Reynolds$ confirm this finding for the considered
surface. However, our results also show a stagnation of the coarsening
over a long time period. The solutions settle in local minima. We will
analyze below that these local minima are induced by geometric
properties of the surface. Only for some instances the solutions
escape from these configurations and further coarsen to the
equilibrium states. For the other compositions, i.e. $65:35$ and
$84:16$, the flow field has only a minor effect and is no longer able
to speed up the coarsening process. We obtain $\interfaceLength \sim
\Time^{-1/3}$ up to late times for all considered $\Reynolds$. This is
in agreement with the results obtained in
Ref.~\onlinecite{ABWPK_PRE_2019} on a torus. Their motivation for
the effect is related to a faster relaxation to a circular shape for
isolated patches under the influence of flow and, further, diffusive
scaling for circular patches. In order to confirm this result, we
analyze the connectivity of the phases and the interfacial shapes of
the isolated patches. This can be done by computing topological and
geometrical measures. The first measure we consider is defined by the
zeroth Betti number $N_{betti}$, which counts isolated patches of the
phase $\phi = 1$, see
Refs.~\onlinecite{Mickelinetal_PRL_2018,RV_PF_2021} for a similar
characterization. This quantity is computed for each time instance and
shown in Fig.~\ref{fig:variationRe}, panels (d)-(f). The Betti number
only weakly depends on $\Reynolds$, but we clearly see a dependency on the
composition. While for the composition $50:50$ the value is low at
early times and further decays over time, for $65:35$ we observe
significantly larger Betti numbers but also a stronger decay in
time. The large Betti numbers result from the formation of isolated
patches and the strong decay can be associated with frequent merging
events at the early stage of the coarsening process. The behavior for
the composition $84:16$ is similar to the case $65:35$, with even
larger Betti numbers at early times and even stronger decay. In this
setting, patches are smaller and more isolated leading to fewer
merging events, and the Betti number is mainly reduced by diffusive
processes. Similar qualitative results on the number of patches have
been considered on spherical unilamellar vesicles
(Ref.~\onlinecite{WPQOM_BBA_2022}). The interfacial shape of the
isolated patches is measured by computing the interfacial shape
distribution (ISD). The ISD is a probability density function for the
probability of finding a point on the interface with specific
principal curvatures. ISD is an established method to geometrically
characterize interfacial morphology in materials science and fluid
mechanics
(Refs.~\onlinecite{MAV_MMT_2003,KTV_PM_2010,HG_PRE_2019,AEVT_PRM_2020}). Adapted
to our context of an interface between two phases on a surface, the method
considers the geodesic curvature ${\cal{H}}_g$ and the corresponding
probability density $P({\cal{H}}_g)$. 

\begin{figure*}
  \centering
  \includegraphics[width=\textwidth]{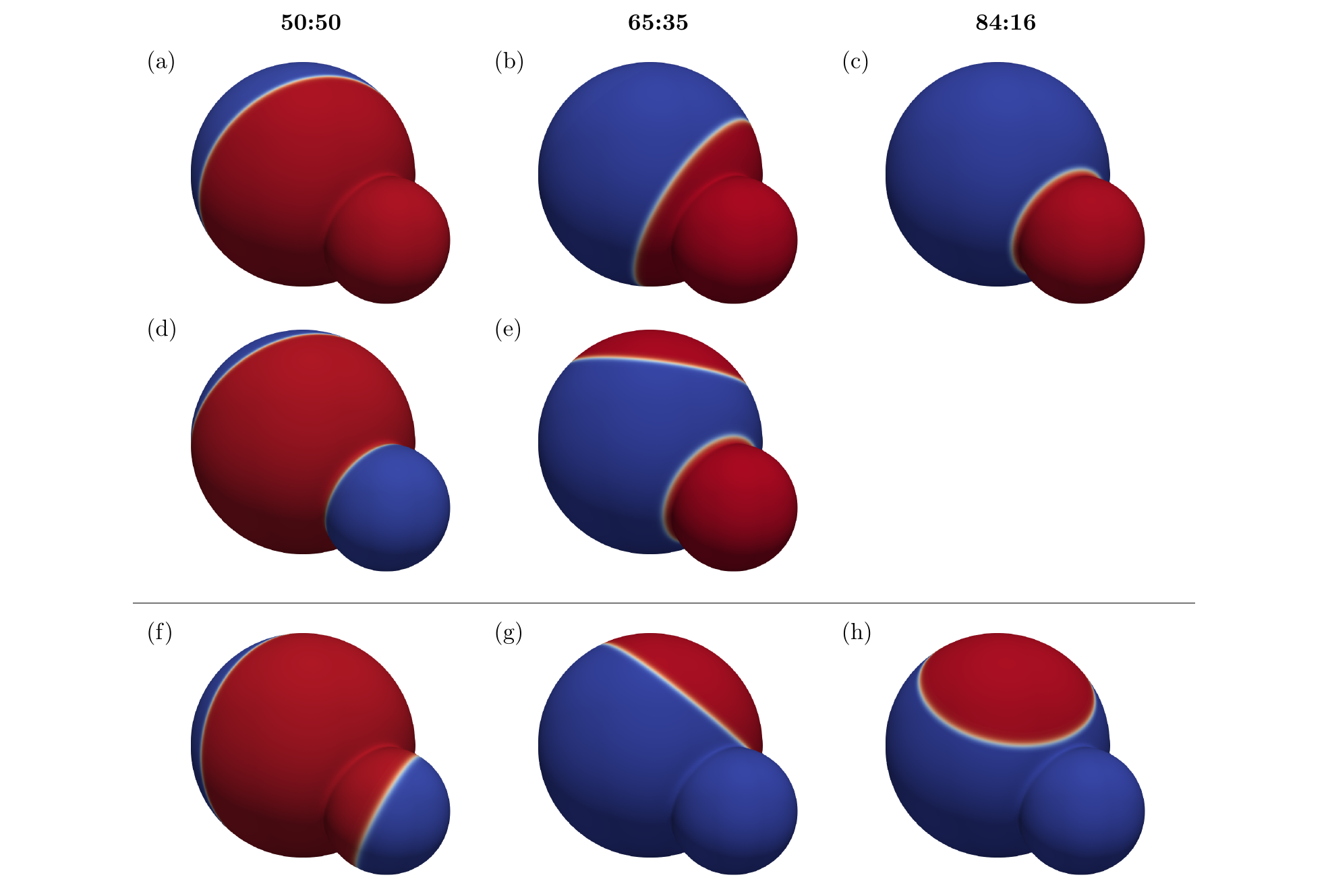}
  \caption{Different final configurations for different
    compositions. (a)-(e) With bending forces. (f)-(h) Without
    bending forces. The configurations depend on the initial
    conditions.}
  \label{fig:limitstates}
\end{figure*}

We use a phase-field approximation to compute ${\cal{H}}_g$. In 2D the mean curvature
${\cal{H}}$ can be approximated by the chemical potential, the
variational derivative of the Cahn-Hilliard energy ${\cal{F}}_{CH} =
\int_\Omega \frac{\epsilon}{2} \|\nabla \phi\|^2 + \frac{1}{\epsilon}
W(\phi) \,d\Omega$, compare the matched asymptotic analysis of a classical
Allen-Cahn equation (Ref.~\onlinecite{Fifeetal_EJDE_1995}). In our
case the chemical potential $\mu$, as considered in
Eq.~\eqref{eq:sCH2} (with $\kappa = {\cal{H}}_0 = 0$ and
$\tilde{\sigma} = \frac{2\sqrt{2}}{3}$), provides a diffuse interface
approximation of the geodesic curvature ${\cal{H}}_g$, see
Ref.~\onlinecite{BKSV_arXiv_2023} for the corresponding matched
asymptotic analysis. The chemical potential $\mu$ can be computed from the phase-field function $\CHsol$ considering a weak formulation. Therefore, we approximate ${\cal{H}}_g$ by the convolution:
\begin{align*}
{\cal{H}}_{g}(\boldsymbol{y})=\int_{\SurfDomain}
&\frac{\epsilon}{2}(\GradSurf\CHsol(\coord),\GradSurf\psi(\coord,\boldsymbol{y}))\\
&+\frac{1}{\epsilon}W(\CHsol(\coord))\psi(\coord,\boldsymbol{y})\,,\mathrm{d}\SurfDomain\,
\end{align*}
where $\psi(\coord,\boldsymbol{y})$ is a mollifier and
$\coord\in\SurfDomain$ and $\boldsymbol{y}\in\{\CHsol=0\}$. We
consider $\Reynolds =1$ and compare $P({\cal{H}}_g)$ at times $t = 1,
10, 100$, see Fig.~\ref{fig:variationRe}, panels (g)-(i). For all
compositions, the ISD sharpens over time and the mean value evolves to
small positive values, indicating an evolution toward large circular
shapes. However, we also see differences with respect to the different
compositions. For the composition of $50:50$, the ISD has a
significant portion for negative ${\cal{H}}_g$ at early times, and
therefore concave parts, which indicates a more bicontinuous-like
structure. In contrast, the ISD at early times for the composition
$84:16$ is fully located in regions for positive ${\cal{H}}_g$, and
therefore convex parts, which is a signature of isolated circular
islands. These results confirm the argument in
Ref.~\onlinecite{ABWPK_PRE_2019}. We would like to remark that the
strong peak for $t = 100$ and composition $50:50$ in
Fig.~\ref{fig:variationRe}-(g) corresponds to the plateau in
Fig.~\ref{fig:variationRe}-(a) (blue curve), which results from a
configuration with circular islands of equal size. The very strong
increase at late times in Fig.~\ref{fig:variationRe}-(c) corresponds
to the alignment of the interface with the neck. For the composition
$84:16$ the equilibrium configuration corresponds to the occupation of the
phase $\phi = -1$ on the large sphere and the phase $\phi = 1$ on the
small sphere, as shown in Figs.~\ref{fig:exp_sim} and
\ref{fig:sim_evo}.

\begin{figure*}
  \centering
  \includegraphics[width=\textwidth]{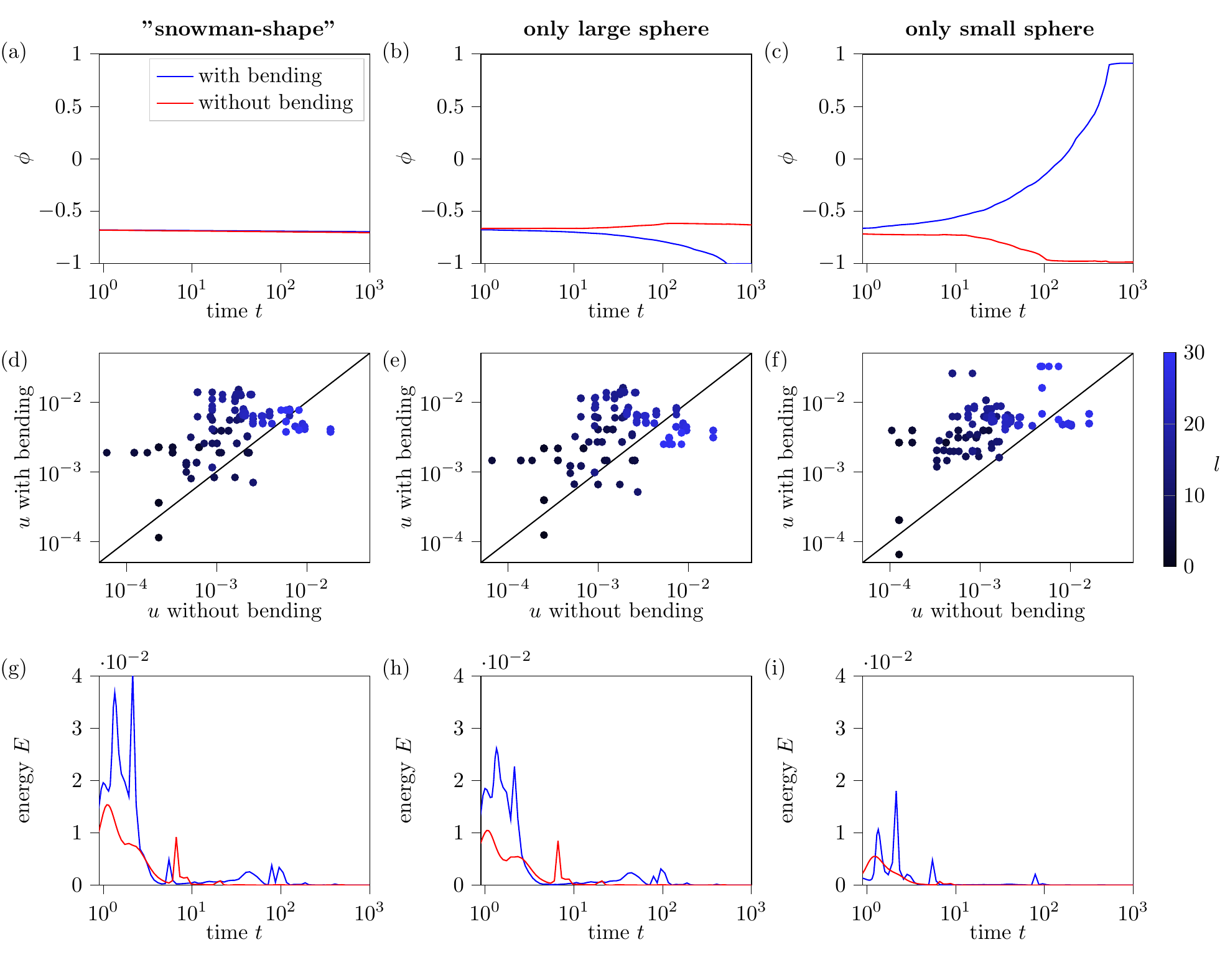}
  \caption{Influence of the bending forces on the evolution for the
    distribution $84:16$ on the ``snowman-shape''. Results are shown
    on the whole geometry, only on the large and only on the small
    sphere, respectively. We compare simulations with and without
    bending forces. (a)-(b)-(c) Mean value of $\CHsol$ over
    time. (d)-(e)-(f) Comparison of the magnitude of the mean velocity
    $u$ as a function of the interface length $l$. The values without
    bending terms are plotted against the corresponding values with
    bending terms and the interface length of the data points is color
    coded. (g)-(h)-(i) Evolution of the kinetic energy $E$ with and
    without bending terms.}
  \label{fig:influenceBending}
\end{figure*}

Fig.~\ref{fig:limitstates} shows some different possibilities for the
final configuration. They depend on the initial condition. While in
the benchmark problems on a sphere in Section~\ref{sec:sCH} any
rotation of the final solution has the same energy, the considered
``snowman-shape'' geometry breaks this symmetry. The interface length of the
final configuration depends on its position, and this dependency is most
prominent for the composition $84:16$. In
Fig.~\ref{fig:limitstates}, panels (c) and (h) show two
configurations, one with bending terms and one without. The optimal
solution is obtained where the interface is along the neck and it is reached
for the case with bending terms, while a geodesic
circle on the large sphere is formed in the case without bending.
The length of the interface $\interfaceLength_{\interfaceparam}$ is
measured by the Cahn-Hilliard part of the energy. From the
computation of
the length values as in Section~\ref{sec:sCH}, we obtain
$\interfaceLength_{\interfaceparam} = 3.1$ (with bending,
Fig.~\ref{fig:limitstates}-(c)) and
$\interfaceLength_{\interfaceparam} = 4.6$ (without bending,
Fig.~\ref{fig:limitstates}-(h)). Thus, the bending terms favor the optimal
configuration. The energy barrier between these configurations is huge
as the geometry requires to enlarge the interface in order to
transform the solution in Fig.~\ref{fig:limitstates}-(h) into the one
in Fig.~\ref{fig:limitstates}-(c). Also for the other compositions,
the neck plays a significant role. In Fig.~\ref{fig:limitstates},
panels (a)-(b)-(e), the small sphere is fully covered by the phase
$\phi = 1$, which is energetically favorable due to the lower bending
rigidity. In Fig.~\ref{fig:limitstates}-(d) we see the opposite: the
small sphere is fully covered by the phase $\phi = -1$ and part of the
interface is along the neck, and the phase $\phi = 1$ is on the
energetically unfavorable large sphere. However, to further reduce the
energy, either by reducing the interface or redistributing the phases,
requires to move the interface away from the neck and this can only be
realized by first enlarging it.
So again we have an energetic barrier
induced by the geometry.
These configurations explain the plateaus
observed in Fig.~\ref{fig:variationRe}-(a). Without bending, we do not
favor any spatial configurations. The final configurations are solely
determined by minimizing the length of the interface. However, also
this process is influenced by the geometry, as the length corresponds
to the geodesic length. Computing the length of the interface
$\interfaceLength_{\interfaceparam}$ for the corresponding
compositions we have:  $\interfaceLength_{\interfaceparam} = 6.3$ (with bending,
  Fig.~\ref{fig:limitstates}-(a)), $\interfaceLength_{\interfaceparam}
  = 9.14$ (with bending, Fig.~\ref{fig:limitstates}-(d)) and
  $\interfaceLength_{\interfaceparam} = 3.4$ (without bending,
  Fig.~\ref{fig:limitstates}-(f)) for the $50:50$ composition; 
$\interfaceLength_{\interfaceparam} = 5.5$ (with bending,
Fig.~\ref{fig:limitstates}-(b)), $\interfaceLength_{\interfaceparam} =
8.2$ (with bending, Fig.~\ref{fig:limitstates}-(e)) and
$\interfaceLength_{\interfaceparam} = 6.1$ (without bending,
Fig.~\ref{fig:limitstates}-(g)) for the $65:35$ composition.
These examples show that the bending
terms can also lead to solutions that are not optimal if only the
interface length is considered. This observation is in agreement with
Ref.~\onlinecite{RFGK_NC_2020}, where the complex interplay of
composition and geometric shape has been explored for the equilibrium
configurations.

We now discuss the influence of the bending terms on 
evolution. Not only the final spatial arrangement is influenced by the
geometric properties, but also the temporal evolution. To see this, we
consider the coarsening process on the subdomains of the two different
spheres of the ``snowman-shape''. Fig.~\ref{fig:influenceBending}
shows simulation results for the composition $84:16$ with $\Reynolds=1$, with and
without bending terms.
Fig.~\ref{fig:influenceBending}-(a) shows the mean value of
$\CHsol$ on the ``snowman-shape'', which essentially just confirms the
conserved evolution, and panels (b) and (c) show the mean value on the
large and the small sphere, respectively. We compare the simulations with and
without bending terms. With (without)
bending the solution on the large sphere converges to a state with
$\CHsol = -1$ (essentially remains at its mean value) and the solution
on the small sphere converges to a state with $\CHsol = 1$
($\CHsol = -1$). The last is due to the larger area of
the large sphere, which is related to a larger probability to locate the
phase $\phi = 1$ there. In Fig.~\ref{fig:influenceBending}, panels (d)-(f)
consider the influence of the bending terms on the velocity
$\mathbf{u}$. We compute the mean velocity on the whole domain, on the
large sphere, and on the small sphere at each time step. To be able to
compare the results with and without bending, we consider time
instances that have the same interface length $l$ and compare these
values. Fig.~\ref{fig:influenceBending}-(d)-(f) shows the comparison
for the magnitude of the averaged velocities $u$ related to the
interface length $l$. Points above the diagonal line indicate an
increased velocity due to bending forces. This is significant in all
three plots but most enhanced on the small sphere. The color of the
data points corresponds to the interface length $l$. For both
situations, with and without bending terms, longer interfaces,
corresponding to earlier times, are related to larger velocities. At
least qualitatively, these results are confirmed by
Fig.~\ref{fig:influenceBending}-(g)-(i), which show the evolution of
the kinetic energy $E$. With bending forces, the kinetic energy is
significantly larger than without bending forces, but in both
situations the kinetic energy decays in time and converges to zero.

\section{Discussion}
\label{sec:4}

We consider a detailed computational study of the role of membrane
curvature on spatial arrangements of lipid phases and their evolution
in idealized biomembranes. We consider multicomponent scaffolded lipid
vesicles (SLV) and investigate the effect of membrane viscosity,
phase-dependent bending rigidity, and phase-dependent spontaneous
curvature on the coarsening process and the emerging equilibrium
configurations. The mathematical model results from a phase-field
approximation of a J\"ulicher-Lipowski energy
(Ref.~\onlinecite{JL_PRL_1993}) with surface flow. It is considered on
a stationary surface and can be viewed as a surface
Navier-Stokes-Cahn-Hilliard model with additional phase-dependent
bending terms. We solve these equations by SFEM
(Refs.~\onlinecite{DE_AN_2013,NNV_JCP_2019}) taking into account
numerical analysis results for the surface Stokes equations to ensure
convergence. We carefully validate the algorithm by convergence
studies of the involved subproblems. For the coupling of these
subproblems, we rely on established strategies in 2D. For the physical
investigations, we consider a ``snowman-shape'' geometry as in the experiments
in Ref.~\onlinecite{RFGK_NC_2020}. We identify the influence of the
Reynolds number $\Reynolds$ and the composition on the scaling
behavior, consider topological and geometrical measures to explain the
results, and identify the effect of the bending terms on the evolution
and the reached equilibrium shape. Concerning the scaling behavior,
the influence of hydrodynamics and compositions known from 2D can also
be found on curved surfaces, at least qualitatively, and explained by
similar arguments. However, quantitatively local geometric features,
such as the considered neck in the ``snowman-shape'', can lead to
various local minima and extended plateaus in the coarsening
process. In situations where the composition allows for a realization
of the geometrically favored configuration, the hydrodynamic enhances
the convergence into this configuration. In our setting, this
corresponds to a composition of $84:16$ with the phase $\CHsol = 1$ on
the small sphere, the phase $\CHsol = -1$ on the large sphere, and the
interface $\phi = 0$ along the neck. The phase-dependent properties in
the bending rigidity and in the spontaneous curvature influence the
coarsening process. These properties guide the lipid phase with lower
bending rigidity toward regions of higher mean curvature, the lipid
phase with higher bending rigidity toward regions of lower mean
curvature, as well as phases with spontaneous curvature toward regions
with the corresponding curvature values. These bending terms are additional
driving forces for the flow field and lead to the observed increase in
kinetic energy.

While the considered ``snowman-shape'' might seem a very special case and the
observed phenomena constructed, we expect these effects to have a broader
impact as soon as the surface is allowed to evolve. In the evolving case, the
``snowman-shape'' can be related to morphologies observed in endocytosis and
exocytosis. The hydrodynamically enhanced phase evolution might be a
key mechanical factor in these processes. However, while endocytosis
and exocytosis are addressed with phase-field models, surface
viscosity is not taken into account. Numerical tools to solve the surface
Navier-Stokes equations on evolving surfaces, so-called fluid
deformable surfaces, have been developed
(Refs.~\onlinecite{Torres_Sanchezetal_JFM_2019,Reutheretal_JFM_2020,KV_arxiv_2022}). These approaches need to be extend to two-phase flows. However, the derivation of a corresponding
thermodynamically consistent two-phase flow problem on an evolving
surface requires some additional thought, see
Refs.~\onlinecite{Al-Izzietal_SCDB_2021,Hoffmannetal_SA_2022,NitschkeVoigt_JoGaP_2022,NitschkeSadikVoigt_A_2022}.

\begin{acknowledgments}
This work was supported by the German Research Foundation (DFG) within
the Research Unit ``Vector- and Tensor-Valued Surface PDEs'' (FOR
3013). We further acknowledge computing resources provided by ZIH at
TU Dresden and by JSC at FZ J\"ulich, within projects WIR and PFAMDIS,
respectively.
\end{acknowledgments}

\section*{Author declarations}
\subsection*{Author contributions}
E.B. and V.K. implemented the codes, performed all simulations, analyzed data, and contributed to conceptual development and manuscript writing. A.V. supervised the research and contributed to conceptual development, data analysis, and manuscript writing. 

\subsection*{Competing interests} 
The authors declare no competing interests. 

\section*{Data availability}
The data that support the findings of this study are available
from the corresponding author upon reasonable request.

\bibliography{strings,biblio}

\begin{thebibliography}{85}%
\makeatletter
\providecommand \@ifxundefined [1]{%
 \@ifx{#1\undefined}
}%
\providecommand \@ifnum [1]{%
 \ifnum #1\expandafter \@firstoftwo
 \else \expandafter \@secondoftwo
 \fi
}%
\providecommand \@ifx [1]{%
 \ifx #1\expandafter \@firstoftwo
 \else \expandafter \@secondoftwo
 \fi
}%
\providecommand \natexlab [1]{#1}%
\providecommand \enquote  [1]{``#1''}%
\providecommand \bibnamefont  [1]{#1}%
\providecommand \bibfnamefont [1]{#1}%
\providecommand \citenamefont [1]{#1}%
\providecommand \href@noop [0]{\@secondoftwo}%
\providecommand \href [0]{\begingroup \@sanitize@url \@href}%
\providecommand \@href[1]{\@@startlink{#1}\@@href}%
\providecommand \@@href[1]{\endgroup#1\@@endlink}%
\providecommand \@sanitize@url [0]{\catcode `\\12\catcode `\$12\catcode
  `\&12\catcode `\#12\catcode `\^12\catcode `\_12\catcode `\%12\relax}%
\providecommand \@@startlink[1]{}%
\providecommand \@@endlink[0]{}%
\providecommand \url  [0]{\begingroup\@sanitize@url \@url }%
\providecommand \@url [1]{\endgroup\@href {#1}{\urlprefix }}%
\providecommand \urlprefix  [0]{URL }%
\providecommand \Eprint [0]{\href }%
\providecommand \doibase [0]{http://dx.doi.org/}%
\providecommand \selectlanguage [0]{\@gobble}%
\providecommand \bibinfo  [0]{\@secondoftwo}%
\providecommand \bibfield  [0]{\@secondoftwo}%
\providecommand \translation [1]{[#1]}%
\providecommand \BibitemOpen [0]{}%
\providecommand \bibitemStop [0]{}%
\providecommand \bibitemNoStop [0]{.\EOS\space}%
\providecommand \EOS [0]{\spacefactor3000\relax}%
\providecommand \BibitemShut  [1]{\csname bibitem#1\endcsname}%
\let\auto@bib@innerbib\@empty
\bibitem [{\citenamefont {McMahon}\ and\ \citenamefont
  {Gallop}(2005)}]{MG_N_2005}%
  \BibitemOpen
  \bibfield  {author} {\bibinfo {author} {\bibfnamefont {H.~T.}\ \bibnamefont
  {McMahon}}\ and\ \bibinfo {author} {\bibfnamefont {J.~L.}\ \bibnamefont
  {Gallop}},\ }\bibfield  {title} {\enquote {\bibinfo {title} {Membrane
  curvature and mechanisms of dynamic cell membrane remodelling},}\ }\href@noop
  {} {\bibfield  {journal} {\bibinfo  {journal} {Nature}\ }\textbf {\bibinfo
  {volume} {438}},\ \bibinfo {pages} {590--596} (\bibinfo {year}
  {2005})}\BibitemShut {NoStop}%
\bibitem [{\citenamefont {Stanich}\ \emph {et~al.}(2013)\citenamefont
  {Stanich}, \citenamefont {Honerkamp-Smith}, \citenamefont {Putzel},
  \citenamefont {Warth}, \citenamefont {Lamprecht}, \citenamefont {Mandal},
  \citenamefont {Mann}, \citenamefont {Hua},\ and\ \citenamefont
  {Keller}}]{Stanichetal_BPJ_2013}%
  \BibitemOpen
  \bibfield  {author} {\bibinfo {author} {\bibfnamefont {C.~A.}\ \bibnamefont
  {Stanich}}, \bibinfo {author} {\bibfnamefont {A.~R.}\ \bibnamefont
  {Honerkamp-Smith}}, \bibinfo {author} {\bibfnamefont {G.~G.}\ \bibnamefont
  {Putzel}}, \bibinfo {author} {\bibfnamefont {C.~S.}\ \bibnamefont {Warth}},
  \bibinfo {author} {\bibfnamefont {A.~K.}\ \bibnamefont {Lamprecht}}, \bibinfo
  {author} {\bibfnamefont {P.}~\bibnamefont {Mandal}}, \bibinfo {author}
  {\bibfnamefont {E.}~\bibnamefont {Mann}}, \bibinfo {author} {\bibfnamefont
  {T.-A.~D.}\ \bibnamefont {Hua}}, \ and\ \bibinfo {author} {\bibfnamefont
  {S.~L.}\ \bibnamefont {Keller}},\ }\bibfield  {title} {\enquote {\bibinfo
  {title} {Coarsening dynamics of domains in lipid membranes},}\ }\href@noop {}
  {\bibfield  {journal} {\bibinfo  {journal} {Biophys. J.}\ }\textbf {\bibinfo
  {volume} {105}},\ \bibinfo {pages} {444--454} (\bibinfo {year}
  {2013})}\BibitemShut {NoStop}%
\bibitem [{\citenamefont {Fan}, \citenamefont {Han},\ and\ \citenamefont
  {Haataja}(2010)}]{FHH_JCP_2010}%
  \BibitemOpen
  \bibfield  {author} {\bibinfo {author} {\bibfnamefont {J.}~\bibnamefont
  {Fan}}, \bibinfo {author} {\bibfnamefont {T.}~\bibnamefont {Han}}, \ and\
  \bibinfo {author} {\bibfnamefont {M.}~\bibnamefont {Haataja}},\ }\bibfield
  {title} {\enquote {\bibinfo {title} {Hydrodynamic effects on spinodal
  decomposition kinetics in planar lipid bilayer membranes},}\ }\href@noop {}
  {\bibfield  {journal} {\bibinfo  {journal} {J. Chem. Phys.}\ }\textbf
  {\bibinfo {volume} {133}},\ \bibinfo {pages} {235101} (\bibinfo {year}
  {2010})}\BibitemShut {NoStop}%
\bibitem [{\citenamefont {Camlay}\ and\ \citenamefont
  {Brown}(2011)}]{CB_JCP_2011}%
  \BibitemOpen
  \bibfield  {author} {\bibinfo {author} {\bibfnamefont {B.~A.}\ \bibnamefont
  {Camlay}}\ and\ \bibinfo {author} {\bibfnamefont {F.~L.~H.}\ \bibnamefont
  {Brown}},\ }\bibfield  {title} {\enquote {\bibinfo {title} {Dynamic scaling
  in phase separation kinetics for quasi-two-dimensional membranes},}\
  }\href@noop {} {\bibfield  {journal} {\bibinfo  {journal} {J. Chem. Phys.}\
  }\textbf {\bibinfo {volume} {135}},\ \bibinfo {pages} {225106} (\bibinfo
  {year} {2011})}\BibitemShut {NoStop}%
\bibitem [{\citenamefont {Saffman}\ and\ \citenamefont
  {Delbr\"uck}(1975)}]{SD_PNAS_1975}%
  \BibitemOpen
  \bibfield  {author} {\bibinfo {author} {\bibfnamefont {P.~G.}\ \bibnamefont
  {Saffman}}\ and\ \bibinfo {author} {\bibfnamefont {M.}~\bibnamefont
  {Delbr\"uck}},\ }\bibfield  {title} {\enquote {\bibinfo {title} {Brownian
  motion in biological membranes},}\ }\href@noop {} {\bibfield  {journal}
  {\bibinfo  {journal} {Proc. Natl. Acad. Sci. USA}\ }\textbf {\bibinfo
  {volume} {72}},\ \bibinfo {pages} {3111--3113} (\bibinfo {year}
  {1975})}\BibitemShut {NoStop}%
\bibitem [{\citenamefont {Baumgart}, \citenamefont {Hess},\ and\ \citenamefont
  {Webb}(2003)}]{BHW_N_2003}%
  \BibitemOpen
  \bibfield  {author} {\bibinfo {author} {\bibfnamefont {T.}~\bibnamefont
  {Baumgart}}, \bibinfo {author} {\bibfnamefont {S.~T.}\ \bibnamefont {Hess}},
  \ and\ \bibinfo {author} {\bibfnamefont {W.~W.}\ \bibnamefont {Webb}},\
  }\bibfield  {title} {\enquote {\bibinfo {title} {Imaging coexisting fluid
  domains in biomembrane models coupling curvature and line tension},}\
  }\href@noop {} {\bibfield  {journal} {\bibinfo  {journal} {Nature}\ }\textbf
  {\bibinfo {volume} {425}},\ \bibinfo {pages} {821--824} (\bibinfo {year}
  {2003})}\BibitemShut {NoStop}%
\bibitem [{\citenamefont {Baumgart}\ \emph {et~al.}(2005)\citenamefont
  {Baumgart}, \citenamefont {Das}, \citenamefont {Webb},\ and\ \citenamefont
  {Jenkins}}]{BDWJ_BPJ_2005}%
  \BibitemOpen
  \bibfield  {author} {\bibinfo {author} {\bibfnamefont {T.}~\bibnamefont
  {Baumgart}}, \bibinfo {author} {\bibfnamefont {S.}~\bibnamefont {Das}},
  \bibinfo {author} {\bibfnamefont {W.~W.}\ \bibnamefont {Webb}}, \ and\
  \bibinfo {author} {\bibfnamefont {J.~T.}\ \bibnamefont {Jenkins}},\
  }\bibfield  {title} {\enquote {\bibinfo {title} {Membrane elasticity in giant
  vesicles with fluid phase coexistence},}\ }\href@noop {} {\bibfield
  {journal} {\bibinfo  {journal} {Biophys. J.}\ }\textbf {\bibinfo {volume}
  {89}},\ \bibinfo {pages} {1067--1080} (\bibinfo {year} {2005})}\BibitemShut
  {NoStop}%
\bibitem [{\citenamefont {Wand}\ and\ \citenamefont {Du}(2008)}]{WD_JMB_2008}%
  \BibitemOpen
  \bibfield  {author} {\bibinfo {author} {\bibfnamefont {X.}~\bibnamefont
  {Wand}}\ and\ \bibinfo {author} {\bibfnamefont {Q.}~\bibnamefont {Du}},\
  }\bibfield  {title} {\enquote {\bibinfo {title} {Modelling and simulations of
  multi-component lipid membranes and open membranes via diffuse interface
  approaches},}\ }\href@noop {} {\bibfield  {journal} {\bibinfo  {journal} {:
  Math. Bio.}\ }\textbf {\bibinfo {volume} {56}},\ \bibinfo {pages} {347--371}
  (\bibinfo {year} {2008})}\BibitemShut {NoStop}%
\bibitem [{\citenamefont {Lowengrub}, \citenamefont {R\"atz},\ and\
  \citenamefont {Voigt}(2009)}]{LRV_PRE_2009}%
  \BibitemOpen
  \bibfield  {author} {\bibinfo {author} {\bibfnamefont {J.~S.}\ \bibnamefont
  {Lowengrub}}, \bibinfo {author} {\bibfnamefont {A.}~\bibnamefont {R\"atz}}, \
  and\ \bibinfo {author} {\bibfnamefont {A.}~\bibnamefont {Voigt}},\ }\bibfield
   {title} {\enquote {\bibinfo {title} {Phase-field modeling of the dynamics of
  multicomponent vesicles: Spinodal decomposition, coarsening, budding, and
  fission},}\ }\href@noop {} {\bibfield  {journal} {\bibinfo  {journal} {Phys.
  Rev. E}\ }\textbf {\bibinfo {volume} {79}},\ \bibinfo {pages} {031926}
  (\bibinfo {year} {2009})}\BibitemShut {NoStop}%
\bibitem [{\citenamefont {Elliott}\ and\ \citenamefont
  {Stinner}(2010)}]{ES_SIAMJAM_2010}%
  \BibitemOpen
  \bibfield  {author} {\bibinfo {author} {\bibfnamefont {C.~M.}\ \bibnamefont
  {Elliott}}\ and\ \bibinfo {author} {\bibfnamefont {B.}~\bibnamefont
  {Stinner}},\ }\bibfield  {title} {\enquote {\bibinfo {title} {A surface phase
  field model for two-phase biological membranes},}\ }\href@noop {} {\bibfield
  {journal} {\bibinfo  {journal} {SIAM J. Appl. Math.}\ }\textbf {\bibinfo
  {volume} {70}},\ \bibinfo {pages} {2904--2928} (\bibinfo {year}
  {2010})}\BibitemShut {NoStop}%
\bibitem [{\citenamefont {Elson}\ \emph {et~al.}(2010)\citenamefont {Elson},
  \citenamefont {Fried}, \citenamefont {Dolbow},\ and\ \citenamefont
  {Genin}}]{EFDG_ARBP_2010}%
  \BibitemOpen
  \bibfield  {author} {\bibinfo {author} {\bibfnamefont {E.~L.}\ \bibnamefont
  {Elson}}, \bibinfo {author} {\bibfnamefont {E.}~\bibnamefont {Fried}},
  \bibinfo {author} {\bibfnamefont {J.~E.}\ \bibnamefont {Dolbow}}, \ and\
  \bibinfo {author} {\bibfnamefont {G.~M.}\ \bibnamefont {Genin}},\ }\bibfield
  {title} {\enquote {\bibinfo {title} {Phase separation in biological
  membranes: Integration of theory and experiment},}\ }\href@noop {} {\bibfield
   {journal} {\bibinfo  {journal} {Annual Review of Biophysics}\ }\textbf
  {\bibinfo {volume} {39}},\ \bibinfo {pages} {207--266} (\bibinfo {year}
  {2010})}\BibitemShut {NoStop}%
\bibitem [{\citenamefont {Garcke}\ \emph {et~al.}(2016)\citenamefont {Garcke},
  \citenamefont {Kampmann}, \citenamefont {R\"atz},\ and\ \citenamefont
  {R\"oger}}]{GKRR_MMMAS_2016}%
  \BibitemOpen
  \bibfield  {author} {\bibinfo {author} {\bibfnamefont {H.}~\bibnamefont
  {Garcke}}, \bibinfo {author} {\bibfnamefont {J.}~\bibnamefont {Kampmann}},
  \bibinfo {author} {\bibfnamefont {A.}~\bibnamefont {R\"atz}}, \ and\ \bibinfo
  {author} {\bibfnamefont {M.}~\bibnamefont {R\"oger}},\ }\bibfield  {title}
  {\enquote {\bibinfo {title} {A coupled surface-{C}ahn-{H}illiard
  bulk-diffusion system modeling lipid raft formation in cell membranes},}\
  }\href@noop {} {\bibfield  {journal} {\bibinfo  {journal} {Math. Models Meth.
  Appl. Sci.}\ }\textbf {\bibinfo {volume} {26}},\ \bibinfo {pages}
  {1149--1189} (\bibinfo {year} {2016})}\BibitemShut {NoStop}%
\bibitem [{\citenamefont {Gera}\ and\ \citenamefont
  {Salac}(2018)}]{GS_CF_2018}%
  \BibitemOpen
  \bibfield  {author} {\bibinfo {author} {\bibfnamefont {P.}~\bibnamefont
  {Gera}}\ and\ \bibinfo {author} {\bibfnamefont {D.}~\bibnamefont {Salac}},\
  }\bibfield  {title} {\enquote {\bibinfo {title} {Modeling of multicomponent
  three-dimensional vesicles},}\ }\href@noop {} {\bibfield  {journal} {\bibinfo
   {journal} {Computers \& Fluids}\ }\textbf {\bibinfo {volume} {172}},\
  \bibinfo {pages} {362--383} (\bibinfo {year} {2018})}\BibitemShut {NoStop}%
\bibitem [{\citenamefont {Zimmermann}\ \emph {et~al.}(2019)\citenamefont
  {Zimmermann}, \citenamefont {Toshniwal}, \citenamefont {Landis},
  \citenamefont {Hughes}, \citenamefont {Mandadapu},\ and\ \citenamefont
  {Sauer}}]{Zimmermannetal_CMAME_2019}%
  \BibitemOpen
  \bibfield  {author} {\bibinfo {author} {\bibfnamefont {C.}~\bibnamefont
  {Zimmermann}}, \bibinfo {author} {\bibfnamefont {D.}~\bibnamefont
  {Toshniwal}}, \bibinfo {author} {\bibfnamefont {C.~M.}\ \bibnamefont
  {Landis}}, \bibinfo {author} {\bibfnamefont {T.~J.}\ \bibnamefont {Hughes}},
  \bibinfo {author} {\bibfnamefont {K.~K.}\ \bibnamefont {Mandadapu}}, \ and\
  \bibinfo {author} {\bibfnamefont {R.~A.}\ \bibnamefont {Sauer}},\ }\bibfield
  {title} {\enquote {\bibinfo {title} {An isogeometric finite element
  formulation for phase transitions on deforming surfaces},}\ }\href@noop {}
  {\bibfield  {journal} {\bibinfo  {journal} {Computer Methods in Applied
  Mechanics and Engineering}\ }\textbf {\bibinfo {volume} {351}},\ \bibinfo
  {pages} {441--477} (\bibinfo {year} {2019})}\BibitemShut {NoStop}%
\bibitem [{\citenamefont {Fonda}\ \emph {et~al.}(2018)\citenamefont {Fonda},
  \citenamefont {Rinaldin}, \citenamefont {Kraft},\ and\ \citenamefont
  {Giomi}}]{FRKG_PRE_2018}%
  \BibitemOpen
  \bibfield  {author} {\bibinfo {author} {\bibfnamefont {P.}~\bibnamefont
  {Fonda}}, \bibinfo {author} {\bibfnamefont {M.}~\bibnamefont {Rinaldin}},
  \bibinfo {author} {\bibfnamefont {D.~J.}\ \bibnamefont {Kraft}}, \ and\
  \bibinfo {author} {\bibfnamefont {L.}~\bibnamefont {Giomi}},\ }\bibfield
  {title} {\enquote {\bibinfo {title} {Interface geometry of binary mixtures on
  curved substrates},}\ }\href@noop {} {\bibfield  {journal} {\bibinfo
  {journal} {Phys. Rev. E}\ }\textbf {\bibinfo {volume} {98}},\ \bibinfo
  {pages} {032801} (\bibinfo {year} {2018})}\BibitemShut {NoStop}%
\bibitem [{\citenamefont {Fonda}\ \emph {et~al.}(2019)\citenamefont {Fonda},
  \citenamefont {Rinaldin}, \citenamefont {Kraft},\ and\ \citenamefont
  {Giomi}}]{FRKG_PRE_2019}%
  \BibitemOpen
  \bibfield  {author} {\bibinfo {author} {\bibfnamefont {P.}~\bibnamefont
  {Fonda}}, \bibinfo {author} {\bibfnamefont {M.}~\bibnamefont {Rinaldin}},
  \bibinfo {author} {\bibfnamefont {D.~J.}\ \bibnamefont {Kraft}}, \ and\
  \bibinfo {author} {\bibfnamefont {L.}~\bibnamefont {Giomi}},\ }\bibfield
  {title} {\enquote {\bibinfo {title} {Thermodynamic equilibrium of binary
  mixtures on curved surfaces},}\ }\href@noop {} {\bibfield  {journal}
  {\bibinfo  {journal} {Phys. Rev. E}\ }\textbf {\bibinfo {volume} {100}},\
  \bibinfo {pages} {032604} (\bibinfo {year} {2019})}\BibitemShut {NoStop}%
\bibitem [{\citenamefont {Rinaldin}\ \emph {et~al.}(2020)\citenamefont
  {Rinaldin}, \citenamefont {Fonda}, \citenamefont {Giomi},\ and\ \citenamefont
  {Kraft}}]{RFGK_NC_2020}%
  \BibitemOpen
  \bibfield  {author} {\bibinfo {author} {\bibfnamefont {M.}~\bibnamefont
  {Rinaldin}}, \bibinfo {author} {\bibfnamefont {P.}~\bibnamefont {Fonda}},
  \bibinfo {author} {\bibfnamefont {L.}~\bibnamefont {Giomi}}, \ and\ \bibinfo
  {author} {\bibfnamefont {D.~J.}\ \bibnamefont {Kraft}},\ }\bibfield  {title}
  {\enquote {\bibinfo {title} {Geometric pinning and antimixing in scaffolded
  lipid vesicles},}\ }\href@noop {} {\bibfield  {journal} {\bibinfo  {journal}
  {Nature Commun.}\ }\textbf {\bibinfo {volume} {11}},\ \bibinfo {pages} {4314}
  (\bibinfo {year} {2020})}\BibitemShut {NoStop}%
\bibitem [{\citenamefont {Arroyo}\ and\ \citenamefont
  {DeSimone}(2009)}]{Arroyoetal_PRE_2009}%
  \BibitemOpen
  \bibfield  {author} {\bibinfo {author} {\bibfnamefont {M.}~\bibnamefont
  {Arroyo}}\ and\ \bibinfo {author} {\bibfnamefont {A.}~\bibnamefont
  {DeSimone}},\ }\bibfield  {title} {\enquote {\bibinfo {title} {Relaxation
  dynamics of fluid membranes},}\ }\href@noop {} {\bibfield  {journal}
  {\bibinfo  {journal} {Phys. Rev. E}\ }\textbf {\bibinfo {volume} {79}},\
  \bibinfo {pages} {031915} (\bibinfo {year} {2009})}\BibitemShut {NoStop}%
\bibitem [{\citenamefont {Nitschke}, \citenamefont {Voigt},\ and\ \citenamefont
  {Wensch}(2012)}]{Nitschkeetal_JFM_2012}%
  \BibitemOpen
  \bibfield  {author} {\bibinfo {author} {\bibfnamefont {I.}~\bibnamefont
  {Nitschke}}, \bibinfo {author} {\bibfnamefont {A.}~\bibnamefont {Voigt}}, \
  and\ \bibinfo {author} {\bibfnamefont {J.}~\bibnamefont {Wensch}},\
  }\bibfield  {title} {\enquote {\bibinfo {title} {A finite element approach to
  incompressible two-phase flow on manifolds},}\ }\href@noop {} {\bibfield
  {journal} {\bibinfo  {journal} {J. Fluid Mech.}\ }\textbf {\bibinfo {volume}
  {708}},\ \bibinfo {pages} {418--438} (\bibinfo {year} {2012})}\BibitemShut
  {NoStop}%
\bibitem [{\citenamefont {Reuther}\ and\ \citenamefont
  {Voigt}(2018{\natexlab{a}})}]{Reutheretal_PF_2018}%
  \BibitemOpen
  \bibfield  {author} {\bibinfo {author} {\bibfnamefont {S.}~\bibnamefont
  {Reuther}}\ and\ \bibinfo {author} {\bibfnamefont {A.}~\bibnamefont
  {Voigt}},\ }\bibfield  {title} {\enquote {\bibinfo {title} {Solving the
  incompressible surface {N}avier-{S}tokes equation by surface finite
  elements},}\ }\href@noop {} {\bibfield  {journal} {\bibinfo  {journal} {Phys.
  Fluids}\ }\textbf {\bibinfo {volume} {30}},\ \bibinfo {pages} {012107}
  (\bibinfo {year} {2018}{\natexlab{a}})}\BibitemShut {NoStop}%
\bibitem [{\citenamefont {Fries}(2018)}]{Fries_IJNMF_2018}%
  \BibitemOpen
  \bibfield  {author} {\bibinfo {author} {\bibfnamefont {T.-P.}\ \bibnamefont
  {Fries}},\ }\bibfield  {title} {\enquote {\bibinfo {title} {Higher-order
  surface {FEM} for incompressible {Navier–Stokes} flows on manifolds},}\
  }\href@noop {} {\bibfield  {journal} {\bibinfo  {journal} {Int. J. Numer.
  Meth. Fluids}\ }\textbf {\bibinfo {volume} {88}},\ \bibinfo {pages} {55--78}
  (\bibinfo {year} {2018})}\BibitemShut {NoStop}%
\bibitem [{\citenamefont {Ambrus}\ \emph {et~al.}(2019)\citenamefont {Ambrus},
  \citenamefont {Busuioc}, \citenamefont {Wagner}, \citenamefont {Paillusson},\
  and\ \citenamefont {Kusumaatmaja}}]{ABWPK_PRE_2019}%
  \BibitemOpen
  \bibfield  {author} {\bibinfo {author} {\bibfnamefont {V.~E.}\ \bibnamefont
  {Ambrus}}, \bibinfo {author} {\bibfnamefont {S.}~\bibnamefont {Busuioc}},
  \bibinfo {author} {\bibfnamefont {A.~J.}\ \bibnamefont {Wagner}}, \bibinfo
  {author} {\bibfnamefont {F.}~\bibnamefont {Paillusson}}, \ and\ \bibinfo
  {author} {\bibfnamefont {H.}~\bibnamefont {Kusumaatmaja}},\ }\bibfield
  {title} {\enquote {\bibinfo {title} {Multicomponent flow on curved surfaces:
  A vielbein lattice boltzmann approach},}\ }\href@noop {} {\bibfield
  {journal} {\bibinfo  {journal} {Phys. Rev. E}\ }\textbf {\bibinfo {volume}
  {100}},\ \bibinfo {pages} {063306} (\bibinfo {year} {2019})}\BibitemShut
  {NoStop}%
\bibitem [{\citenamefont {Olshanskii}, \citenamefont {Palzhanov},\ and\
  \citenamefont {Quaini}(2022)}]{Olshanskiietal_VJM_2022}%
  \BibitemOpen
  \bibfield  {author} {\bibinfo {author} {\bibfnamefont {M.}~\bibnamefont
  {Olshanskii}}, \bibinfo {author} {\bibfnamefont {Y.}~\bibnamefont
  {Palzhanov}}, \ and\ \bibinfo {author} {\bibfnamefont {A.}~\bibnamefont
  {Quaini}},\ }\bibfield  {title} {\enquote {\bibinfo {title} {A comparison of
  cahn-hilliard and navier-stokes-cahn-hilliard models on manifolds},}\
  }\href@noop {} {\bibfield  {journal} {\bibinfo  {journal} {Vietnam J. Math.}\
  }\textbf {\bibinfo {volume} {50}},\ \bibinfo {pages} {929--945} (\bibinfo
  {year} {2022})}\BibitemShut {NoStop}%
\bibitem [{\citenamefont {J\"ulicher}\ and\ \citenamefont
  {Lipowsky}(1993)}]{JL_PRL_1993}%
  \BibitemOpen
  \bibfield  {author} {\bibinfo {author} {\bibfnamefont {F.}~\bibnamefont
  {J\"ulicher}}\ and\ \bibinfo {author} {\bibfnamefont {R.}~\bibnamefont
  {Lipowsky}},\ }\bibfield  {title} {\enquote {\bibinfo {title} {Domain-induced
  budding of vesicles},}\ }\href@noop {} {\bibfield  {journal} {\bibinfo
  {journal} {Phys. Rev. Lett.}\ }\textbf {\bibinfo {volume} {70}},\ \bibinfo
  {pages} {2964--2967} (\bibinfo {year} {1993})}\BibitemShut {NoStop}%
\bibitem [{\citenamefont {Canham}(1970)}]{C_JTB_1970}%
  \BibitemOpen
  \bibfield  {author} {\bibinfo {author} {\bibfnamefont {P.~B.}\ \bibnamefont
  {Canham}},\ }\bibfield  {title} {\enquote {\bibinfo {title} {Minimum energy
  of bending as a possible explanation of biconcave shape of human red blood
  cell},}\ }\href@noop {} {\bibfield  {journal} {\bibinfo  {journal} {J. Theo.
  Biol.}\ }\textbf {\bibinfo {volume} {26}},\ \bibinfo {pages} {61--76}
  (\bibinfo {year} {1970})}\BibitemShut {NoStop}%
\bibitem [{\citenamefont {Helfrich}(1973)}]{H_ZN_1973}%
  \BibitemOpen
  \bibfield  {author} {\bibinfo {author} {\bibfnamefont {W.}~\bibnamefont
  {Helfrich}},\ }\bibfield  {title} {\enquote {\bibinfo {title} {Elastic
  properties of lipid bilayers - theory and possible experiments},}\
  }\href@noop {} {\bibfield  {journal} {\bibinfo  {journal} {Zeitschrift f\"ur
  Naturforschung}\ }\textbf {\bibinfo {volume} {C 28}},\ \bibinfo {pages}
  {693--703} (\bibinfo {year} {1973})}\BibitemShut {NoStop}%
\bibitem [{\citenamefont {Elliott}, \citenamefont {Hatcher},\ and\
  \citenamefont {Stinner}(2022)}]{EHS_IFB_2022}%
  \BibitemOpen
  \bibfield  {author} {\bibinfo {author} {\bibfnamefont {C.~M.}\ \bibnamefont
  {Elliott}}, \bibinfo {author} {\bibfnamefont {L.}~\bibnamefont {Hatcher}}, \
  and\ \bibinfo {author} {\bibfnamefont {B.}~\bibnamefont {Stinner}},\
  }\bibfield  {title} {\enquote {\bibinfo {title} {On the sharp interface limit
  of a phase field model for near spherical two phase biomembranes},}\
  }\href@noop {} {\bibfield  {journal} {\bibinfo  {journal} {Interf. Free
  Bound.}\ }\textbf {\bibinfo {volume} {24}},\ \bibinfo {pages} {263--286}
  (\bibinfo {year} {2022})}\BibitemShut {NoStop}%
\bibitem [{\citenamefont {Aland}\ \emph {et~al.}(2014)\citenamefont {Aland},
  \citenamefont {Egerer}, \citenamefont {Lowengrub},\ and\ \citenamefont
  {Voigt}}]{AEV_JCP_2014}%
  \BibitemOpen
  \bibfield  {author} {\bibinfo {author} {\bibfnamefont {S.}~\bibnamefont
  {Aland}}, \bibinfo {author} {\bibfnamefont {S.}~\bibnamefont {Egerer}},
  \bibinfo {author} {\bibfnamefont {J.}~\bibnamefont {Lowengrub}}, \ and\
  \bibinfo {author} {\bibfnamefont {A.}~\bibnamefont {Voigt}},\ }\bibfield
  {title} {\enquote {\bibinfo {title} {Diffuse interface models of locally
  inextensible vesicles in a viscous fluid},}\ }\href@noop {} {\bibfield
  {journal} {\bibinfo  {journal} {J. Comput. Phys.}\ }\textbf {\bibinfo
  {volume} {277}},\ \bibinfo {pages} {32--47} (\bibinfo {year}
  {2014})}\BibitemShut {NoStop}%
\bibitem [{\citenamefont {Hau{\ss}er}\ \emph {et~al.}(2013)\citenamefont
  {Hau{\ss}er}, \citenamefont {Marth}, \citenamefont {Li}, \citenamefont
  {Lowengrub}, \citenamefont {R\"atz},\ and\ \citenamefont
  {Voigt}}]{HMLLRV_IJBB_2013}%
  \BibitemOpen
  \bibfield  {author} {\bibinfo {author} {\bibfnamefont {F.}~\bibnamefont
  {Hau{\ss}er}}, \bibinfo {author} {\bibfnamefont {W.}~\bibnamefont {Marth}},
  \bibinfo {author} {\bibfnamefont {S.}~\bibnamefont {Li}}, \bibinfo {author}
  {\bibfnamefont {J.~S.}\ \bibnamefont {Lowengrub}}, \bibinfo {author}
  {\bibfnamefont {A.}~\bibnamefont {R\"atz}}, \ and\ \bibinfo {author}
  {\bibfnamefont {A.}~\bibnamefont {Voigt}},\ }\bibfield  {title} {\enquote
  {\bibinfo {title} {Thermodynamically consistent models for two component
  vesicles},}\ }\href@noop {} {\bibfield  {journal} {\bibinfo  {journal} {Int.
  J. Biomath. Biostat.}\ }\textbf {\bibinfo {volume} {2}},\ \bibinfo {pages}
  {19--48} (\bibinfo {year} {2013})}\BibitemShut {NoStop}%
\bibitem [{\citenamefont {Lowengrub}, \citenamefont {Xu},\ and\ \citenamefont
  {Voigt}(2007)}]{LXV_FDMP_2007}%
  \BibitemOpen
  \bibfield  {author} {\bibinfo {author} {\bibfnamefont {J.~S.}\ \bibnamefont
  {Lowengrub}}, \bibinfo {author} {\bibfnamefont {J.-J.}\ \bibnamefont {Xu}}, \
  and\ \bibinfo {author} {\bibfnamefont {A.}~\bibnamefont {Voigt}},\ }\bibfield
   {title} {\enquote {\bibinfo {title} {Surface phase separation and flow in a
  simple model of multicomponent drops and vesicles},}\ }\href@noop {}
  {\bibfield  {journal} {\bibinfo  {journal} {FDMP-Fluid Dyn. Mat. Proc.}\
  }\textbf {\bibinfo {volume} {3}},\ \bibinfo {pages} {1--19} (\bibinfo {year}
  {2007})}\BibitemShut {NoStop}%
\bibitem [{\citenamefont {Garcke}\ and\ \citenamefont
  {Nurnberg}(2021)}]{GN_IMAJNA_2021}%
  \BibitemOpen
  \bibfield  {author} {\bibinfo {author} {\bibfnamefont {H.}~\bibnamefont
  {Garcke}}\ and\ \bibinfo {author} {\bibfnamefont {R.}~\bibnamefont
  {Nurnberg}},\ }\bibfield  {title} {\enquote {\bibinfo {title}
  {Structure-preserving discretizations of gradient flows for axisymmetric
  two-phase biomembranes},}\ }\href@noop {} {\bibfield  {journal} {\bibinfo
  {journal} {IMA J. Num. Anal.}\ }\textbf {\bibinfo {volume} {41}},\ \bibinfo
  {pages} {1899--1940} (\bibinfo {year} {2021})}\BibitemShut {NoStop}%
\bibitem [{\citenamefont {Reuther}\ and\ \citenamefont
  {Voigt}(2015)}]{Reutheretal_MMS_2015}%
  \BibitemOpen
  \bibfield  {author} {\bibinfo {author} {\bibfnamefont {S.}~\bibnamefont
  {Reuther}}\ and\ \bibinfo {author} {\bibfnamefont {A.}~\bibnamefont
  {Voigt}},\ }\bibfield  {title} {\enquote {\bibinfo {title} {The interplay of
  curvature and vortices in flow on curved surfaces},}\ }\href@noop {}
  {\bibfield  {journal} {\bibinfo  {journal} {Multiscale Model. Sim.}\ }\textbf
  {\bibinfo {volume} {13}},\ \bibinfo {pages} {632--643} (\bibinfo {year}
  {2015})}\BibitemShut {NoStop}%
\bibitem [{\citenamefont {Nitschke}, \citenamefont {Reuther},\ and\
  \citenamefont {Voigt}(2017)}]{Nitschkeetal_TFI_2017}%
  \BibitemOpen
  \bibfield  {author} {\bibinfo {author} {\bibfnamefont {I.}~\bibnamefont
  {Nitschke}}, \bibinfo {author} {\bibfnamefont {S.}~\bibnamefont {Reuther}}, \
  and\ \bibinfo {author} {\bibfnamefont {A.}~\bibnamefont {Voigt}},\ }\bibfield
   {title} {\enquote {\bibinfo {title} {Discrete exterior calculus {(DEC)} for
  the surface {Navier-Stokes} equation},}\ }in\ \href@noop {} {\emph {\bibinfo
  {booktitle} {Transport Processes at Fluidic Interfaces}}},\ \bibinfo {editor}
  {edited by\ \bibinfo {editor} {\bibfnamefont {D.}~\bibnamefont {Bothe}}\ and\
  \bibinfo {editor} {\bibfnamefont {A.}~\bibnamefont {Reusken}}}\ (\bibinfo
  {publisher} {Springer International Publishing},\ \bibinfo {year} {2017})\
  pp.\ \bibinfo {pages} {177--197}\BibitemShut {NoStop}%
\bibitem [{\citenamefont {Reuther}\ and\ \citenamefont
  {Voigt}(2018{\natexlab{b}})}]{Reutheretal_MMS_2018}%
  \BibitemOpen
  \bibfield  {author} {\bibinfo {author} {\bibfnamefont {S.}~\bibnamefont
  {Reuther}}\ and\ \bibinfo {author} {\bibfnamefont {A.}~\bibnamefont
  {Voigt}},\ }\bibfield  {title} {\enquote {\bibinfo {title} {Erratum: "{T}he
  interplay of curvature and vortices in flow on curved surfaces"},}\
  }\href@noop {} {\bibfield  {journal} {\bibinfo  {journal} {Multiscale Model.
  Sim.}\ }\textbf {\bibinfo {volume} {16}},\ \bibinfo {pages} {1448--1453}
  (\bibinfo {year} {2018}{\natexlab{b}})}\BibitemShut {NoStop}%
\bibitem [{\citenamefont {Jankuhn}, \citenamefont {Olshanskii},\ and\
  \citenamefont {Reusken}(2018)}]{Jankuhnetal_IFB_2018}%
  \BibitemOpen
  \bibfield  {author} {\bibinfo {author} {\bibfnamefont {T.}~\bibnamefont
  {Jankuhn}}, \bibinfo {author} {\bibfnamefont {M.~A.}\ \bibnamefont
  {Olshanskii}}, \ and\ \bibinfo {author} {\bibfnamefont {A.}~\bibnamefont
  {Reusken}},\ }\bibfield  {title} {\enquote {\bibinfo {title} {{Incompressible
  fluid problems on embedded surfaces: Modeling and variational
  formulations}},}\ }\href@noop {} {\bibfield  {journal} {\bibinfo  {journal}
  {Interf. Free Bound.}\ }\textbf {\bibinfo {volume} {{20}}},\ \bibinfo {pages}
  {{353--377}} (\bibinfo {year} {{2018}})}\BibitemShut {NoStop}%
\bibitem [{\citenamefont {Torres-S{\'a}nchez}, \citenamefont {Mill{\'a}n},\
  and\ \citenamefont {Arroyo}(2019)}]{Torres_Sanchezetal_JFM_2019}%
  \BibitemOpen
  \bibfield  {author} {\bibinfo {author} {\bibfnamefont {A.}~\bibnamefont
  {Torres-S{\'a}nchez}}, \bibinfo {author} {\bibfnamefont {D.}~\bibnamefont
  {Mill{\'a}n}}, \ and\ \bibinfo {author} {\bibfnamefont {M.}~\bibnamefont
  {Arroyo}},\ }\bibfield  {title} {\enquote {\bibinfo {title} {Modelling fluid
  deformable surfaces with an emphasis on biological interfaces},}\ }\href@noop
  {} {\bibfield  {journal} {\bibinfo  {journal} {J. Fluid Mech.}\ }\textbf
  {\bibinfo {volume} {872}},\ \bibinfo {pages} {218--271} (\bibinfo {year}
  {2019})}\BibitemShut {NoStop}%
\bibitem [{\citenamefont {Reuther}, \citenamefont {Nitschke},\ and\
  \citenamefont {Voigt}(2020)}]{Reutheretal_JFM_2020}%
  \BibitemOpen
  \bibfield  {author} {\bibinfo {author} {\bibfnamefont {S.}~\bibnamefont
  {Reuther}}, \bibinfo {author} {\bibfnamefont {I.}~\bibnamefont {Nitschke}}, \
  and\ \bibinfo {author} {\bibfnamefont {A.}~\bibnamefont {Voigt}},\ }\bibfield
   {title} {\enquote {\bibinfo {title} {A numerical approach for fluid
  deformable surfaces},}\ }\href@noop {} {\bibfield  {journal} {\bibinfo
  {journal} {J. Fluid Mech.}\ }\textbf {\bibinfo {volume} {900}},\ \bibinfo
  {pages} {R8} (\bibinfo {year} {2020})}\BibitemShut {NoStop}%
\bibitem [{\citenamefont {Lederer}, \citenamefont {Lehrenfeld},\ and\
  \citenamefont {Sch\"{o}berl}(2020)}]{Ledereretal_IJNME_2020}%
  \BibitemOpen
  \bibfield  {author} {\bibinfo {author} {\bibfnamefont {P.~L.}\ \bibnamefont
  {Lederer}}, \bibinfo {author} {\bibfnamefont {C.}~\bibnamefont {Lehrenfeld}},
  \ and\ \bibinfo {author} {\bibfnamefont {J.}~\bibnamefont {Sch\"{o}berl}},\
  }\bibfield  {title} {\enquote {\bibinfo {title} {Divergence-free tangential
  finite element methods for incompressible flows on surfaces},}\ }\href@noop
  {} {\bibfield  {journal} {\bibinfo  {journal} {Int. J. Numer. Meth. Eng.}\
  }\textbf {\bibinfo {volume} {121}},\ \bibinfo {pages} {2503--2533} (\bibinfo
  {year} {2020})}\BibitemShut {NoStop}%
\bibitem [{\citenamefont {Yavari}, \citenamefont {Ozakin},\ and\ \citenamefont
  {Sadik}(2016)}]{Yavarietal_JNS_2016}%
  \BibitemOpen
  \bibfield  {author} {\bibinfo {author} {\bibfnamefont {A.}~\bibnamefont
  {Yavari}}, \bibinfo {author} {\bibfnamefont {A.}~\bibnamefont {Ozakin}}, \
  and\ \bibinfo {author} {\bibfnamefont {S.}~\bibnamefont {Sadik}},\ }\bibfield
   {title} {\enquote {\bibinfo {title} {Nonlinear elasticity in a deforming
  ambient space},}\ }\href@noop {} {\bibfield  {journal} {\bibinfo  {journal}
  {J. Nonli. Sci.}\ }\textbf {\bibinfo {volume} {26}},\ \bibinfo {pages}
  {1651--1692} (\bibinfo {year} {2016})}\BibitemShut {NoStop}%
\bibitem [{\citenamefont {Koba}, \citenamefont {Liu},\ and\ \citenamefont
  {Giga}(2017)}]{Kobaetal_QAM_2017}%
  \BibitemOpen
  \bibfield  {author} {\bibinfo {author} {\bibfnamefont {H.}~\bibnamefont
  {Koba}}, \bibinfo {author} {\bibfnamefont {C.}~\bibnamefont {Liu}}, \ and\
  \bibinfo {author} {\bibfnamefont {Y.}~\bibnamefont {Giga}},\ }\bibfield
  {title} {\enquote {\bibinfo {title} {{Energetic variational approaches for
  incompressible fluid systems on an evolving surface}},}\ }\href@noop {}
  {\bibfield  {journal} {\bibinfo  {journal} {Quart. Appl. Math.}\ }\textbf
  {\bibinfo {volume} {75}},\ \bibinfo {pages} {359--389} (\bibinfo {year}
  {2017})}\BibitemShut {NoStop}%
\bibitem [{\citenamefont {Miura}(2018)}]{Miura_QAM_2018}%
  \BibitemOpen
  \bibfield  {author} {\bibinfo {author} {\bibfnamefont {T.-H.}\ \bibnamefont
  {Miura}},\ }\bibfield  {title} {\enquote {\bibinfo {title} {On singular limit
  equations for incompressible fluids in moving thin domains},}\ }\href@noop {}
  {\bibfield  {journal} {\bibinfo  {journal} {Quart. Appl. Math.}\ }\textbf
  {\bibinfo {volume} {76}},\ \bibinfo {pages} {215--251} (\bibinfo {year}
  {2018})}\BibitemShut {NoStop}%
\bibitem [{\citenamefont {Koba}, \citenamefont {Liu},\ and\ \citenamefont
  {Giga}(2018)}]{Kobaetal_QAM_2018}%
  \BibitemOpen
  \bibfield  {author} {\bibinfo {author} {\bibfnamefont {H.}~\bibnamefont
  {Koba}}, \bibinfo {author} {\bibfnamefont {C.}~\bibnamefont {Liu}}, \ and\
  \bibinfo {author} {\bibfnamefont {Y.}~\bibnamefont {Giga}},\ }\bibfield
  {title} {\enquote {\bibinfo {title} {{Errata to “Energetic variational
  approaches for incompressible fluid systems on an evolving surface"}},}\
  }\href@noop {} {\bibfield  {journal} {\bibinfo  {journal} {Quart. Appl.
  Math.}\ }\textbf {\bibinfo {volume} {76}},\ \bibinfo {pages} {174--152}
  (\bibinfo {year} {2018})}\BibitemShut {NoStop}%
\bibitem [{\citenamefont {Pr\"uss}, \citenamefont {Simonett},\ and\
  \citenamefont {Wilke}(2021)}]{Pruessetal_JEE_2021}%
  \BibitemOpen
  \bibfield  {author} {\bibinfo {author} {\bibfnamefont {J.}~\bibnamefont
  {Pr\"uss}}, \bibinfo {author} {\bibfnamefont {G.}~\bibnamefont {Simonett}}, \
  and\ \bibinfo {author} {\bibfnamefont {M.}~\bibnamefont {Wilke}},\ }\bibfield
   {title} {\enquote {\bibinfo {title} {On the {Navier–Stokes} equations on
  surfaces},}\ }\href@noop {} {\bibfield  {journal} {\bibinfo  {journal} {J.
  Evol. Eq.}\ }\textbf {\bibinfo {volume} {21}},\ \bibinfo {pages} {3153--3179}
  (\bibinfo {year} {2021})}\BibitemShut {NoStop}%
\bibitem [{\citenamefont {Hohenberg}\ and\ \citenamefont
  {Halperin}(1977)}]{HH_RMP_1977}%
  \BibitemOpen
  \bibfield  {author} {\bibinfo {author} {\bibfnamefont {P.~C.}\ \bibnamefont
  {Hohenberg}}\ and\ \bibinfo {author} {\bibfnamefont {B.~I.}\ \bibnamefont
  {Halperin}},\ }\bibfield  {title} {\enquote {\bibinfo {title} {Theory of
  dynamic critical phenomena},}\ }\href@noop {} {\bibfield  {journal} {\bibinfo
   {journal} {Rev. Mod. Phys.}\ }\textbf {\bibinfo {volume} {49}},\ \bibinfo
  {pages} {435--479} (\bibinfo {year} {1977})}\BibitemShut {NoStop}%
\bibitem [{\citenamefont {Gurtin}, \citenamefont {Polignone},\ and\
  \citenamefont {Vinals}(1996)}]{GPV_MMMAS_1996}%
  \BibitemOpen
  \bibfield  {author} {\bibinfo {author} {\bibfnamefont {M.}~\bibnamefont
  {Gurtin}}, \bibinfo {author} {\bibfnamefont {D.}~\bibnamefont {Polignone}}, \
  and\ \bibinfo {author} {\bibfnamefont {J.}~\bibnamefont {Vinals}},\
  }\bibfield  {title} {\enquote {\bibinfo {title} {Two-phase binary fluids and
  immiscible fluids described by an order parameter},}\ }\href@noop {}
  {\bibfield  {journal} {\bibinfo  {journal} {Math. Modles Meth. Appl. Sci.}\
  }\textbf {\bibinfo {volume} {6}},\ \bibinfo {pages} {815--831} (\bibinfo
  {year} {1996})}\BibitemShut {NoStop}%
\bibitem [{\citenamefont {Aland}\ and\ \citenamefont
  {Voigt}(2012)}]{AV_IJNMF_2012}%
  \BibitemOpen
  \bibfield  {author} {\bibinfo {author} {\bibfnamefont {S.}~\bibnamefont
  {Aland}}\ and\ \bibinfo {author} {\bibfnamefont {A.}~\bibnamefont {Voigt}},\
  }\bibfield  {title} {\enquote {\bibinfo {title} {Benchmark computations of
  diffuse interface models for two-dimensional bubble dynamics},}\ }\href@noop
  {} {\bibfield  {journal} {\bibinfo  {journal} {Int. J. Numer. Meth. Fluids}\
  }\textbf {\bibinfo {volume} {69}},\ \bibinfo {pages} {747--761} (\bibinfo
  {year} {2012})}\BibitemShut {NoStop}%
\bibitem [{\citenamefont {Abels}, \citenamefont {Garcke},\ and\ \citenamefont
  {Gr\"un}(2012)}]{AGG_MMMAS_2012}%
  \BibitemOpen
  \bibfield  {author} {\bibinfo {author} {\bibfnamefont {H.}~\bibnamefont
  {Abels}}, \bibinfo {author} {\bibfnamefont {H.}~\bibnamefont {Garcke}}, \
  and\ \bibinfo {author} {\bibfnamefont {G.}~\bibnamefont {Gr\"un}},\
  }\bibfield  {title} {\enquote {\bibinfo {title} {Thermodynamically
  consistent, frame indifferent diffuse interface models for incompressible
  two-phase flows with different densities},}\ }\href@noop {} {\bibfield
  {journal} {\bibinfo  {journal} {Math. Modles Meth. Appl. Sci.}\ }\textbf
  {\bibinfo {volume} {22}},\ \bibinfo {pages} {1150013} (\bibinfo {year}
  {2012})}\BibitemShut {NoStop}%
\bibitem [{\citenamefont {Abels}(2009)}]{A_ARMA_2009}%
  \BibitemOpen
  \bibfield  {author} {\bibinfo {author} {\bibfnamefont {H.}~\bibnamefont
  {Abels}},\ }\bibfield  {title} {\enquote {\bibinfo {title} {On a diffuse
  interface model for two-phase flows of viscous, incompressible fluids with
  matched densities},}\ }\href@noop {} {\bibfield  {journal} {\bibinfo
  {journal} {Arch. Rat. Mech. Anal.}\ }\textbf {\bibinfo {volume} {194}},\
  \bibinfo {pages} {463--506} (\bibinfo {year} {2009})}\BibitemShut {NoStop}%
\bibitem [{\citenamefont {Magaletti}\ \emph {et~al.}(2013)\citenamefont
  {Magaletti}, \citenamefont {Picano}, \citenamefont {Chinappi}, \citenamefont
  {Marino},\ and\ \citenamefont {Casciola}}]{MPCMC_JFM_2013}%
  \BibitemOpen
  \bibfield  {author} {\bibinfo {author} {\bibfnamefont {F.}~\bibnamefont
  {Magaletti}}, \bibinfo {author} {\bibfnamefont {F.}~\bibnamefont {Picano}},
  \bibinfo {author} {\bibfnamefont {M.}~\bibnamefont {Chinappi}}, \bibinfo
  {author} {\bibfnamefont {L.}~\bibnamefont {Marino}}, \ and\ \bibinfo {author}
  {\bibfnamefont {C.~M.}\ \bibnamefont {Casciola}},\ }\bibfield  {title}
  {\enquote {\bibinfo {title} {The sharp-interface limit of the
  {Cahn-Hilliard/Navier-Stokes} model for binary fluids},}\ }\href@noop {}
  {\bibfield  {journal} {\bibinfo  {journal} {J. Fluid Mech.}\ }\textbf
  {\bibinfo {volume} {714}},\ \bibinfo {pages} {95--126} (\bibinfo {year}
  {2013})}\BibitemShut {NoStop}%
\bibitem [{\citenamefont {Dziuk}\ and\ \citenamefont
  {Elliott}(2013)}]{DE_AN_2013}%
  \BibitemOpen
  \bibfield  {author} {\bibinfo {author} {\bibfnamefont {G.}~\bibnamefont
  {Dziuk}}\ and\ \bibinfo {author} {\bibfnamefont {C.~M.}\ \bibnamefont
  {Elliott}},\ }\bibfield  {title} {\enquote {\bibinfo {title} {Finite element
  methods for surfaces {PDEs}},}\ }\href@noop {} {\bibfield  {journal}
  {\bibinfo  {journal} {Acta Num.}\ }\textbf {\bibinfo {volume} {22}},\
  \bibinfo {pages} {289--396} (\bibinfo {year} {2013})}\BibitemShut {NoStop}%
\bibitem [{\citenamefont {Nestler}, \citenamefont {Nitschke},\ and\
  \citenamefont {Voigt}(2019)}]{NNV_JCP_2019}%
  \BibitemOpen
  \bibfield  {author} {\bibinfo {author} {\bibfnamefont {M.}~\bibnamefont
  {Nestler}}, \bibinfo {author} {\bibfnamefont {I.}~\bibnamefont {Nitschke}}, \
  and\ \bibinfo {author} {\bibfnamefont {A.}~\bibnamefont {Voigt}},\ }\bibfield
   {title} {\enquote {\bibinfo {title} {A finite element approach for vector-
  and tensor-valued surface {PDEs}},}\ }\href@noop {} {\bibfield  {journal}
  {\bibinfo  {journal} {J. Comput. Phys.}\ }\textbf {\bibinfo {volume} {389}},\
  \bibinfo {pages} {48--61} (\bibinfo {year} {2019})}\BibitemShut {NoStop}%
\bibitem [{\citenamefont {Mickelin}\ \emph {et~al.}(2018)\citenamefont
  {Mickelin}, \citenamefont {Slomka}, \citenamefont {Burns}, \citenamefont
  {Lecoanet}, \citenamefont {Vasil}, \citenamefont {Faria},\ and\ \citenamefont
  {Dunkel}}]{Mickelinetal_PRL_2018}%
  \BibitemOpen
  \bibfield  {author} {\bibinfo {author} {\bibfnamefont {O.}~\bibnamefont
  {Mickelin}}, \bibinfo {author} {\bibfnamefont {J.}~\bibnamefont {Slomka}},
  \bibinfo {author} {\bibfnamefont {K.~J.}\ \bibnamefont {Burns}}, \bibinfo
  {author} {\bibfnamefont {D.}~\bibnamefont {Lecoanet}}, \bibinfo {author}
  {\bibfnamefont {G.~M.}\ \bibnamefont {Vasil}}, \bibinfo {author}
  {\bibfnamefont {L.~M.}\ \bibnamefont {Faria}}, \ and\ \bibinfo {author}
  {\bibfnamefont {J.}~\bibnamefont {Dunkel}},\ }\bibfield  {title} {\enquote
  {\bibinfo {title} {Anomalous chained turbulence in actively driven flows on
  spheres},}\ }\href@noop {} {\bibfield  {journal} {\bibinfo  {journal} {Phys.
  Rev. Lett.}\ }\textbf {\bibinfo {volume} {120}},\ \bibinfo {pages} {164503}
  (\bibinfo {year} {2018})}\BibitemShut {NoStop}%
\bibitem [{\citenamefont {Rank}\ and\ \citenamefont
  {Voigt}(2021)}]{RV_PF_2021}%
  \BibitemOpen
  \bibfield  {author} {\bibinfo {author} {\bibfnamefont {M.}~\bibnamefont
  {Rank}}\ and\ \bibinfo {author} {\bibfnamefont {A.}~\bibnamefont {Voigt}},\
  }\bibfield  {title} {\enquote {\bibinfo {title} {Active flows on curved
  surfaces},}\ }\href@noop {} {\bibfield  {journal} {\bibinfo  {journal} {Phys.
  Fluids}\ }\textbf {\bibinfo {volume} {33}},\ \bibinfo {pages} {072110}
  (\bibinfo {year} {2021})}\BibitemShut {NoStop}%
\bibitem [{\citenamefont {Mendoza}, \citenamefont {Alkemper},\ and\
  \citenamefont {Voorhees}(2003)}]{MAV_MMT_2003}%
  \BibitemOpen
  \bibfield  {author} {\bibinfo {author} {\bibfnamefont {R.}~\bibnamefont
  {Mendoza}}, \bibinfo {author} {\bibfnamefont {J.}~\bibnamefont {Alkemper}}, \
  and\ \bibinfo {author} {\bibfnamefont {P.~W.}\ \bibnamefont {Voorhees}},\
  }\bibfield  {title} {\enquote {\bibinfo {title} {The morphological evolution
  of dendritic microstructures during coarsening},}\ }\href@noop {} {\bibfield
  {journal} {\bibinfo  {journal} {Metall. Mater. Trans.}\ }\textbf {\bibinfo
  {volume} {34}},\ \bibinfo {pages} {481} (\bibinfo {year} {2003})}\BibitemShut
  {NoStop}%
\bibitem [{\citenamefont {Kwon}, \citenamefont {Thornton},\ and\ \citenamefont
  {Voorhees}(2010)}]{KTV_PM_2010}%
  \BibitemOpen
  \bibfield  {author} {\bibinfo {author} {\bibfnamefont {Y.}~\bibnamefont
  {Kwon}}, \bibinfo {author} {\bibfnamefont {K.}~\bibnamefont {Thornton}}, \
  and\ \bibinfo {author} {\bibfnamefont {P.~W.}\ \bibnamefont {Voorhees}},\
  }\bibfield  {title} {\enquote {\bibinfo {title} {Morphology and topology in
  coarsening of domains via non-conserved and conserved dynamics},}\
  }\href@noop {} {\bibfield  {journal} {\bibinfo  {journal} {Phil. Mag.}\
  }\textbf {\bibinfo {volume} {317-335}},\ \bibinfo {pages} {90} (\bibinfo
  {year} {2010})}\BibitemShut {NoStop}%
\bibitem [{\citenamefont {Demlow}(2009)}]{Demlow2009}%
  \BibitemOpen
  \bibfield  {author} {\bibinfo {author} {\bibfnamefont {A.}~\bibnamefont
  {Demlow}},\ }\bibfield  {title} {\enquote {\bibinfo {title} {Higher-order
  finite element methods and pointwise error estimates for elliptic problems on
  surfaces},}\ }\href {\doibase 10.1137/070708135} {\bibfield  {journal}
  {\bibinfo  {journal} {SIAM J. Num. Anal.}\ }\textbf {\bibinfo {volume}
  {47}},\ \bibinfo {pages} {805--827} (\bibinfo {year} {2009})}\BibitemShut
  {NoStop}%
\bibitem [{\citenamefont {Praetorius}\ and\ \citenamefont
  {Stenger}(2020)}]{praetorius2020dunecurvedgrid}%
  \BibitemOpen
  \bibfield  {author} {\bibinfo {author} {\bibfnamefont {S.}~\bibnamefont
  {Praetorius}}\ and\ \bibinfo {author} {\bibfnamefont {F.}~\bibnamefont
  {Stenger}},\ }\bibfield  {title} {\enquote {\bibinfo {title}
  {{DUNE-CurvedGrid--A DUNE module for surface parametrization}},}\ }\href@noop
  {} {\bibfield  {journal} {\bibinfo  {journal} {Arch. Num. Software}\ }\textbf
  {\bibinfo {volume} {22}},\ \bibinfo {pages} {1--22} (\bibinfo {year}
  {2020})}\BibitemShut {NoStop}%
\bibitem [{\citenamefont {Nestler}\ \emph {et~al.}(2018)\citenamefont
  {Nestler}, \citenamefont {Nitschke}, \citenamefont {Praetorius},\ and\
  \citenamefont {Voigt}}]{NNPV_JNS_2018}%
  \BibitemOpen
  \bibfield  {author} {\bibinfo {author} {\bibfnamefont {M.}~\bibnamefont
  {Nestler}}, \bibinfo {author} {\bibfnamefont {I.}~\bibnamefont {Nitschke}},
  \bibinfo {author} {\bibfnamefont {S.}~\bibnamefont {Praetorius}}, \ and\
  \bibinfo {author} {\bibfnamefont {A.}~\bibnamefont {Voigt}},\ }\bibfield
  {title} {\enquote {\bibinfo {title} {Orientational order on surfaces: The
  coupling of topology, geometry, and dynamics},}\ }\href@noop {} {\bibfield
  {journal} {\bibinfo  {journal} {J. Nonl. Sci.}\ }\textbf {\bibinfo {volume}
  {28}},\ \bibinfo {pages} {147--191} (\bibinfo {year} {2018})}\BibitemShut
  {NoStop}%
\bibitem [{\citenamefont {Hansbo}, \citenamefont {Larson},\ and\ \citenamefont
  {Larsson}(2020)}]{HLL_IMAJNA_2020}%
  \BibitemOpen
  \bibfield  {author} {\bibinfo {author} {\bibfnamefont {P.}~\bibnamefont
  {Hansbo}}, \bibinfo {author} {\bibfnamefont {M.~G.}\ \bibnamefont {Larson}},
  \ and\ \bibinfo {author} {\bibfnamefont {K.}~\bibnamefont {Larsson}},\
  }\bibfield  {title} {\enquote {\bibinfo {title} {Analysis of finite element
  methods for vector laplacians on surfaces},}\ }\href@noop {} {\bibfield
  {journal} {\bibinfo  {journal} {IMA J. Num. Anal.}\ }\textbf {\bibinfo
  {volume} {40}},\ \bibinfo {pages} {1652--1701} (\bibinfo {year}
  {2020})}\BibitemShut {NoStop}%
\bibitem [{\citenamefont {Brandner}\ \emph {et~al.}(2022)\citenamefont
  {Brandner}, \citenamefont {Jankuhn}, \citenamefont {Praetorius},
  \citenamefont {Reusken},\ and\ \citenamefont
  {Voigt}}]{Brandneretal_SIAMJSC_2022}%
  \BibitemOpen
  \bibfield  {author} {\bibinfo {author} {\bibfnamefont {P.}~\bibnamefont
  {Brandner}}, \bibinfo {author} {\bibfnamefont {T.}~\bibnamefont {Jankuhn}},
  \bibinfo {author} {\bibfnamefont {S.}~\bibnamefont {Praetorius}}, \bibinfo
  {author} {\bibfnamefont {A.}~\bibnamefont {Reusken}}, \ and\ \bibinfo
  {author} {\bibfnamefont {A.}~\bibnamefont {Voigt}},\ }\bibfield  {title}
  {\enquote {\bibinfo {title} {Finite element discretization methods for
  velocity-pressure and stream function formulations of surface stokes
  equations},}\ }\href@noop {} {\bibfield  {journal} {\bibinfo  {journal} {SIAM
  J. Sci. Comput.}\ }\textbf {\bibinfo {volume} {44}},\ \bibinfo {pages}
  {A1807--A1832} (\bibinfo {year} {2022})}\BibitemShut {NoStop}%
\bibitem [{\citenamefont {Nestler}\ and\ \citenamefont
  {Voigt}(2022)}]{NV_CICP_2022}%
  \BibitemOpen
  \bibfield  {author} {\bibinfo {author} {\bibfnamefont {M.}~\bibnamefont
  {Nestler}}\ and\ \bibinfo {author} {\bibfnamefont {A.}~\bibnamefont
  {Voigt}},\ }\bibfield  {title} {\enquote {\bibinfo {title} {Active
  nematodynamics on curved surfaces-the influence of geometric forces on motion
  patterns of topological defects},}\ }\href@noop {} {\bibfield  {journal}
  {\bibinfo  {journal} {Commun. Comput. Phys.}\ }\textbf {\bibinfo {volume}
  {31}},\ \bibinfo {pages} {947--965} (\bibinfo {year} {2022})}\BibitemShut
  {NoStop}%
\bibitem [{\citenamefont {Hardering}\ and\ \citenamefont
  {Praetorius}()}]{HP_unpublished}%
  \BibitemOpen
  \bibfield  {author} {\bibinfo {author} {\bibfnamefont {H.}~\bibnamefont
  {Hardering}}\ and\ \bibinfo {author} {\bibfnamefont {S.}~\bibnamefont
  {Praetorius}},\ }\href@noop {} {\enquote {\bibinfo {title} {personal
  communication},}\ }\BibitemShut {NoStop}%
\bibitem [{\citenamefont {Jankuhn}\ and\ \citenamefont
  {Reusken}(2021)}]{JR_IMAJNA_2021}%
  \BibitemOpen
  \bibfield  {author} {\bibinfo {author} {\bibfnamefont {T.}~\bibnamefont
  {Jankuhn}}\ and\ \bibinfo {author} {\bibfnamefont {A.}~\bibnamefont
  {Reusken}},\ }\bibfield  {title} {\enquote {\bibinfo {title} {Trace finite
  element methods for surface vector-laplace equations},}\ }\href@noop {}
  {\bibfield  {journal} {\bibinfo  {journal} {IMA J. Num. Anal.}\ }\textbf
  {\bibinfo {volume} {41}},\ \bibinfo {pages} {48--83} (\bibinfo {year}
  {2021})}\BibitemShut {NoStop}%
\bibitem [{\citenamefont {Jankuhn}\ \emph {et~al.}(2021)\citenamefont
  {Jankuhn}, \citenamefont {Olshanskii}, \citenamefont {Reusken},\ and\
  \citenamefont {Zhiliakov}}]{JORZ_JNM_2021}%
  \BibitemOpen
  \bibfield  {author} {\bibinfo {author} {\bibfnamefont {T.}~\bibnamefont
  {Jankuhn}}, \bibinfo {author} {\bibfnamefont {M.~A.}\ \bibnamefont
  {Olshanskii}}, \bibinfo {author} {\bibfnamefont {A.}~\bibnamefont {Reusken}},
  \ and\ \bibinfo {author} {\bibfnamefont {A.}~\bibnamefont {Zhiliakov}},\
  }\bibfield  {title} {\enquote {\bibinfo {title} {Error analysis of higher
  order trace finite element methods for the surface stokes equation},}\
  }\href@noop {} {\bibfield  {journal} {\bibinfo  {journal} {J. Num. Math.}\
  }\textbf {\bibinfo {volume} {29}},\ \bibinfo {pages} {245--267} (\bibinfo
  {year} {2021})}\BibitemShut {NoStop}%
\bibitem [{\citenamefont {Hardering}\ and\ \citenamefont
  {Praetorius}(2022)}]{HP_IMAJNA_2022}%
  \BibitemOpen
  \bibfield  {author} {\bibinfo {author} {\bibfnamefont {H.}~\bibnamefont
  {Hardering}}\ and\ \bibinfo {author} {\bibfnamefont {S.}~\bibnamefont
  {Praetorius}},\ }\bibfield  {title} {\enquote {\bibinfo {title} {Tangential
  errors of tensor surface finite elements},}\ }\href {\doibase
  10.1093/imanum/drac015} {\bibfield  {journal} {\bibinfo  {journal} {IMA J.
  Num. Anal.}\ } (\bibinfo {year} {2022}),\ 10.1093/imanum/drac015}\BibitemShut
  {NoStop}%
\bibitem [{\citenamefont {Vey}\ and\ \citenamefont
  {Voigt}(2007)}]{Vey_CVS_2007}%
  \BibitemOpen
  \bibfield  {author} {\bibinfo {author} {\bibfnamefont {S.}~\bibnamefont
  {Vey}}\ and\ \bibinfo {author} {\bibfnamefont {A.}~\bibnamefont {Voigt}},\
  }\bibfield  {title} {\enquote {\bibinfo {title} {{AMDiS}: adaptive
  multidimensional simulations},}\ }\href@noop {} {\bibfield  {journal}
  {\bibinfo  {journal} {Comput. Vis. Sci.}\ }\textbf {\bibinfo {volume} {10}},\
  \bibinfo {pages} {57--67} (\bibinfo {year} {2007})}\BibitemShut {NoStop}%
\bibitem [{\citenamefont {Witkowski}\ \emph {et~al.}(2015)\citenamefont
  {Witkowski}, \citenamefont {Ling}, \citenamefont {Praetorius},\ and\
  \citenamefont {Voigt}}]{Witkowski_ACM_2015}%
  \BibitemOpen
  \bibfield  {author} {\bibinfo {author} {\bibfnamefont {T.}~\bibnamefont
  {Witkowski}}, \bibinfo {author} {\bibfnamefont {S.}~\bibnamefont {Ling}},
  \bibinfo {author} {\bibfnamefont {S.}~\bibnamefont {Praetorius}}, \ and\
  \bibinfo {author} {\bibfnamefont {A.}~\bibnamefont {Voigt}},\ }\bibfield
  {title} {\enquote {\bibinfo {title} {Software concepts and numerical
  algorithms for a scalable adaptive parallel finite element method},}\
  }\href@noop {} {\bibfield  {journal} {\bibinfo  {journal} {Adv. Comput.
  Math.}\ }\textbf {\bibinfo {volume} {41}},\ \bibinfo {pages} {1145--1177}
  (\bibinfo {year} {2015})}\BibitemShut {NoStop}%
\bibitem [{\citenamefont {R\"atz}(2016)}]{R_AML_2016}%
  \BibitemOpen
  \bibfield  {author} {\bibinfo {author} {\bibfnamefont {A.}~\bibnamefont
  {R\"atz}},\ }\bibfield  {title} {\enquote {\bibinfo {title} {A benchmark for
  the surface {Cahn–Hilliard} equation},}\ }\href {\doibase
  10.1016/j.aml.2015.12.008} {\bibfield  {journal} {\bibinfo  {journal} {Appl.
  Math. Lett.}\ }\textbf {\bibinfo {volume} {56}},\ \bibinfo {pages} {65--71}
  (\bibinfo {year} {2016})}\BibitemShut {NoStop}%
\bibitem [{\citenamefont {Feng}\ and\ \citenamefont
  {Prohl}(2003)}]{FP_NM_2003}%
  \BibitemOpen
  \bibfield  {author} {\bibinfo {author} {\bibfnamefont {X.}~\bibnamefont
  {Feng}}\ and\ \bibinfo {author} {\bibfnamefont {A.}~\bibnamefont {Prohl}},\
  }\bibfield  {title} {\enquote {\bibinfo {title} {Numerical analysis of the
  {Allen-Cahn} equation and approximation for mean curvature flows},}\
  }\href@noop {} {\bibfield  {journal} {\bibinfo  {journal} {Num. Math.}\
  }\textbf {\bibinfo {volume} {94}},\ \bibinfo {pages} {33--65} (\bibinfo
  {year} {2003})}\BibitemShut {NoStop}%
\bibitem [{\citenamefont {Jokisaari}\ \emph {et~al.}(2017)\citenamefont
  {Jokisaari}, \citenamefont {Voorhees}, \citenamefont {Guyer}, \citenamefont
  {Warren},\ and\ \citenamefont {Heinonen}}]{Jokisaarietal_CMATS_2017}%
  \BibitemOpen
  \bibfield  {author} {\bibinfo {author} {\bibfnamefont {A.}~\bibnamefont
  {Jokisaari}}, \bibinfo {author} {\bibfnamefont {P.}~\bibnamefont {Voorhees}},
  \bibinfo {author} {\bibfnamefont {J.}~\bibnamefont {Guyer}}, \bibinfo
  {author} {\bibfnamefont {J.}~\bibnamefont {Warren}}, \ and\ \bibinfo {author}
  {\bibfnamefont {O.}~\bibnamefont {Heinonen}},\ }\bibfield  {title} {\enquote
  {\bibinfo {title} {Benchmark problems for numerical implementations of phase
  field models},}\ }\href {\doibase 10.1016/j.commatsci.2016.09.022} {\bibfield
   {journal} {\bibinfo  {journal} {Comput. Mater. Sci.}\ }\textbf {\bibinfo
  {volume} {126}},\ \bibinfo {pages} {139--151} (\bibinfo {year}
  {2017})}\BibitemShut {NoStop}%
\bibitem [{\citenamefont {R\"atz}\ and\ \citenamefont
  {Voigt}(2006)}]{RV_CMS_2006}%
  \BibitemOpen
  \bibfield  {author} {\bibinfo {author} {\bibfnamefont {A.}~\bibnamefont
  {R\"atz}}\ and\ \bibinfo {author} {\bibfnamefont {A.}~\bibnamefont {Voigt}},\
  }\bibfield  {title} {\enquote {\bibinfo {title} {{PDE}'s on surfaces - a
  diffuse interface approach},}\ }\href@noop {} {\bibfield  {journal} {\bibinfo
   {journal} {Commun. Math. Sci.}\ }\textbf {\bibinfo {volume} {4}},\ \bibinfo
  {pages} {575--590} (\bibinfo {year} {2006})}\BibitemShut {NoStop}%
\bibitem [{\citenamefont {Witkowski}, \citenamefont {Backofen},\ and\
  \citenamefont {Voigt}(2012)}]{WBV_PCCP_2012}%
  \BibitemOpen
  \bibfield  {author} {\bibinfo {author} {\bibfnamefont {T.}~\bibnamefont
  {Witkowski}}, \bibinfo {author} {\bibfnamefont {R.}~\bibnamefont {Backofen}},
  \ and\ \bibinfo {author} {\bibfnamefont {A.}~\bibnamefont {Voigt}},\
  }\bibfield  {title} {\enquote {\bibinfo {title} {The influence of membrane
  bound proteins on phase separation and coarsening in cell membranes},}\
  }\href@noop {} {\bibfield  {journal} {\bibinfo  {journal} {Phys. Chem. Chem.
  Phys.}\ }\textbf {\bibinfo {volume} {14}},\ \bibinfo {pages} {14509--14515}
  (\bibinfo {year} {2012})}\BibitemShut {NoStop}%
\bibitem [{\citenamefont {Gross}\ and\ \citenamefont
  {Atzberger}(2018)}]{Grossetal_JCP_2018}%
  \BibitemOpen
  \bibfield  {author} {\bibinfo {author} {\bibfnamefont {B.}~\bibnamefont
  {Gross}}\ and\ \bibinfo {author} {\bibfnamefont {P.~J.}\ \bibnamefont
  {Atzberger}},\ }\bibfield  {title} {\enquote {\bibinfo {title} {Hydrodynamic
  flows on curved surfaces: {S}pectral numerical methods for radial manifold
  shapes},}\ }\href@noop {} {\bibfield  {journal} {\bibinfo  {journal} {J.
  Comput. Phys.}\ }\textbf {\bibinfo {volume} {371}},\ \bibinfo {pages}
  {663--689} (\bibinfo {year} {2018})}\BibitemShut {NoStop}%
\bibitem [{\citenamefont {Krause}\ and\ \citenamefont
  {Voigt}(2022)}]{KV_arxiv_2022}%
  \BibitemOpen
  \bibfield  {author} {\bibinfo {author} {\bibfnamefont {V.}~\bibnamefont
  {Krause}}\ and\ \bibinfo {author} {\bibfnamefont {A.}~\bibnamefont {Voigt}},\
  }\bibfield  {title} {\enquote {\bibinfo {title} {A numerical approach for
  fluid deformable surfaces with conserved enclosed volume},}\ }\href {\doibase
  10.48550/ARXIV.2210.03585} {\bibfield  {journal} {\bibinfo  {journal}
  {arXiv}\ } (\bibinfo {year} {2022}),\ 10.48550/ARXIV.2210.03585}\BibitemShut
  {NoStop}%
\bibitem [{Note1()}]{Note1}%
  \BibitemOpen
  \bibinfo {note} {We note a typing error in Ref.~\protect \rev@citealpnum
  {Nitschkeetal_TFI_2017} which has been confirmed by the authors.}\BibitemShut
  {Stop}%
\bibitem [{\citenamefont {Wagner}\ and\ \citenamefont
  {Yeomany}(1998)}]{WY_PRL_1998}%
  \BibitemOpen
  \bibfield  {author} {\bibinfo {author} {\bibfnamefont {A.}~\bibnamefont
  {Wagner}}\ and\ \bibinfo {author} {\bibfnamefont {J.}~\bibnamefont
  {Yeomany}},\ }\bibfield  {title} {\enquote {\bibinfo {title} {Breakdown of
  scale invariance in the coarsening of phase-separating binary fluids},}\
  }\href@noop {} {\bibfield  {journal} {\bibinfo  {journal} {Phys. Rev. Lett.}\
  }\textbf {\bibinfo {volume} {80}},\ \bibinfo {pages} {1429} (\bibinfo {year}
  {1998})}\BibitemShut {NoStop}%
\bibitem [{\citenamefont {Wang}\ \emph {et~al.}(2022)\citenamefont {Wang},
  \citenamefont {Palzhanov}, \citenamefont {Quaini}, \citenamefont
  {Olshanskii},\ and\ \citenamefont {Majd}}]{WPQOM_BBA_2022}%
  \BibitemOpen
  \bibfield  {author} {\bibinfo {author} {\bibfnamefont {Y.}~\bibnamefont
  {Wang}}, \bibinfo {author} {\bibfnamefont {Y.}~\bibnamefont {Palzhanov}},
  \bibinfo {author} {\bibfnamefont {A.}~\bibnamefont {Quaini}}, \bibinfo
  {author} {\bibfnamefont {M.}~\bibnamefont {Olshanskii}}, \ and\ \bibinfo
  {author} {\bibfnamefont {S.}~\bibnamefont {Majd}},\ }\bibfield  {title}
  {\enquote {\bibinfo {title} {Lipid domain coarsening and fluidity in
  multicomponent lipid vesicles: A continuum based model and its experimental
  validation},}\ }\href@noop {} {\bibfield  {journal} {\bibinfo  {journal}
  {Biochimica et Biophysica Acta (BBA)-Biomembranes}\ }\textbf {\bibinfo
  {volume} {1864}},\ \bibinfo {pages} {183898} (\bibinfo {year}
  {2022})}\BibitemShut {NoStop}%
\bibitem [{\citenamefont {Henry}\ and\ \citenamefont
  {Tegze}(2019)}]{HG_PRE_2019}%
  \BibitemOpen
  \bibfield  {author} {\bibinfo {author} {\bibfnamefont {H.}~\bibnamefont
  {Henry}}\ and\ \bibinfo {author} {\bibfnamefont {G.}~\bibnamefont {Tegze}},\
  }\bibfield  {title} {\enquote {\bibinfo {title} {Kinetics of coarsening have
  dramatic effects on the microstructure: Self-similarity breakdown induced by
  viscosity contrast},}\ }\href@noop {} {\bibfield  {journal} {\bibinfo
  {journal} {Phys. Rev. E}\ }\textbf {\bibinfo {volume} {100}},\ \bibinfo
  {pages} {013116} (\bibinfo {year} {2019})}\BibitemShut {NoStop}%
\bibitem [{\citenamefont {Andrews}\ \emph {et~al.}(2020)\citenamefont
  {Andrews}, \citenamefont {Elder}, \citenamefont {Voorhees},\ and\
  \citenamefont {Thornton}}]{AEVT_PRM_2020}%
  \BibitemOpen
  \bibfield  {author} {\bibinfo {author} {\bibfnamefont {W.~B.}\ \bibnamefont
  {Andrews}}, \bibinfo {author} {\bibfnamefont {K.~L.~M.}\ \bibnamefont
  {Elder}}, \bibinfo {author} {\bibfnamefont {P.~W.}\ \bibnamefont {Voorhees}},
  \ and\ \bibinfo {author} {\bibfnamefont {K.}~\bibnamefont {Thornton}},\
  }\bibfield  {title} {\enquote {\bibinfo {title} {Effect of transport
  mechanism on the coarsening of bicontinuous structures: A comparison between
  bulk and surface diffusion},}\ }\href@noop {} {\bibfield  {journal} {\bibinfo
   {journal} {Phys. Rev. Mat.}\ }\textbf {\bibinfo {volume} {4}},\ \bibinfo
  {pages} {013401} (\bibinfo {year} {2020})}\BibitemShut {NoStop}%
\bibitem [{\citenamefont {Fife}\ and\ \citenamefont
  {Penrose}(1995)}]{Fifeetal_EJDE_1995}%
  \BibitemOpen
  \bibfield  {author} {\bibinfo {author} {\bibfnamefont {P.~C.}\ \bibnamefont
  {Fife}}\ and\ \bibinfo {author} {\bibfnamefont {O.}~\bibnamefont {Penrose}},\
  }\bibfield  {title} {\enquote {\bibinfo {title} {Interfacial dynamics for
  thermodynamically consistent phasefield models with nonconserved order
  parameter},}\ }\href@noop {} {\bibfield  {journal} {\bibinfo  {journal}
  {Electr. J. Diff. Eq.}\ }\textbf {\bibinfo {volume} {16}},\ \bibinfo {pages}
  {1--49} (\bibinfo {year} {1995})}\BibitemShut {NoStop}%
\bibitem [{\citenamefont {Benes}\ \emph {et~al.}(2023)\citenamefont {Benes},
  \citenamefont {Kolar}, \citenamefont {Sischka},\ and\ \citenamefont
  {Voigt}}]{BKSV_arXiv_2023}%
  \BibitemOpen
  \bibfield  {author} {\bibinfo {author} {\bibfnamefont {M.}~\bibnamefont
  {Benes}}, \bibinfo {author} {\bibfnamefont {M.}~\bibnamefont {Kolar}},
  \bibinfo {author} {\bibfnamefont {J.~M.}\ \bibnamefont {Sischka}}, \ and\
  \bibinfo {author} {\bibfnamefont {A.}~\bibnamefont {Voigt}},\ }\bibfield
  {title} {\enquote {\bibinfo {title} {Degenerate area preserving surface
  allen-cahn equation and its sharp interface limit},}\ }\href@noop {}
  {\bibfield  {journal} {\bibinfo  {journal} {arXiv}\ } (\bibinfo {year}
  {2023})}\BibitemShut {NoStop}%
\bibitem [{\citenamefont {Al-Izzi}\ and\ \citenamefont
  {Morris}(2021)}]{Al-Izzietal_SCDB_2021}%
  \BibitemOpen
  \bibfield  {author} {\bibinfo {author} {\bibfnamefont {S.~C.}\ \bibnamefont
  {Al-Izzi}}\ and\ \bibinfo {author} {\bibfnamefont {R.~G.}\ \bibnamefont
  {Morris}},\ }\bibfield  {title} {\enquote {\bibinfo {title} {Active flows and
  deformable surfaces in development},}\ }\href@noop {} {\bibfield  {journal}
  {\bibinfo  {journal} {Seminars in Cell \& Developmental Biology}\ }\textbf
  {\bibinfo {volume} {120}},\ \bibinfo {pages} {44--52} (\bibinfo {year}
  {2021})}\BibitemShut {NoStop}%
\bibitem [{\citenamefont {Hoffmann}\ \emph {et~al.}(2022)\citenamefont
  {Hoffmann}, \citenamefont {Carenza}, \citenamefont {Eckert},\ and\
  \citenamefont {Giomi}}]{Hoffmannetal_SA_2022}%
  \BibitemOpen
  \bibfield  {author} {\bibinfo {author} {\bibfnamefont {L.~A.}\ \bibnamefont
  {Hoffmann}}, \bibinfo {author} {\bibfnamefont {L.~N.}\ \bibnamefont
  {Carenza}}, \bibinfo {author} {\bibfnamefont {J.}~\bibnamefont {Eckert}}, \
  and\ \bibinfo {author} {\bibfnamefont {L.}~\bibnamefont {Giomi}},\ }\bibfield
   {title} {\enquote {\bibinfo {title} {{Theory of defect-mediated
  morphogenesis}},}\ }\href@noop {} {\bibfield  {journal} {\bibinfo  {journal}
  {{Sci. Adv.}}\ }\textbf {\bibinfo {volume} {{8}}},\ \bibinfo {pages}
  {{eabk2712}} (\bibinfo {year} {{2022}})}\BibitemShut {NoStop}%
\bibitem [{\citenamefont {Nitschke}\ and\ \citenamefont
  {Voigt}(2022)}]{NitschkeVoigt_JoGaP_2022}%
  \BibitemOpen
  \bibfield  {author} {\bibinfo {author} {\bibfnamefont {I.}~\bibnamefont
  {Nitschke}}\ and\ \bibinfo {author} {\bibfnamefont {A.}~\bibnamefont
  {Voigt}},\ }\bibfield  {title} {\enquote {\bibinfo {title}
  {Observer-invariant time derivatives on moving surfaces},}\ }\href@noop {}
  {\bibfield  {journal} {\bibinfo  {journal} {J. Geom. Phys.}\ }\textbf
  {\bibinfo {volume} {173}},\ \bibinfo {pages} {104428} (\bibinfo {year}
  {2022})}\BibitemShut {NoStop}%
\bibitem [{\citenamefont {Nitschke}, \citenamefont {Sadik},\ and\ \citenamefont
  {Voigt}(2022)}]{NitschkeSadikVoigt_A_2022}%
  \BibitemOpen
  \bibfield  {author} {\bibinfo {author} {\bibfnamefont {I.}~\bibnamefont
  {Nitschke}}, \bibinfo {author} {\bibfnamefont {S.}~\bibnamefont {Sadik}}, \
  and\ \bibinfo {author} {\bibfnamefont {A.}~\bibnamefont {Voigt}},\ }\bibfield
   {title} {\enquote {\bibinfo {title} {Tangential tensor fields on deformable
  surfaces -- how to derive consistent {$L^2$}-gradient flows},}\ }\href
  {\doibase 10.48550/arXiv.2209.13272} {\bibfield  {journal} {\bibinfo
  {journal} {arXiv}\ } (\bibinfo {year} {2022}),\
  10.48550/arXiv.2209.13272}\BibitemShut {NoStop}%
\end{thebibliography}%

\end{document}